\documentclass[aps,showpacs,twocolumn,superscriptaddress]{revtex4-1}

\usepackage{dcolumn}% Align table columns on decimal point
\usepackage{appendix}
\usepackage{bm}% bold math
\usepackage{color}
\usepackage{array}
\usepackage{mathrsfs}
\usepackage{ulem} % sout
\usepackage{amsmath}
\usepackage{amssymb}
\usepackage{url}
\usepackage{epsfig}

\newcommand{\orcid}[1]{\href{https://orcid.org/#1}{\includegraphics[width=8pt]{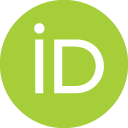}}}

\begin{document}
	
	\title{Spin liquid in twisted homobilayers group-VI gichalcogenides}

	\author{Mohammad-Hossein Zare \orcid{0000-0003-4638-7987}}
	\email{zare@qut.ac.ir}
	\affiliation{Department of Physics, Qom University of Technology, Qom 37181-46645, Iran}
	
	\author{Hamid Mosadeq\orcid{0000-0001-8150-0821}}
	\affiliation{Department of physics, Faculty of Science, Shahrekord University, Shahrekord 88186-34141, Iran }

	\date{\today}
	
	\begin{abstract}
     Twisted transition metal dichalcogenide (TMD) homobilayers have recently emerged as a powerful platform for studying correlated insulating states.
     In the strongly correlated limit, we construct an effective spin Hamiltonian on a honeycomb lattice that includes the Heisenberg interaction and nonsymmetric interactions such as a Dzyaloshinskii-Moriya interaction and a Kane-Mele coupling for the Mott-insulating phase at half-filling. 
     For the twisted TMD homobilayers, 
     the spin-orbit coupling in the Hubbard model, which is expected to induce the antisymmetric exchange couplings in the effective spin Hamiltonian, is a highly tunable and experimentally accessible quantity that can be tuned by an applied electric field.
     In this study, we investigate classical and quantum phase diagrams of the effective spin Hamiltonian using analytical and numerical methods.
     We show that the model exhibits a rich classical phase diagram including an antiferromagnetic (AFM) phase, a planar spiral ordered phase with high classical degeneracy, a $z$-AFM phase, a noncoplanar phase, a noncollinear phase, and a 120$^{\circ}$-AFM  phase.
     In the quantum treatment, we calculate low-energy magnon excitation spectrum, ground state energy, and static spin structure factor using linear spin-wave theory and density matrix renormalization group methods to compose the quantum phase diagram of the effective spin Hamiltonian.
    Beyond the Heisenberg interaction, we find that the existence of these antisymmetric couplings is responsible for the quantum spin liquid, $z$-AFM, noncoplanar, and 120$^{\circ}$ phases. 
     Twisted TMD homobilayers, therefore, offer rich platforms for realizing rich phases of matter such as quantum spin liquid, noncoplanar, and 120$^{\circ}$, resulting from the spin-orbit coupling.
    \end{abstract}
	

	\maketitle

	\section{Introduction}
	Correlated quantum spin states, described by the term quantum spin liquid, are an exotic state of matter treating key features such as long-range entanglement, intrinsic topological order, and fractional quasiparticle excitations~\cite{ANDERSON_1973,Lee_2008,Balents_2010,Savary_2016,Zhou_2017,Knolle_2019,Takagi_2019} in spin systems wherein interacting spins refuse to order even at absolute zero temperature.           %

	Long-distance entanglement in the ground state of such systems leads to the emergence of excitations that carry fractional quantum numbers. 
	These fractionalized excitations are non-Abelian Majorana states that support the quantum spin liquids (QSLs) as desirable practical quasiparticles in the field of quantum information and topological quantum computation~\cite{Kitaev_2003,Kitaev_2006,Jiang_2012}.

	Aside from the fascinating physics of QSLs, according to the resonating valence bond (RVB) theory of Anderson, QSLs are the parent state for high-temperature unconventional superconductors~\cite{Anderson_1987}. 
	A variety of quantum materials are proposed to host this nontrivial collective phase due to extreme quantum fluctuations where frustration induces a macroscopically-degenerate ground-state at the classical level.    
	It should be noted that QSLs originating from geometrical frustration in triangular, kagome, and pyrochlore lattices or quantum spin models with frustration can generate a macroscopic ground state degeneracy resulting in strong quantum fluctuations.
	 These frustration driven spin configurations behave as liquid and do not exhibit long-range magnetic order
	 % but remain liquid-like
	 even at absolute zero of temperature.

	The promising QSL candidates can be classified into two categories: (i) geometrically frustrated materials including layered compounds of the two-dimensional systems such as organic salts~\cite{Yamashita_2008a,Yamashita_2008,Yamashita_2010,Yamashita_2011,Powell_2011} and YbMgGaO$_4$~\cite{Li_2015,Li_2015a,Paddison_2016,Shen_2016,Li_2016,Li_2017,Li_2017a} with an underlying triangular lattice, and material herbertsmithite an underlying kagome lattice~\cite{Shores_2005,Bert_2007,Olariu_2008,Vries_2009,Mendels_2010,Helton_2010,Han_2012,Jeschke_2013,Pilon_2013}. 
	(ii) honeycomb lattice Kitaev materials in which bond-dependent exchange interactions between spins induce strong quantum fluctuations and frustrate magnetic.           
	Kitaev-QSLs may exist in spin-orbit coupled Mott insulators due to the interplay between spin-orbit coupling and three-fold rotational symmetry of a honeycomb lattice giving rise to the bond dependent Ising-type interactions (Kitaev-type exchange interaction) between nearest-neighbors~\cite{Witczak_2014}.
	It is worth noting that these Kitaev materials with dominant bond-dependent interactions exhibit a long-range magnetic order at low temperatures due to additional spin exchange interactions~\cite{Rau_2016,Winter_2017,Trebst_2017}.
	Currently, the search for Kitaev-QSLs has been focused on heavy 4$d$ and 5$d$ transition metal compounds with partially filled $d$-orbitals such as H$_3$LiIr$_2$O$_6$~\cite{Kitagawa_2018,Yadav_2018,Li_2018,Knolle_2019a}, Na$_2$IrO$_3$, $\alpha$-Li$_2$IrO$_3$ and $\alpha$-RuCl$_3$~\cite{HwanChun_2015,Rau_2016,Winter_2017} due to small ordering temperature and uncommon magnetic excitation spectrum experimentally observed	~\cite{HwanChun_2015,Takayama_2015,Sandilands_2015,Banerjee_2016,Cao_2016,Banerjee_2017,Revelli_2019}.
	Physics of the Kitaev materials under application of external factors such as strain~\cite{Takayama_2015,Breznay_2017,Wang_2018,Clancy_2018,Yadav_2018a,Simutis_2018} and magnetic field~\cite{Revelli_2019,Yadav_2016,Sears_2017,Kasahara_2018,Balz_2019} is especially interesting owing to  tuning exchange interactions and suppressing of long-range magnetic order.      

	Twisted bilayer systems are van der Waals heterostructures of semiconductors or semimetals~\cite{Hunt_2013,Dean_2013,Wang_2015,Spanton_2018},
	which have attracted tremendous attention both from fundamental and applied research point of view in physics ~\cite{Lopes_2007,Li_2009,Bistritzer_2010,Mele_2010,Luican_2011,Lopes_2012,Jung_2014,Wong_2015,Kim_2017,Nam_2017,Efimkin_2018}. 
	Recent experimental observations on magic angle twisted bilayer graphene have revealed a dissipationless electron current, wherein the electron density can be tuned by a gate voltage ~\cite{Cao_2018,Cao_2018a}.
	Theoretical analysis shows the existence of nearly flat bands in the magic angle twisted bilayer graphene that can be completely isolated from the rest of the spectrum~\cite{Bistritzer_2010}.
	The Mott insulator behavior for the twisted-bilayer graphene is experimentally demonstrated for filling $1/4$ or $3/4$ at low temperatures~\cite{Cao_2018}.
	Inducing the charge carriers in the twisted-bilayer graphene via the gate voltage produces an unconventional superconducting phase at low temperatures ~\cite{Cao_2018a,Xu_2018,Isobe_2018,Wu_2018,You_2019,Roy_2019}.
	These facts indicate the possibility for engineering interesting strong correlation physics in the twisted bilayer graphene by tuning the band dispersion.

	Apart from the twisted bilayer graphene, theoretical studies show that the twisted bilayers of group-VI transition metal dichalcogenides (TMDs) is a promising platform to realize the flat bands at small twist angles for both heterobilayers~\cite{Ruiz_2019,Wu_2019a} and homobilayers~\cite{Wu_2019,Naik_2018}.
	In contrast to the twisted bilayer graphene in which the flat bands appear only within a narrow window ($\pm0.1^{\circ}$) around twist angle $1.1^{\circ}$, in TMDs the flat bands occur in a larger range of twist angles in the twisted TMDs.
	Most recently, the experimental discovery of correlated insulating states and superconductivity in heterobilayers~\cite{Tang_2020,Regan_2020} and -homobilayers~\cite{Zhang_2020,Wang_2020} TMDs promotes them to promising playground to realize intriguing correlated quantum phases~\cite{Wu_2019a,Wu_2019}.

	The aim of this research is to present analytical and numerical investigations on the zero-temperature magnetic phase diagram of the TMD homobilayers in the strong correlation limit. 
	To this end, we pay special attention to the large-spin limit and study the effect of quantum corrections on the classical phase diagram.         
%

%%%%%%%%%%%%%%%%%====== figure ======%%%%%%%%%%%%%%%%%
\begin{figure}[t]
	\includegraphics[trim=0cm 0cm 0cm 0cm, clip, scale=0.32]{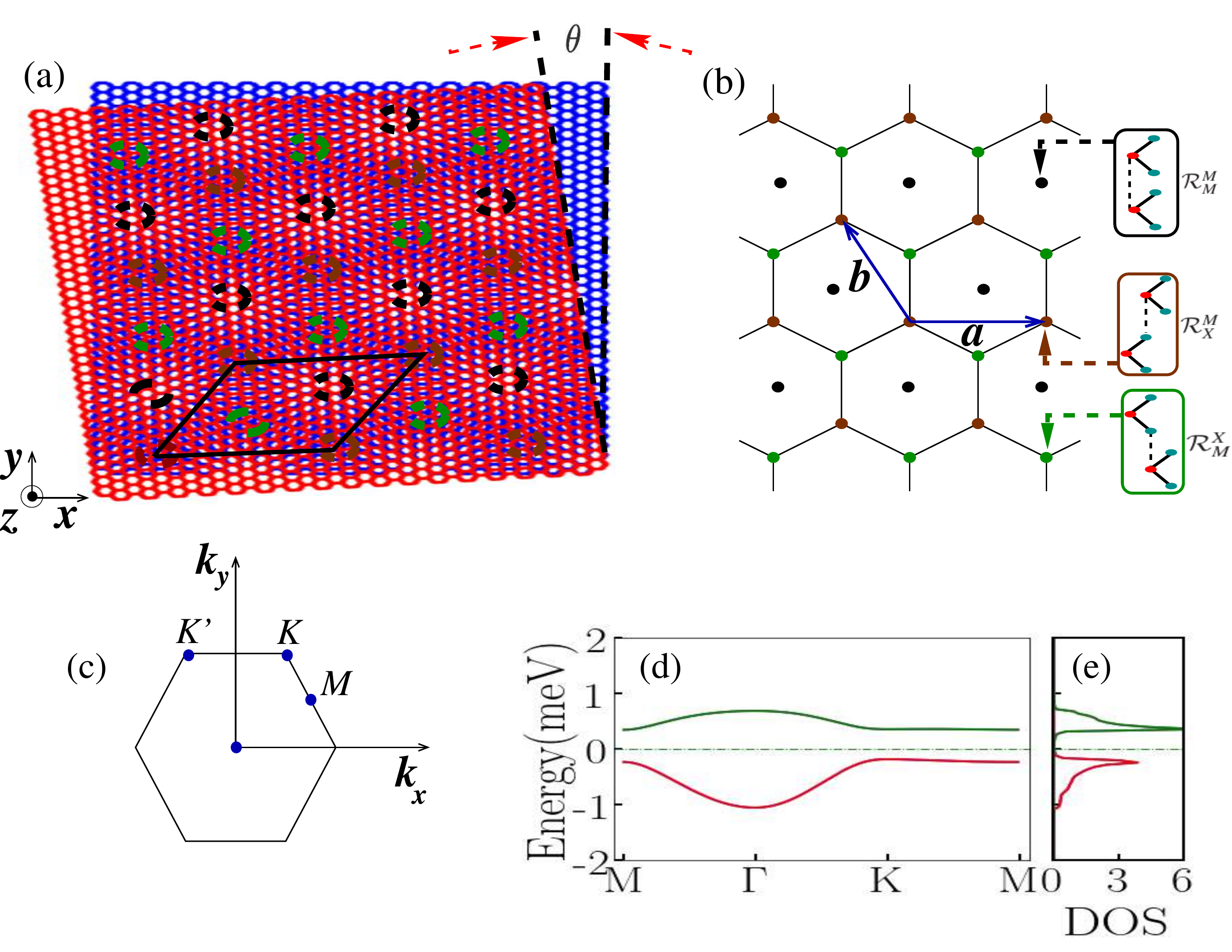}
	\caption{ (a) Schematic representation of twisted TMD homobilayers with a small twist angle $\theta$ wherein the high-symmetry sites are marked by dashed-circles.
	(b) The effective lattice model which results from $\mathcal{R}_{M}^{X}$ and $\mathcal{R}_{X}^{M}$ as effective sites form the real space honeycomb lattice with two basis vectors $\mathbf{a}$ and $\mathbf{b}$.
	 (c)The first Brillouin zone and the high symmetry points of two-dimensional honeycomb lattice
	  (d) Moir\'e band structure and (e) DOS of the twisted TMD homobilayer obtained by using the effective tight-binding model in Eq.~(\ref{eq:1}) where the hopping integral parameters are chosen as $t_1\approx 0.29$~meV and $t_2\approx
      0.06$~meV~\cite{Wu_2019}.
  }
	\label{fig:1}
\end{figure}		
	
	\section{ model Hamiltonian}
	 Twisted homobilayers between binary van der Walls semiconducting TMDs form a moir\'e superlattice for small twist angles ($\theta$) near 0$^{\circ}$ and 180$^{\circ}$~\cite{Wu_2019}.
	Here, we consider a homobilayer with a small twist angle close to 0$^{\circ}$ ($\theta>0$). A schematic representation of this moir\'e pattern is depicted in Fig.~\ref{fig:1}(a). 
	The moir\'e superlattice is obtained by a slight counterclockwise rotation of the top layer around the $z$-axis with respect to the bottom layer.
	As shown in Fig.~\ref{fig:1}(a), in each moir\'e unit cell of this AA stacking, there are three types of high-symmetry points:     
    $\mathcal{R}_{M}^{X}$,~$\mathcal{R}_{X}^{M}$ and $\mathcal{R}_{M}^{M}$, where $X$ and $M$ indicate chalcogen and transition metal atoms, respectively.
	Here, we define $\mathcal{R}_{\beta}^{\alpha}$ at which the $\alpha$ atoms in the top layer are vertically aligned with the $\beta$ atoms in the bottom layer.    
	Recently, two experimental studies report signatures of nearly flat bands with narrow bandwidths in twisted WSe$_2$/WSe$_2$ homobilayers~\cite{Zhang_2020,Wang_2020}.
	Theoretical studies illustrate that these isolated flat moir\'e bands can be described by effective tight-binding models on triangular or honeycomb lattices depending on the electron density distribution in real-space~\cite{Wu_2019,Pan_2020}.
	This finding shows a good agreement with the interesting experimental observations~\cite{Zhang_2020,Wang_2020}.
	The one orbital tight-binding model on the triangular lattice is proposed in the case where large electron densities are located around the $\mathcal{R}_{M}^{M}$ sites. While the effective honeycomb lattice model can be considered in the case where the real-space distribution of the electron density is concentrated close to the  $\mathcal{R}_{M}^{X}$/$\mathcal{R}_{X}^{M}$ positions [Fig.~\ref{fig:1} (b)].     

    Here, we focus on the corresponding tight-binding model 
    of the moir\'e honeycomb lattice by including two topological flat bands which show good consistency with the moir\'e bands obtained by using a low energy continuum model~\cite{Wu_2019,Pan_2020}.	
	Two topological moir\'e valence bands with opposite Chern numbers in TMD homobilayers with very small twist angles can be described with a generalized Kane-Mele (KM) model~\cite{Kane_2005} with an extra effective gauge potential due to the momentum shift between the twisted layers~\cite{Wu_2019,devakul_2021} as follows:
	
	%---------------------------------------------------------------------------------------------------
	\begin{align}
		\begin{aligned}
			&
			{\cal H}_{k} \!=\!
			t{_1} \sum_{\langle ij\rangle,\alpha} 
			c^{\dagger}_{i,\alpha}c^{}_{j,\alpha} 
			+ 
			t{_2} \sum_{\langle\langle ij\rangle\rangle,\alpha}
			e^{ \frac{2\pi {\rm i} }{3}\nu_{ij}}
			c^{\dagger}_{i,\alpha} c^{}_{j,\alpha}.
			\label{eq:1}
		\end{aligned}
	\end{align}
	%---------------------------------------------------------------------------------------------------
	%
	where $c^{\dagger}_{i,\alpha}$ ($c_{i\alpha}$) stands for the creation (annihilation) operator of an electron with spin-$\alpha$ ($\alpha=\uparrow,\downarrow$) on site $i$. The first term represents the nearest-neighbor hopping with amplitude $t_1$ associated with interlayer hopping of the twisted homobilayer between orbitals concentrated near $\mathcal{R}_{M}^{X}$/$\mathcal{R}_{X}^{M}$ positions. 
	The second term captures intralayer bond-dependent hopping with $\nu_{ij}=(\hat{\mathbf{d}}_{1}\times\hat{\mathbf{d}}_{2}){_z}$, in which the unit vectors 
	$\hat{\mathbf{d}}_{1}$ and $\hat{\mathbf{d}}_{2}$ connecting the next-nearest neighbors on the honeycomb lattice.
	The complex second nearest neighbor hopping term can  effectively induce a spin-dependent hopping integral, which causes the tight-binding model to contain effective spin-orbit couplings.
	In other words, the moir\'e superlattices in van der Waals heterostructures are a benchmark for studying and controlling of the spin-orbit coupling due to spin-valley locking ~\cite{Bistritzer_2010,Po_2018,Zhang_2019}.
	In Figs.~\ref{fig:1}(d) and ~\ref{fig:1}(e), we show the two valence bands and density of state (DOS) have been obtained from the effective model in Eq.~(\ref{eq:1}), with the tight-binding parameters as $t_1\approx 0.29$ meV and $t_2\approx 0.06$ meV, respectively. 
	The van Hove singularities in the density of states originate from saddle-points in the dispersion which give rise to an enhancement of interaction effects.
	Therefore, many-body effects for electrons in the flat bands are remarkably increased since the kinetic energy is strongly decreased under the suppressed bandwidth condition in the moir\'e bands.
	It should be emphasized that the two flat bands with extremely small bandwidth ($W<1$) exhibit relatively remarkable band dispersions by increasing the twist angle~\cite{devakul_2021}.
	The ratio of interaction strength to bandwidth, $U/W$, of the moi\'re homobilayers is well modified through varying the twist angle and using a three-dimensional dielectric environment~\cite{Cao_2018,Yankowitz_2019,Pan_2020,devakul_2021}. 
	Wu \textit{et al.,} have shown that these moi\'re homobilayers with integer numbers of holes per moi\'re unit cell can be a two-dimensional platform for studying exotic insulator phases such as ferromagnetic insulating state, quantum spin Hall insulator, antiferromagnetic insulator, and fractional topological insulators~\cite{Wu_2019}. 
	Here, in order to study the interaction effects on the twisted TMD homobilayer by a small angle, we only consider on-site repulsive interaction as follows:
	%---------------------------------------------------------------------------------------------------
	\begin{equation}
		{\cal H}_U=  U_{}\sum_{i} n_{i\uparrow} n_{i\downarrow}
		\label{eq:2}
	\end{equation}
	%---------------------------------------------------------------------------------------------------
	%
	where, $n_{i\alpha} = c^{\dagger}_{i\alpha} c_{i\alpha}$ is a number operator
	and $U$ is the strength of the on-site Coulomb interaction.
	Thus, the Hamiltonian for describing the interacting electrons in moi\'re homobilayers, ${\cal H}={\cal H}_k+{\cal H}_U$, is a generalized Kane-Mele-Hubbard model. 

	For these systems with the isolated flat bands, the interaction strength can be much larger than the hopping integrals $t_1$ and $t_2$ in Eq.~(\ref{eq:1}). 
	In the twisted TMD homobilayers, the on-site Coulomb interaction strength is at least one order-of-magnitude greater than the hopping parameters for the small twist angles between layers~\cite{Wu_2019,Pan_2020,devakul_2021}.
	Due to the suppression of charge fluctuations in the limit of the strong on-site Coulomb repulsion, we try to obtain the effective spin Hamiltonian using second- order perturbation theory.
	The perturbation to second-order in $t_1/U$ ($t_2/U$), we find an effective spin Hamiltonian for the half-filling case including the Kane-Mele Heisenberg~\cite{Zare_2013,Zare_2014} and Dzyaloshinskii-Moriya (DM) terms as follows:
	%---------------------------------------------------------------------------------------------------
	\begin{align}
		\begin{aligned}
			{\cal H}\!
			=
			&
			\!
			\sum_{\langle ij \rangle}
			J_1\mathbf{S}_i\cdot\mathbf{S}_j
			+\sum_{\langle\langle ij \rangle\rangle}
			\!\Big[J_2\mathbf{S}_i\cdot\mathbf{S}_j
			\\
			&
			-
			g_2(S^{x}_{i}S^{x}_{j} + S^{y}_{i}S^{y}_{j} - S^{z}_{i}S^{z}_{j})
			-{D}\nu_{ij}(\mathbf{S}_i\times\mathbf{S}_j)_z
			\Big],
			\label{eq:3}
		\end{aligned}
	\end{align}
	%---------------------------------------------------------------------------------------------------
	%
	where the first two terms are isotropic exchange interactions between nearest- and next-nearest-neighbors with amplitudes of
	$J_1=4t^2_1/U$ and $J_2=t^2_2/U$, respectively. In addition, the XXZ and antisymmetric DM exchanges between the next-nearest-neighbors originate from the spin-dependent hopping integral with the magnitudes of 
	$g_2=3J_2$ and $D=2\sqrt{3}J_2$.
	It clearly shows that the KM exchange interaction favors in-plane ferromagnetic order and supports antiferromagnetic order in the ${z}$ direction.
	Since the DM exchange interaction is suppressed when spins are in the same direction, this interaction favors a noncollinear (NCL) magnetic phase, which may lead to the stability of the triplet-pairing correlation in the twisted TMD homobilayers.

   \section{Classical phase diagram}  
	
	\subsection{Luttinger-Tisza method}
	To obtain the classical phase diagram of the generic model Hamiltonian as given in Eq.~(\ref{eq:3}), we first use the Luttinger-Tisza (LT) approximation~\cite{Luttinger_1946,Litvin_1974}. 
	Here, instead of using the constraint of fixed spin length at each site, we apply the weak constraint,
	
    \begin{equation}
    \sum_{i} |S_i|^2~=~NS^2	
    \label{eq:wc}
    \end{equation}
	where $N$ is the total number of lattice points. 
	Performing Fourier transform on Eq.~(\ref{eq:3}) and diagonalizing of its matrix representation gives the stable magnetic configurations for the different constant couplings, as will be discussed in more detail below.

	A honeycomb lattice with two sites per unit cell, $\nu \in \{1,2\}$, can be considered as two interpenetrating triangular Bravais lattices with primitive translational  
	vectors $\mathbf{a}=\hat{\mathbf{x}}$ and $\mathbf{b}=-1/2~\hat{\mathbf{x}}+{\sqrt{3}}/{2}~\hat{\mathbf{y}}$ [Fig.~\ref{fig:1}(b)].
	The Fourier transforms of the spins on each of the sublattices $\nu$ are defined by:
	
	\begin{equation}
		\begin{split}
			S^{\nu}_{\rho}(\mathbf{r}_ i) = \frac{1}{\sqrt{N/2}}\sum_{\mathbf{k}} {S}^{\nu}_{\mathbf{k},{\rho}} e^{i\mathbf{k} \cdot \mathbf{r}_i},
			\label{eq:4}
		\end{split}
	\end{equation}
	where ${\rho}\in\{x,y,z\}$ corresponds to the component of the spin vector and $ N/2 $ is the number of primitive cells. The summation $\sum_{\mathbf{k}}$ is taken on the first Brillouin zone (FBZ) [Fig.~\ref{fig:1}(c)].
	Rewriting the effective spin Hamiltonian (Eq.~\ref{eq:3}) in terms of ${S}^{\nu}_{\mathbf{k},\rho}$, results in

	\begin{equation}
		{\cal H} = \sum_{\mathbf{k}} \sum_{\alpha,\beta \in \{\nu\}} {\tilde{S}}_{-{\mathbf{k}}} ^{\alpha\rm T} {\cal J}_{\mathbf{k}} ^{\alpha\beta} \tilde{S}_{\mathbf{k}} ^{\beta},
		\label{eq:5}
	\end{equation}
	in which 
	$\tilde{S}^{\beta}_{\mathbf{k}} = 
	\begin{pmatrix}S_{{\mathbf{k}},x}^{\beta} \ S_{{\mathbf{k}},y}^{\beta} \ S_{{\mathbf{k}},z}^{\beta} \end{pmatrix}^{\rm T}$,
	Moreover, ${\cal J}_{\mathbf{k}} ^{\alpha\beta}$ is a $6\times6$ matrix whose elements are obtained from the FT of the exchange interaction terms. It has the form of

	\begin{figure}[t]
	\includegraphics[  scale=0.75]{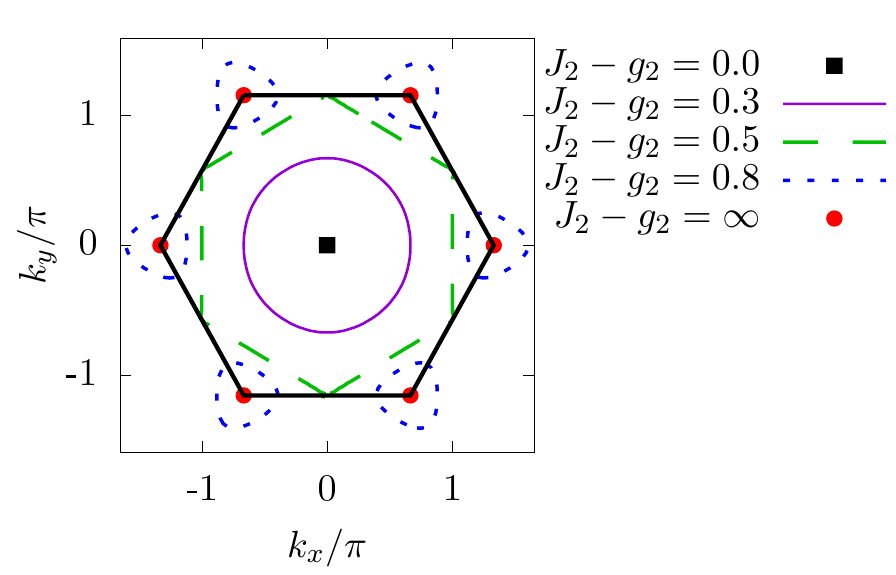}
	\vspace {-0.25cm}
	\caption{ (Color online) Degenerate manifolds of spin spirals in momentum space for different value of $J_2-g_2$.
		The  outer black hexagonal denotes the first Brillouin zone.}
	\label{fig:2}
\end{figure}
	
	\begin{equation}
		{\cal J}_{\mathbf{k}} = \begin{pmatrix}
			\xi_{\mathbf{k}}  &  \gamma_{\mathbf{k}}  \\
			\gamma^{*}_{\mathbf{k}}  &  \xi^{*}_{\mathbf{k}}
			\label{eq:6},
		\end{pmatrix}
	\end{equation}
	where
	\begin{align}
		\begin{aligned}
			&
			\xi_{\mathbf{k}} = \begin{pmatrix} 
				{\cal B}_{\mathbf{k}} & {\cal F}_{\mathbf{k}} & 0 \\ 
				{\cal F}^{*}_{\mathbf{k}} & {\cal B}_{\mathbf{k}} & 0 \\ 
				0 & 0 & {\cal C}_{\mathbf{k}}  \\
			\end{pmatrix},
			~~~~~~
			\gamma_{\mathbf{k}} = \begin{pmatrix}
				{\cal A}_{\mathbf{k}} & 0 & 0 \\ 
				0 & {\cal A}_{\mathbf{k}} & 0 \\ 
				0 & 0 & {\cal A}_{\mathbf{k}}  \\
			\end{pmatrix},
			\label{eq:7}
		\end{aligned}    
	\end{align}
	the elements of matrices $\xi_{\mathbf{k}}$ and $\gamma_{\mathbf{k}}$ are given by
	\begin{align}
		\begin{aligned}
			&
			{\cal A}_{\mathbf{k}}= J_1 [ 1 \!+\! \exp({ik_b}) +\exp({i(k_a+k_b)}) ]  \\ &
			{\cal B}_{\mathbf{q}} = 2(J_{2}\!-\!g_{2}) [ \cos k_{a} \!+\! \cos k_{b} \!+\! \cos(k_{a} \!+\! k_{b}) ] \\ &
			{\cal C}_{\mathbf{q}} = 2(J_{2}\!+\!g_{2})[\cos k_{a}+\cos k_{b}+\cos(k_{a} + k_{b})]  \\ &
			{\cal F}_{\mathbf{k}} = -2i D[\sin k_{a}+\sin k_{b}+\sin(k_{a} + k_{b})].
			\label{eq:8}
		\end{aligned}    
	\end{align}
	with $ k_a=\mathbf{k}\cdot\mathbf{a}  $ and $k_b=\mathbf{k}\cdot\mathbf{b}$.
	Obviously, the interaction matrix, ${\cal J}_{\mathbf{k}}$,
	is Hermitian and has real eigenvalues.
	The normalized eigenmodes of ${\cal J}_{\mathbf{k}}$ generate a $6\times 6$ unitary matrix. Within this orthonormal basis, Eq.~(\ref{eq:5}) can be written as
	\begin{equation}
		{\cal H}=\sum_{\mathbf{k}} \sum^{6}_{\mu=1} 
		\lambda_{\mathbf{k}}^{\mu} |\mathbf{S}_{\mathbf{k}}^{'\mu}|^{2}
		\label{eq:9}
	\end{equation}
	in which $\lambda_{\mathbf{k}}^{\mu}$ represents the $\mu$-th eigenvalue of ${\cal J}_{\mathbf{k}}$ and the corresponding eigenvector satisfies the relation 
	\begin{equation}
		{\cal J}_{\mathbf{k}} w^{\mu}_{\mathbf{k}} =\lambda_{\mathbf{k}}^{\mu}w^{\mu}_{\mathbf{k}},
		\label{eq:10}
	\end{equation} 
	Furthermore, $\mathbf{S}_{\mathbf{k}}^{'\mu}$ is known as spin structure factor, which is defined by
	\begin{equation}
		\mathbf{S}_{\mathbf{k}}^{'\mu}= w^{\mu}_{\mathbf{k}}\tilde{\mathbf{S}}_{\mathbf{k}}.
		\label{eq:11}
	\end{equation}

	Within the LT method, to obtain the magnetic ground state of the Hamiltonian Eq.~(\ref{eq:3}), we need to find a global minimum $\lambda_{0}$.
	Using the weak constraint in Fourier space, the classical energy (Eq.~\ref{eq:9}) can be reexpressed
	as 
	\begin{equation}
		{\cal H}=N\lambda_{0} + \sum_{\mathbf{k}} \sum_{\mu\neq0} 
		(\lambda_{\mathbf{k}}^{\mu}-\lambda_{0}) |\mathbf{S}_{\mathbf{k}}^{'\mu}|^{2}.
		\label{eq:12}
	\end{equation}
	
	To minimize this classical energy, the second term in Eq.~(\ref{eq:12}) should be equal to zero because
	$(\lambda^{\mu}_{\mathbf{k}}-\lambda_{0})>0$.
	For this purpose, the coefficients $\mathbf{S}_{\mathbf{k}}^{'\mu}$ with $\mu\neq0$ must be eliminated so that these generic conditions allow us to realize the possible ground state spin configurations. 
	Here, for the sake of simplicity we assume $J_1=1$.

 \begin{figure}[t]
	\includegraphics[width=1.\linewidth]{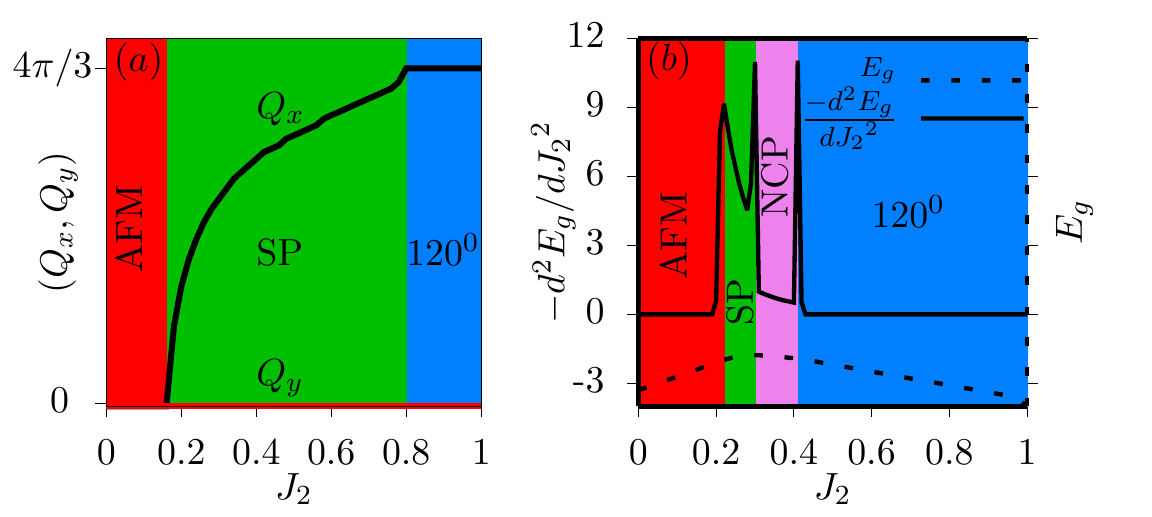}
	\vspace{-0.5cm}
	\caption{ (Color online) (a) 
		The $J_2$ dependence of the 
		ordering vector obtained in the LT analysis for $D=0.1$ and $g_2=0$.
		(b) Classical ground state energy, $E_g$, (dashed line), and  its second derivative, $-d^{2}E_{g}/dJ^{2}_{2}$, (solid line) as a function of $J_2$ which computed by the LT approach, for given $D=0.1$ and $g_2=0.05$.}	%
	\label{fig:3}
\end{figure}

	Zero-temperature phase diagram in the $g_{2}-J_{2}$ plane for the special case with $D=0$ includes three distinct phases \cite{Zare_2013}:
	(i) Planar commensurate  N{\'e}el-type antiferromagnetic state~($ xy $-AFM) for $(J_2-g_2)<1/6$.
	This spin configuration is originated from the KM term reducing the $O_3$ spin symmetry to an $O_2$ symmetry.
	(ii) Classical spiral (SP) spin liquid for $(J_2-g_2)>1/6$, in which $g_2<1/6$, indeed hosts a massive ground-state degeneracy with any wave vector $\mathbf{Q}^{*}_{}$ within the FBZ satisfying the relation: 
	\begin{equation}
		\hspace{-2.mm}\cos(\frac{Q^{*}_x}{2})\cos(\frac{\sqrt{3}}{2}Q^{*}_y) + \frac{1}{2}\cos(Q^{*}_x)= \frac{1}{16(J_2-g_2)^2}-\frac{3}{4},
		\label{eq:13}
	\end{equation}
	For $1/6<(J_2-g_2)<1/2$ and $(J_2-g_2)>1/2$, these wave vectors form contours in the momentum space around the center point (${\bm\Gamma}$) and vertices ($\mathbf{K}$ and $\mathbf{K'}$)  points of FBZ,  respectively [Fig.~\ref{fig:2}].    
	Moreover, the phase shift between the two spins in the same cell for 
	$(J_2-g_2)>1/6$ is entirely defined  by
	
	\begin{align}
		\begin{aligned}
			&
			\cos\varphi^{*}_{}=2(J_2-g_2)[1+\cos(\mathbf{Q}^{*}_b)+\cos(\mathbf{Q}^{*}_a+\mathbf{Q}^{*}_b)], \\ &
			\sin \varphi^{*}_{}=2(J_2-g_2)[\sin(\mathbf{Q}^{*}_b)+\sin(\mathbf{Q}^{*}_a+\mathbf{Q}^{*}_b)],
			\label{eq:phaseshif}
		\end{aligned}
	\end{align}
	in which $\mathbf{Q}^{*}_a=\mathbf{Q}^{*}\cdot \mathbf{a}$ and $\mathbf{Q}^{*}_b=\mathbf{Q}^{*}\cdot \mathbf{b}$.
	(iii) Collinear-$z$ AFM ($z$-AFM) state with commensurate wave vector corresponding to three inequivalent $\mathbf{M}$ points in which the spins aligned along the $z$-direction without any spin canting.
	It is worth mentioning that the spin configuration related to this state is of stripy or zigzag form with three-fold degeneracy.      
	It should be noted that valley splitting in magnon spectrum of transition-metal tricalcogenides, such as MnPS$_3$ and MnPSe$_3$, originated from the $z$-AFM order on the honeycomb lattice~\cite{Li_2013,Sivadas_2015}.
	There are more details for the possible magnetic phases of the special case $D=0$ in our previous paper in Ref.~\cite{Zare_2013}.

\begin{figure}[t]
	\includegraphics[  scale=0.8]{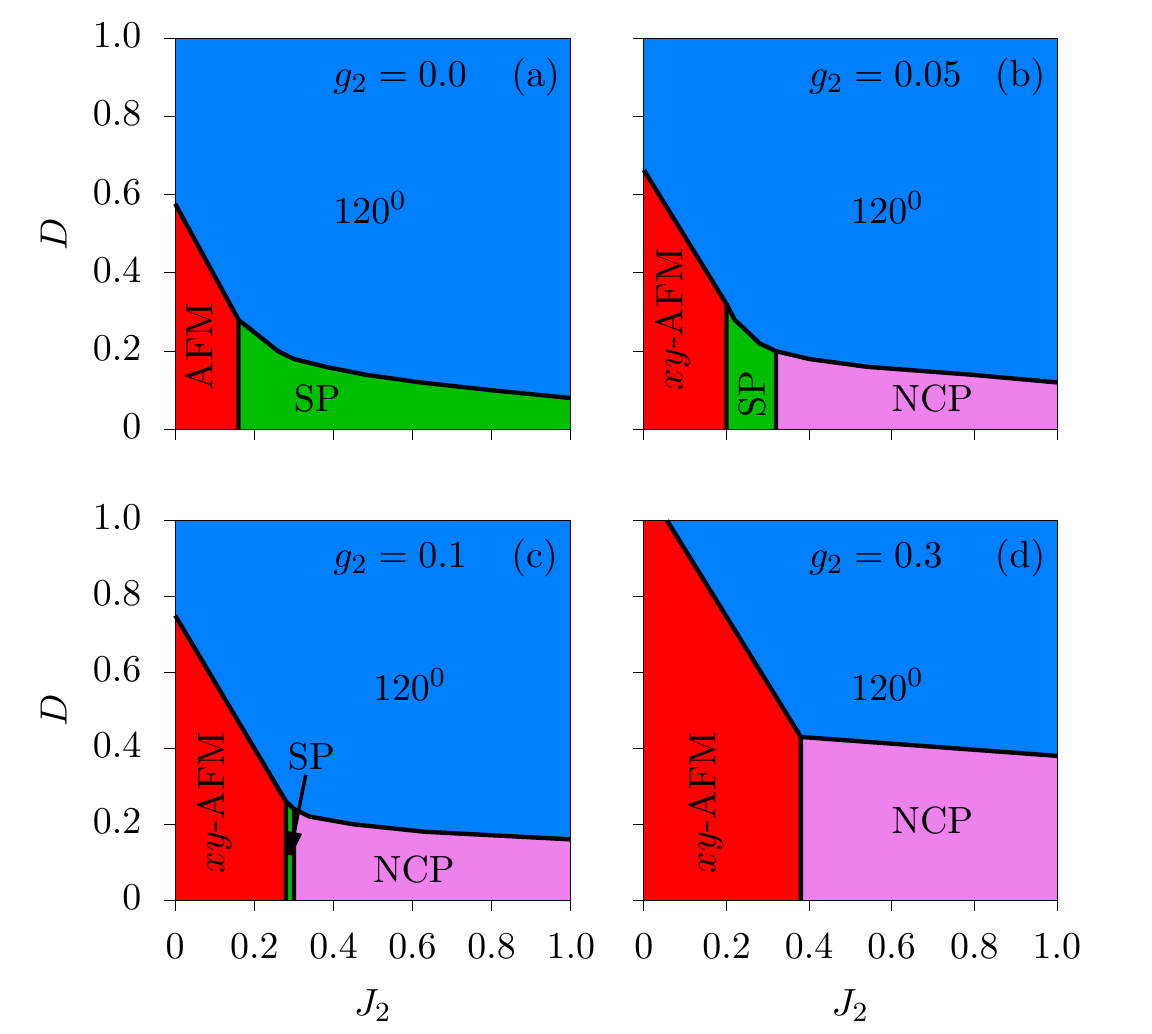}
	\vspace {-0.75cm}
	\caption{ (Color online) Classical phase diagrams obtained from the LT method in the plane of $J_2$ and $D$ for the various values of the KM term: (a) $g_2$= 0, (b) $g_2$= 0.05, (c) $g_2$= 0.1, and (d) $g_2$= 0.3.}
	\label{fig:4}
\end{figure}

 	We proceed to study the classical zero-temperature ground state of the generic equation in the plane of $J_2$ and $D$ for various values of $g_2$ using the LT method.
     First, we examine the case $g_2=0$.
	 The $J_2$ dependence of the ordering wave vector for given $D=0.1$, is shown in Fig.	 \ref{fig:3}(a).
	  For this special case, an AFM phase with ${\mathbf Q}={\mathbf \Gamma}=(0,0)$ appears for $J_2<1/6$.
	   To minimize the classical energy, we find that the minimum energy solutions for $1/6<J_2<0.8$ corresponds to an incommensurate wave vector selected along the ${\bm\Gamma}-\mathbf{K}$ directions in the FBZ.
	   We note that the sixfold degeneracy of this  planar spiral phase is originated from the $\pi/6$ rotation symmetry.
	   The classical honeycomb lattice forms 120$^{\circ}$ order for $J_2>0.8$ described by
	$\mathbf{Q}= \mathbf{K}= (4\pi/3,0)$ or $\mathbf{K'}=(2\pi/3,2\pi/\sqrt{3})$ for which 
	the two sublattices are completely decoupled. 
	The ordering wave vectors related to the AFM order and 120$^{\circ}$-order are smoothly connected to the planar spiral phase, thus the transition between them is of second-order.
	The complete ground state phase diagram in the plane of $J_2-D$ for given $g_2=0$, shown in Fig.~\ref{fig:4}(a).
	For small values of $J_2$, the AFM ordering is stable up to $D\simeq0.4$.
	On further enhancing frustration, i.e., for $J_2>1/6$, we find that the system makes a transition to the planar spiral order with the six-fold degeneracy, which remains stable in the limit of $D<0.2$.
	For large values of $D$, 
	the AFM- and planar spiral-states become 120$^{\circ}$ order with the wave vector $\mathbf{K}$ or $\mathbf{K'}$.    

	Next, we will extend these results by considering some non-zero values of $g_2$.  
   Let us examine the case of $g_2=0.05$.
   Taking into account the ground state energy per site,  $E_g$, and its second derivative, $-d^{2}E_g/d{J_2}^2$, we can determine the classical phase boundaries.
Figure~\ref{fig:3}(b) shows the $J_2$ dependence of $E_g$ and $-d^{2}E_g/d{J_2}^2$, which is calculated by the LT method, for given $D=0.1$ and $g_2=0.05$. 
Here, the second derivative of the ground state energy exhibits three anomalies with increasing $J_2$.  
  Therefore, we expect three possible points where the character of the magnetic phases changes.   
	In Figs. \ref{fig:4}(b)-4(d), we report the evolution of the classical phase diagram with an increase in the strength of the KM interaction. 
	It clearly shows that the KM term favors the stability of an $xy$-AFM order rather than the AFM phase.
	As mentioned earlier, the KM exchange term reduces the $O_3$ symmetry of the Heisenberg model to the $O_2$ symmetry which causes the spins all to lie in the $xy$ plane.
	Moreover, it is found that the stability area of the planar spiral phase in the plane of $J_{2}-D$ begins to decrease with enhancing $g_2$
	and the planar spiral order is destroyed in favor of an incommensurate noncoplanar (NCP) configuration with high classical degeneracy.
	It should be noted that at larger $g_2$ the planar spiral state is  highly unstable and only survives stable for $g_2\lesssim0.1$ in a narrow region about $J_2\approx 0.3$, as indicated in Figs.~\ref{fig:4}(b) and \ref{fig:4}(c).
	For the NCP phase with similar physics to the spiral spin liquid, the spins are oriented out of the $xy$ plane.
	One immediate insight is afforded which the interplay of the DM interaction and KM coupling plays a key role in favoring the NCP states over the planar spiral.
	We have identified the stabilization of the NCP magnetic phases with wave vectors $\mathbf Q^{*}$ satisfying the Eqs.~(\ref{eq:13}) and (\ref{eq:phaseshif}) only by replacing the $J_2-g_2$ with the $J_2+g_2$.
	Within the LT method, our findings indicate that the DM exchange interaction tends to destabilize the $z$-AFM ordering, which we have found for the Kane-Mele Heisenberg model (the special case of Eq.~\ref{eq:3} with $D=0$) in our previous work~\cite{Zare_2013}, in favor of the NCP phase.

	In the next subsection, we introduce a variational optimization (VO) method that studies the ground state of the generic model Hamiltonian (Eq.~\ref{eq:3}).  
	The VO method attempts to find the classical ground state by enforcing the constraint of fixed spin length at each site $i$,
	
	\begin{equation}
   	{\bf S}_i \cdot {\bf S}_i~=~S^2,
   \label{eq:sc}
   \end{equation}   	
	which is termed the strong constraint. For each wave vector ${\bf Q}$, we can rewrite the fixed spin-length constraint on every site in terms of ${\bf S_k}$
	
	\begin{equation}
		\sum_{\bf k} {\bf S_k}\cdot {\bf S_{Q-k}}~=~NS^2\delta_{\bf QG},
		\label{eq:wcq}
	\end{equation}
	where ${\bf G}$ denotes a reciprocal lattice vector.
	It is known that the spin vectors are real, thus we find the relation on the ${\bf S_k}$ as,
		
	\begin{equation}
		{\bf S_k}^{*}~=~{\bf S_{-k}}.
	\end{equation} 
	
	Within the LT method, we consider minimization of the classical energy in Eq.~\ref{eq:9}, under the relaxed constraint (Eq.~\ref{eq:wc}) in the momentum space
	
	\begin{equation}
		\sum_{\bf k} |{\bf S_k}|^2~=~NS^2,
	\end{equation} 
	the minimum energy solution would be the true physical ground state when it satisfies all local constraints (Eq.~\ref{eq:wcq}), i.e., ${\bf Q}={\bf G}$.
	Indeed, for the spin configurations in which spins are ordered at an incommensurate wave vector ${\bf Q}$, the local constraints (Eq.~\ref{eq:wcq}) are not necessarily satisfied, then the LT method failed to give the physical ground state~\cite{Liu_2016a}. Next, we therefore explore such cases in more detail by performing numerical optimization.

	\subsection{Variational optimization method}
	To numerically validate the classical phase diagrams, $T=0$ limit, obtained with the LT method, here we use the variational optimization approach.
	The remarkable point in this method is to parametrize the spin vectors in terms of the variational parameters as wave vector, polar- and azimuthal-angles so that the local length constraint of the unit spin size in each site is fulfilled.
	 Therefore, there would be an ability to search the NCP magnetic ground state in detail.
	To parametrize the spin vectors on the two sublattices of the honeycomb lattice as follows:

	\begin{align}
		\hspace{-0.6 cm}
		\textbf{S}^{\nu}(\mathbf{r}_ i)= \sin(\theta^{\nu}_{i})[\hat{\mathbf{x}} \cos(\varphi^{\nu}_{i})+\hat{\mathbf{y}} \sin(\varphi^{\nu}_{i})]+\hat{\mathbf{z}} \cos(\theta^{\nu}_{i}),
		\label{eq:14}
	\end{align}
	in which
	\begin{align}
		\begin{aligned}
			&
			\varphi^{1}_{i}= \mathbf{Q}\cdot \mathbf{r}_ i,~~~~~~~~~ 
			\theta^{1}_{i}= \mathbf{Q'}\cdot \mathbf{r}_ i + \gamma,
			\\ &
			\varphi^{2}_{i}= \mathbf{Q}\cdot \mathbf{r}_ i + \varphi,~~~
			\theta^{2}_{i}= \mathbf{Q'}\cdot \mathbf{r}_ i + \eta.
			\label{eq:15}
		\end{aligned}    
	\end{align}
	%
	%
	
	%%%%%%%%%%%%%%%%%====== figure ======%%%%%%%%%%%%%%%%%
	\begin{figure}[t]
		\begin{center}
			\vspace{0.10cm}
			\hspace{-0.1cm}
			\includegraphics[width=0.95 \linewidth]{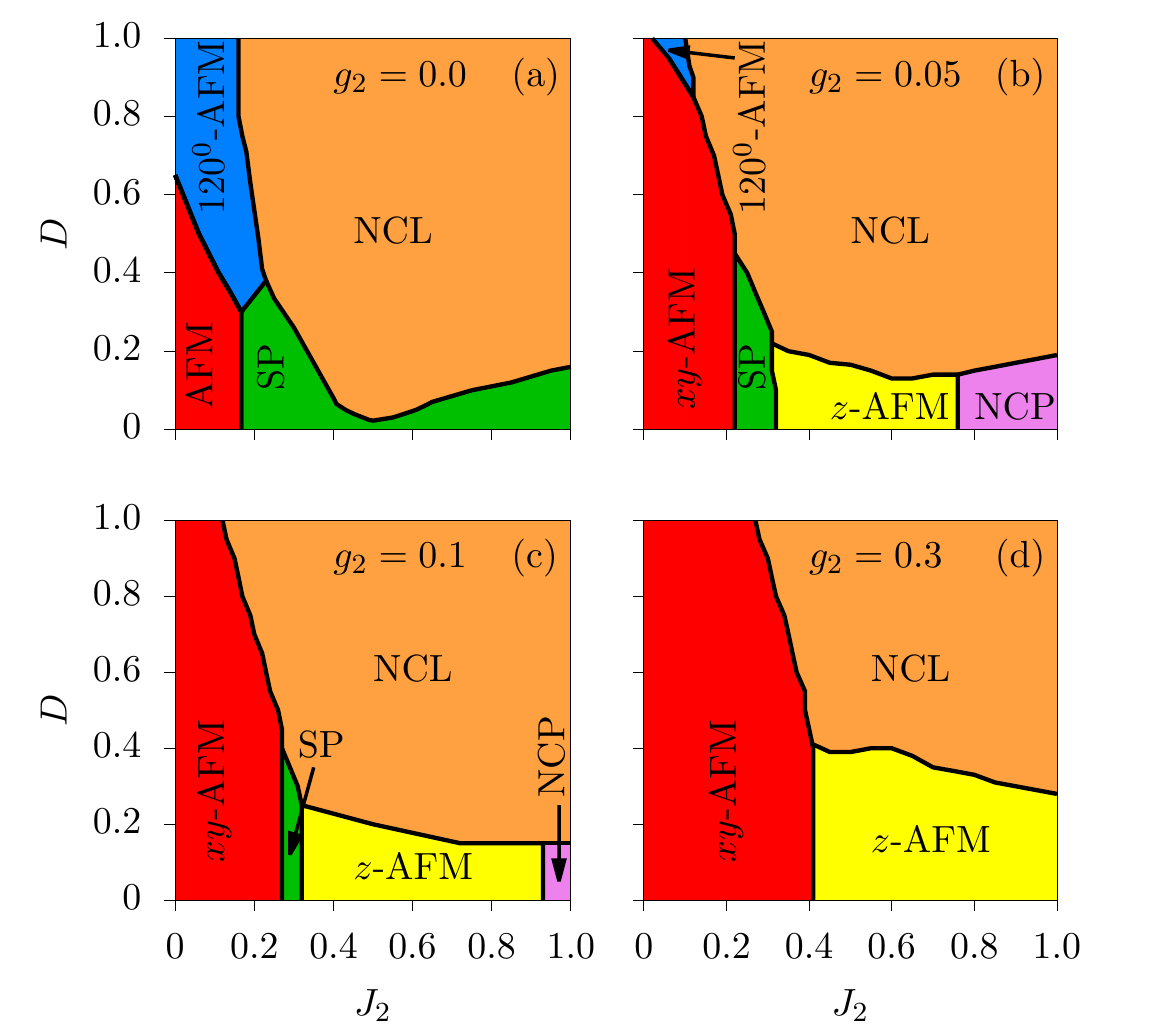}  \hspace{.0cm}
			\includegraphics[width=0.8\linewidth]{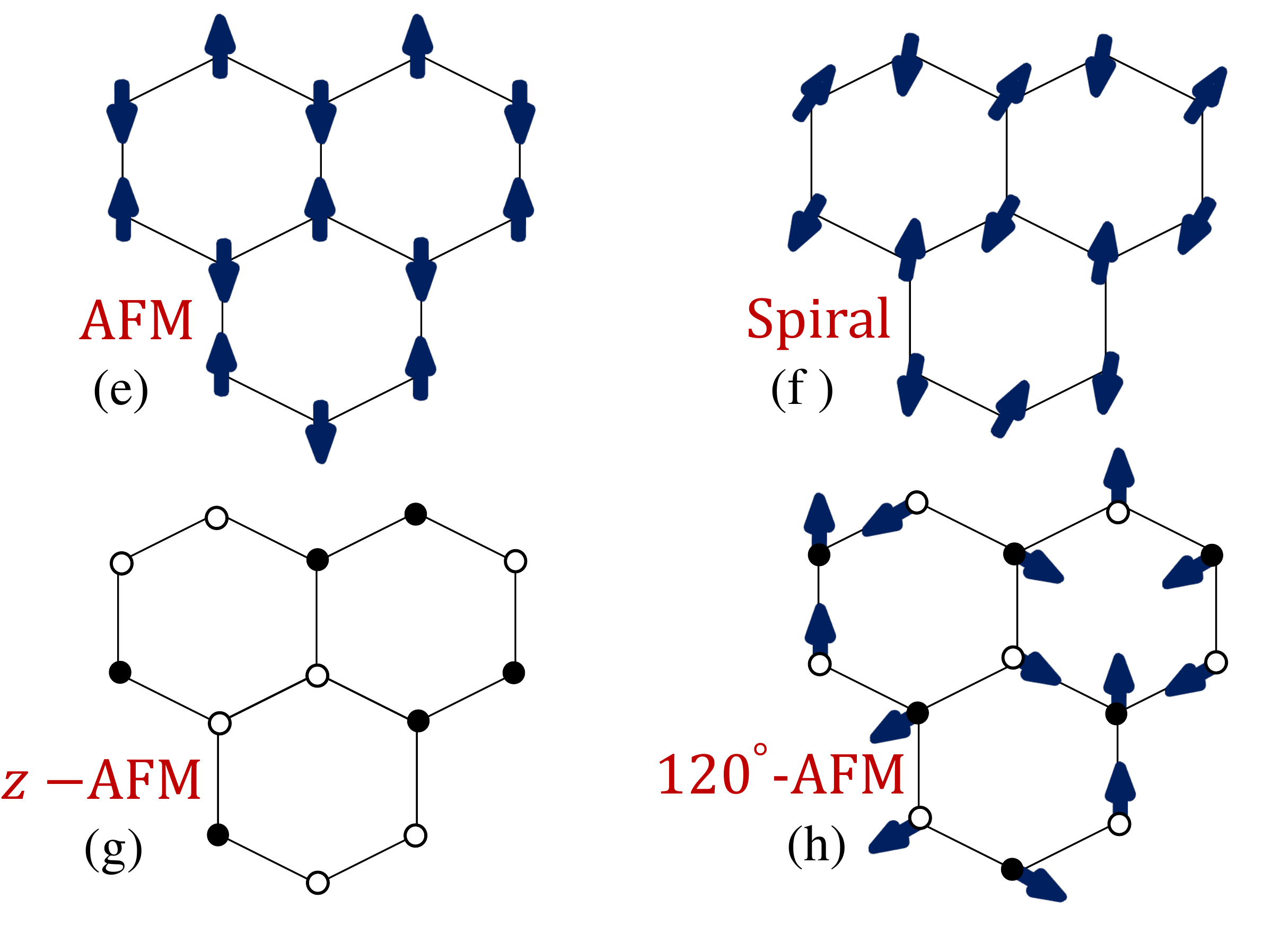} \hspace{.0cm}
			\end{center}
		\vspace{-0.2cm}
		\caption{(Color online) 
			 Classical ground state phase diagram obtained from the VO method in the plane of $J_2$ and $D$ for some values of the KM term: (a) $g_2=0$, (b) $g_2=0.05$, (c) $g_2=0.1$, and (d) $g_2=0.3$. Real-space ordering for some spin configurations in (e) the N\'eel antiferromagnetic, (f) the spiral state, (g) the $z$-AFM ordered phase (filled and open circles indicate the spins aligned along the $z$ and $-z$ directions, respectively), and (h) the 120$^{\circ}$-AFM phase.} 
		\label{fig:5}
		\end{figure}

	We have computed the general variational energy of the magnetic phases with commensurate and incommensurate wave-vectors by substituting the spin components defined by Eq.~(\ref{eq:14}) in the effective spin Hamiltonian (Eq.~\ref{eq:3}) (for more detail, see Appendix:~\ref{appA}).
	There may well be the reason why the classical energy of the incommensurate ones, Eq.~(\ref{eq:a2}), is independent of the DM interaction.
	The main reason can be attributed to the fact that the contribution of the DM term to the energy is as
	$ D\sum_{\langle\langle ij\rangle\rangle}^{} {\nu_{ij}}\cos[\mathbf{Q'}\cdot (\mathbf{r}_{i}+\mathbf{r}_{j})+2\delta],
	$
	with $\mathbf{Q'}$ an incommensurate wave vector.
	Thus, $\mathbf{Q'}\cdot (\mathbf{r}_{i}+\mathbf{r}_{j})\neq 2n\pi$ concludes that this summation for these magnetic phases should vanish.
	To obtain the ground state phase diagram of the effective spin Hamiltonian in Eq.~(\ref{eq:3}) for each set of couplings, the classical energies should be minimized with respect to the variational parameters $Q_x,~Q_y,~ Q'_x,~ Q'_y, ~{\varphi},~ \gamma,$ and $\eta$.
	For the purpose of this work, we numerically minimize the variational energy functions (\ref{eq:a1}) and (\ref{eq:a2}) by using the
	simulated annealing scheme from the MATHEMATICA optimization package~\cite{Mathematica} to determine various magnetic configurations.

\begin{figure}[t]
	\includegraphics[angle=0,origin=c,scale=0.35]{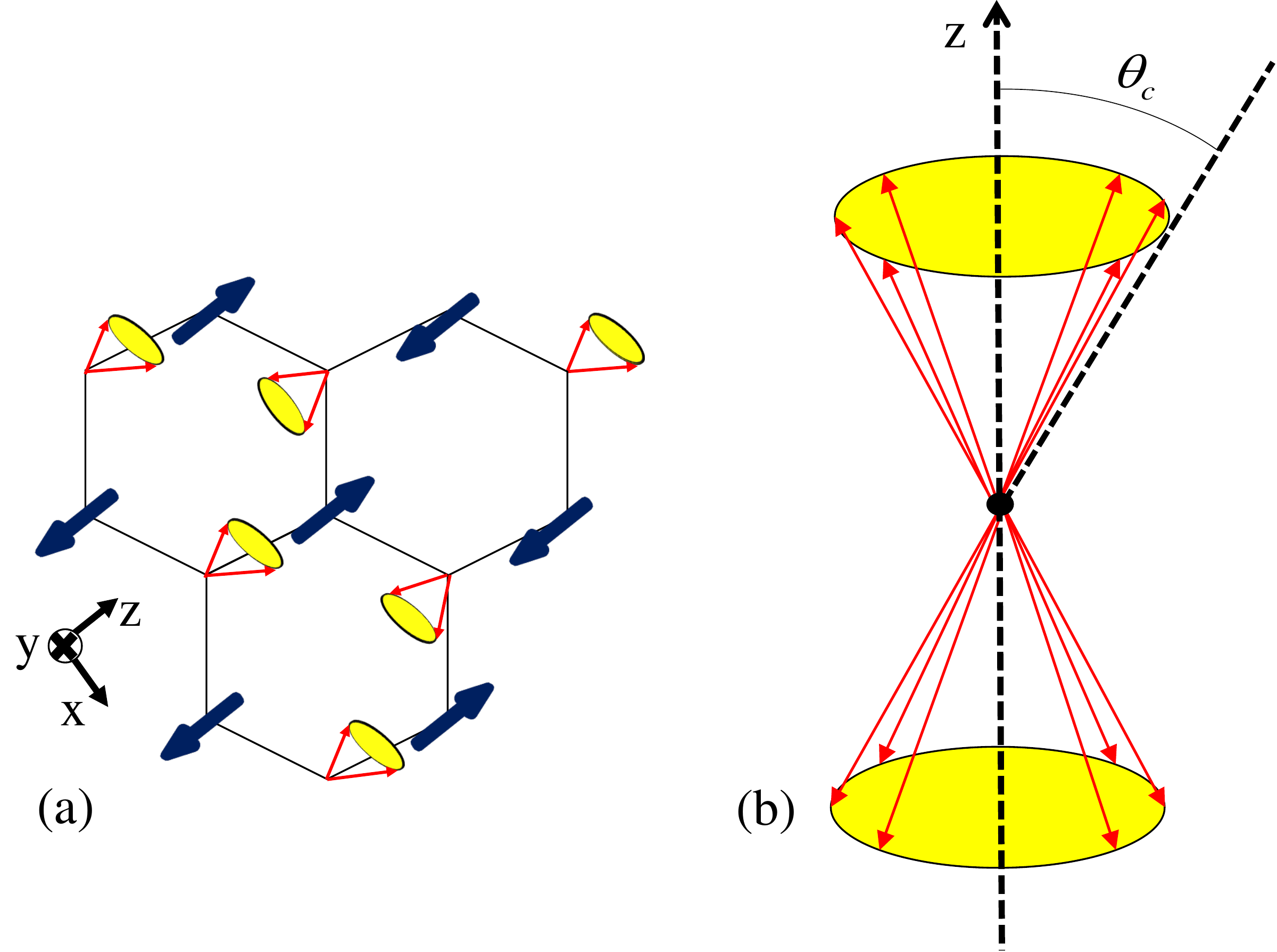}
	\vspace{-.6cm}
	\caption{ (Color online) (a) The real-space spin configuration of the noncollinear phase. (b) The double conical spiral, with the conical angle $\theta_c$ and along the $z$ direction, of a single sublattice plotted in the same origin.} 
	\label{fig:5a}
\end{figure}

	The global classical phase diagrams of the ground state of the spin Hamiltonian, Eq.~(\ref{eq:3}), for the different values of the KM term are indicated in Fig.~\ref{fig:5}.   
	For the special case of $g_2=0$, the ground state phase diagram includes four distinct phases [Fig.~\ref{fig:5}(a)]: two commensurate phases (labelled as AFM and 120$^{\circ}$-AFM) and two incommensurate orders (called spiral and NCL) .
	 Within this notation, two distinct commensurate states may also exist: (i) AFM described by $\mathbf{Q}$ and $\mathbf{Q'}={\bm \Gamma}$ with a phase difference $\pi$,
	(ii) 120$^{\circ}$-AFM phase has in- and normal-to-plane spin components 
	with different wave vectors in reciprocal space characterized with $\mathbf{Q}=\mathbf{K}$ or $\mathbf{K'}$ and $\mathbf{Q}'={\bm \Gamma}$, respectively.       
	Note that the spins on the two sublattices are totally decoupled with an arbitrary phase shift. 
	 In addition, the phase diagram includes a wide region of the two incommensurate phases:
	 (iii) planar spiral order with high classical degeneracy in terms of $xy$ wave vectors satisfying Eq.~(\ref{eq:13}), and a phase shift $\varphi^{*}$ between the two spins within each unit cell defined by the relation Eq.~(\ref{eq:phaseshif}).
	These classically degenerate solutions form manifolds around ${\bm\Gamma}$ for $1/6<J_2<1/2$ and closed contours around $\mathbf{K}$ and $\mathbf{K'}$ for $J_2>1/2$ as shown in Fig.~\ref{fig:2}.
	(iv) In the case of noncollinear state, the in-plane and out-of-plane spin components have different wave vectors in the momentum space.	
	The in-plane spin component has two Fourier components with the incommensurate wave vectors ${\mathbf Q}=\pm(\frac{\pi}{3},-\pi)$,
	while its $z$ component has Fourier momenta of the wave vector $\mathbf{Q}'=\mathbf{M}$.
	In this phase, one sublattice forms a conical spiral phase, while the spins on other sublattice are  aligned along the $z$ direction [Fig.~\ref{fig:5a}(a)]. 
  	This phase includes double conical spiral with conical angle $\theta_c$ around 
  	the $z$ or $-z$ directions, as depicted in Fig.~\ref{fig:5a}(b).	
	It should be noted that the conical angle is strongly dependent on the exchange coupling parameters. 
   At first sight, it is evident that the out-of-plane ordering has a spin configuration similar to the $z$-AFM corresponds to the wave vector $\mathbf{Q}'=\mathbf{M}$.
   Our results show that a subtle interplay between the antiferromagnetic Heisenberg and KM exchange interactions plays a decisive role in the stabilization of the NCL state. 
	By increasing the strength of the KM term,
	the classical spiral spin liquid phase for the small values of $D$ is divided between three phases: the spiral, the $z$-AFM, and the NCP [Fig.~\ref{fig:5}(b)]: (I) The spiral is highly degenerate as previously found in the case of $g_2=0$. The classically degenerate solutions correspond to an infinite set of wave vectors $\mathbf{Q}^{}$ satisfy Eq.~\ref{eq:13},
	(II) The stability of the $z$-AFM magnetic order in the phase diagram has been attributed to the existence of the KM term, $g_2\neq0$, in Eq.~(\ref{eq:3})~\cite{Zare_2013}, and
	(III) The incommensurate NCP phase, which is highly degenerate in terms of both $\mathbf{Q}$ and $\mathbf{Q'}$ (the two ordering wave vectors are inequivalent, $\mathbf{Q}\neq\mathbf{Q'}$).
    The fact distinguishes our NCP spin configuration from the the planar spiral order, which has two order parameters, one associated with a planar spiral, and the other with incommensurate ordering along the $z$ direction.
	Here, we report evidence for the stability of the NCP phase on the non-Bravais lattice due to the interplay of the KM- and DM couplings.
	On the other hand, it should be emphasized that the new magnetic commensurate states ($xy$-AFM and $z$-AFM) that emerge in the ground state phase diagram are due to the $g_2$ term that breaks the global $O_3$ spin rotation symmetry down to the $O_2$ spin rotation symmetry around the $z$-axis.
	As it is clear in Figs.~\ref{fig:5}(b)-(\ref{fig:5}(d), the stability region associated with these two commensurate states begins to extend by enhancing the strength of $g_2$.  
	In contrast, we find that the stability regions of the planar spiral- and 120$^{\circ}$-AFM magnetic phases 
	narrowing by increasing $g_2$. Therefore, these two phases remain stable only 
	for the small values of the $g_2$ [Figs.~\ref{fig:5}(b) and \ref{fig:5}(c)].
	Consequently, the obtained classical phase diagrams address that the DM interaction, allowed by the complex hopping integrals in the second-neighbor coupling \cite{Pan_2020,Wolf_2021}, stabilizes an in-plane spin configuration such as the classical spiral spin liquid phase, and magnetic configurations with an out-of-plane spin component as AFM, 120$^{\circ}$-AFM, and NCL phases in the twisted homobilayer systems at half-filling.
	 Now, we consider how the addition of the KM term modifies the obtained results.

	In what follows, we will study the effects of quantum fluctuations on the stability of the classical ground states of the model (\ref{eq:3}) using a theoretical method such as linear spin-wave theory (LSW), and the density renormalization group (DMRG) is a numerical method.

	\section{Instability of Magnon Spectra }
	Here, we will investigate the stability regions of the classical ground state magnetic structures in contrast to the quantum fluctuations by calculating the excitation spectrum
	using linear spin-wave.
	In this method, instead of working in the laboratory frame $\{x,y,z\}$ on each lattice site, we switch to the local frame $\{\tilde{x},\tilde{y},\tilde{z}\}$ where the $\tilde{z}$ axis is aligned along the spin's quantization direction on that site for a given classical magnetic order.
	Using the rotation matrix ${\mathbf{\hat R}}$ to rotate the spin vector in the laboratory reference frame at the site $i$ to the local one as follows:
	\begin{equation}
		\mathbf{S}^{\nu}_{}(\mathbf{r}_i)={\hat{\mathbf{R}}}_{i}{\tilde{\mathbf{S}}}^{\nu}_{}(\mathbf{r}_i) 
		\label{eq:16}
	\end{equation}

	In the locally rotated coordinates, the spin vectors obtained by a two-stage process
	as 
	${\mathbf{\hat R}}={\mathbf{\hat R}_{\varphi}}\cdot{\mathbf{\hat R}_{\theta}}$.
	Here, ${\mathbf{\hat R}_{\varphi}}$ (${\mathbf{\hat R}_{\theta}}$) represents
	the rotation in the $xy$ ($xz$) plane around the $z$ ($y$)
	axis by angle $\varphi$ ($\theta$).
	Rotation matrix in the $xy$-plane by the azimuthal angle $\varphi$ is given by
	%
	%===========================================================
	\begin{align}
		\begin{aligned}
			{\mathbf{\hat R}_{\varphi}}=
			\begin{pmatrix}
				\cos\varphi & -\sin\varphi & 0
				\\
				\sin\varphi & \cos\varphi & 0
				\\
				0 & 0 & 1  
			\end{pmatrix}
		\end{aligned}.
		\label{eq:R1}
	\end{align}
	%===========================================================
	%
	and the rotation matrix in the $xz$-plane by the polar angle $\theta$ takes the form
	%
	%===========================================================
	\begin{align}
		\begin{aligned}
			{\mathbf{\hat R}_{\theta}}=
			\begin{pmatrix}
				\sin\theta & 0 & \cos\theta
				\\
				0 & 1 & 0
				\\
				-\cos\theta &  0 & \sin\theta  
			\end{pmatrix}
		\end{aligned}.
		\label{eq:R2}
	\end{align}
	%===========================================================
	%
	%
	
	Thus, the rotation transformation between the two reference frames is characterized by
	%===========================================================
	\begin{align}
		\begin{aligned}
			{\mathbf{\hat R}^{\nu}_{i}}=
			\begin{pmatrix}
				\sin\theta^{\nu}_{i}\cos\varphi^{\nu}_i & -\sin \varphi^{\nu}_i &  \cos\theta^{\nu}_{i}\cos\varphi^{\nu}_i
				\\
				\sin\theta^{\nu}_{i}\sin\varphi^{\nu}_i & \cos \varphi^{\nu}_i & \cos\theta^{\nu}_i\sin\varphi^{\nu}_i
				\\
				-\cos\theta^{\nu}_i &  0 & \sin\theta^{\nu}_i  
			\end{pmatrix}
		\end{aligned}.
		\label{eq:RR}
	\end{align}
	%===========================================================  
	%
	
	Now we can reexpress the spin Hamiltonian (Eq.~\ref{eq:3}) in terms of the new spin components as follows: 

	\begin{align}
		\begin{aligned}
			&
			{\cal H}= \sum_{ij} \sum_{\alpha,\beta\in\{\nu\}}\mathbf{S}^{\alpha{\rm T}}_{i} {\cal J}^{\alpha\beta}_{ij} \mathbf{S}^{\alpha}_{j}
			=\sum_{ij} \sum_{\alpha,\beta\in\{\nu\}}
			\tilde{\mathbf{S}}^{\alpha{\rm T}}_{i} \tilde{\cal J}^{\alpha\beta}_{ij} \tilde{\mathbf{S}}^{\alpha}_{j}
			\label{eq:20}
		\end{aligned}
	\end{align}
	in which the exchange matrix in the rotated frame is defined as $\tilde{\cal J}^{\alpha\beta}_{ij}={\mathbf{\hat R}^{}_{i}}{\cal J}^{\alpha\beta}_{ij}{\mathbf{\hat R}^{}_{j}}$.

	We proceed to perform the well-known Holstein-Primakoff (HP) transformation on the rotated spin Hamiltonian (Eq.~\ref{eq:20}) to obtain the magnon spectrum.
	Here, we use an LSW theory that rewrites the rotated spin model (Eq.~\ref{eq:20}) in terms of the HP boson operators.
	In this case, it can be turned into
	${\cal H}= {\cal E}_{cl}+{\cal H}_{0}+{\cal H}_{1}+{\cal H}_{2}+...$. The first term represents the classical energy of the spin configurations. In the following, we only keep the quadratic terms in bosonic operators, i.e. the harmonic order of the spin wave theory.
	This means that we try to find the classical lattice vibrations and therefore to quantize the spin model (Eq.~\ref{eq:20}) as harmonic oscillations in the LSW theory.
	By diagonalizing the harmonic term using Bogoliubov transformation, one can obtain the magnon spectrum and first quantum corrections to the zero-point energy related to the classical spin configurations.    

	For a given classical magnetic structure, one can consider its stability boundaries as a function of the exchange coupling parameters using the magnon spectra.
	Notice that the appearance of soft modes for a magnon branch in one or at a set of wave vectors may result in imaginary spectrum.
	This feature is well known as magnon instability allowing us to identify a magnetic phase transition. 
	Moreover, we can obtain more information regarding the nature of phase transition and the existence of intermediate phases that could not been verified within classical calculations {\cite{Chubukov_1991,Maksimov_2019}}.
	The overlap of the magnon stability regions of the adjacent magnetic phases illustrates the existence of the first-transition between them. 
	Note that if the magnon instability threshold takes place before the obtained classical boundaries, this result indicates that the phase transition occurs via an intermediate phase.

	\subsection{Coplanar phases}
	Here, we start to investigate the magnon stability regions of the in-plane magnetic orders.
	The ordering vector related to different states determines the number of sublattices within a magnetic unit cell.
	For the in-plane spin configurations, the rotation matrix can be obtained by fixing ${\theta}^{\nu}_i=0$. As a result, the spin components are reexpressed in the local frame as follows:

	\begin{align}
		\begin{aligned}
			&
			S^{\nu}_{i,x}= \cos\varphi^{\nu}_{i} {\tilde S}^{\nu}_{i,z}-\sin\varphi^{\nu}_{i} {\tilde S}^{\nu}_{i,y},
			\\ & 
			S^{\nu}_{i,y}= \sin\varphi^{\nu}_{i} {\tilde S}^{\nu}_{i,z} + \cos\varphi^{\nu}_{i} {\tilde S}^{\nu}_{i,y},
			\\ &
			S^{\nu}_{i,z}= -{\tilde S}^{\nu}_{i,x},
			\label{eq:21}
		\end{aligned}
	\end{align}
	in which $\varphi^{\nu}_i=\mathbf{Q}\cdot \mathbf{r}_i+\phi^{\nu}_{i}$. 
	After this rotation, the spin model (Eq.~\ref{eq:3}) turns into  
	
	\begin{widetext}
		\begin{align}
			\begin{aligned}
				&
				{\cal H}= J_1 \sum_{\langle ij \rangle}
				\cos(\mathbf{Q}\cdot\bm{\delta}_1-\phi)[{\tilde S}^{1}_{i,y}{\tilde S}^{2}_{j,y}+{\tilde S}^{1}_{i,z}{\tilde S}^{2}_{j,z}]+
				\sin(\mathbf{Q}\cdot\bm{\delta}_1-\phi)[{\tilde S}^{1}_{i,y}{\tilde S}^{2}_{j,y} - {\tilde S}^{1}_{i,z}{\tilde S}^{2}_{j,z}]+{\tilde S}^{1}_{i,x}{\tilde S}^{2}_{j,x}
				\\ & \hspace{0.7 cm}
				+\sum_{\langle\langle ij \rangle\rangle}
				(J_2-g_2)\cos(\mathbf{Q}\cdot\bm{\delta}_2)[{\tilde S}^{1}_{i,y}{\tilde S}^{1}_{j,y}+{\tilde S}^{1}_{i,z}{\tilde S}^{1}_{j,z}]+ (J_2+g_2)
				{\tilde S}^{1}_{i,x}{\tilde S}^{1}_{j,x}+ (1\leftrightarrow 2)
				\\ & \hspace{0.7 cm}
				- \sum_{\langle\langle ij \rangle\rangle}
				D\nu_{ij}\sin(\mathbf{Q}\cdot\bm{\delta}_2)[{\tilde S}^{1}_{i,y}{\tilde S}^{1}_{j,y}+{\tilde S}^{1}_{i,z}{\tilde S}^{1}_{j,z}]
				-D\nu_{ij}\cos(\mathbf{Q}\cdot\bm{\delta}_2)[{\tilde S}^{1}_{i,y}{\tilde S}^{1}_{j,z}-{\tilde S}^{1}_{i,z}{\tilde S}^{1}_{j,y}]+ (1\leftrightarrow 2).
				\label{eq:22}
			\end{aligned}
		\end{align}
	\end{widetext}

	\begin{figure}[t]
		\begin{center}
			\vspace{0.10cm}
			\hspace{-0.1cm}
			\includegraphics[width=0.48 \linewidth]{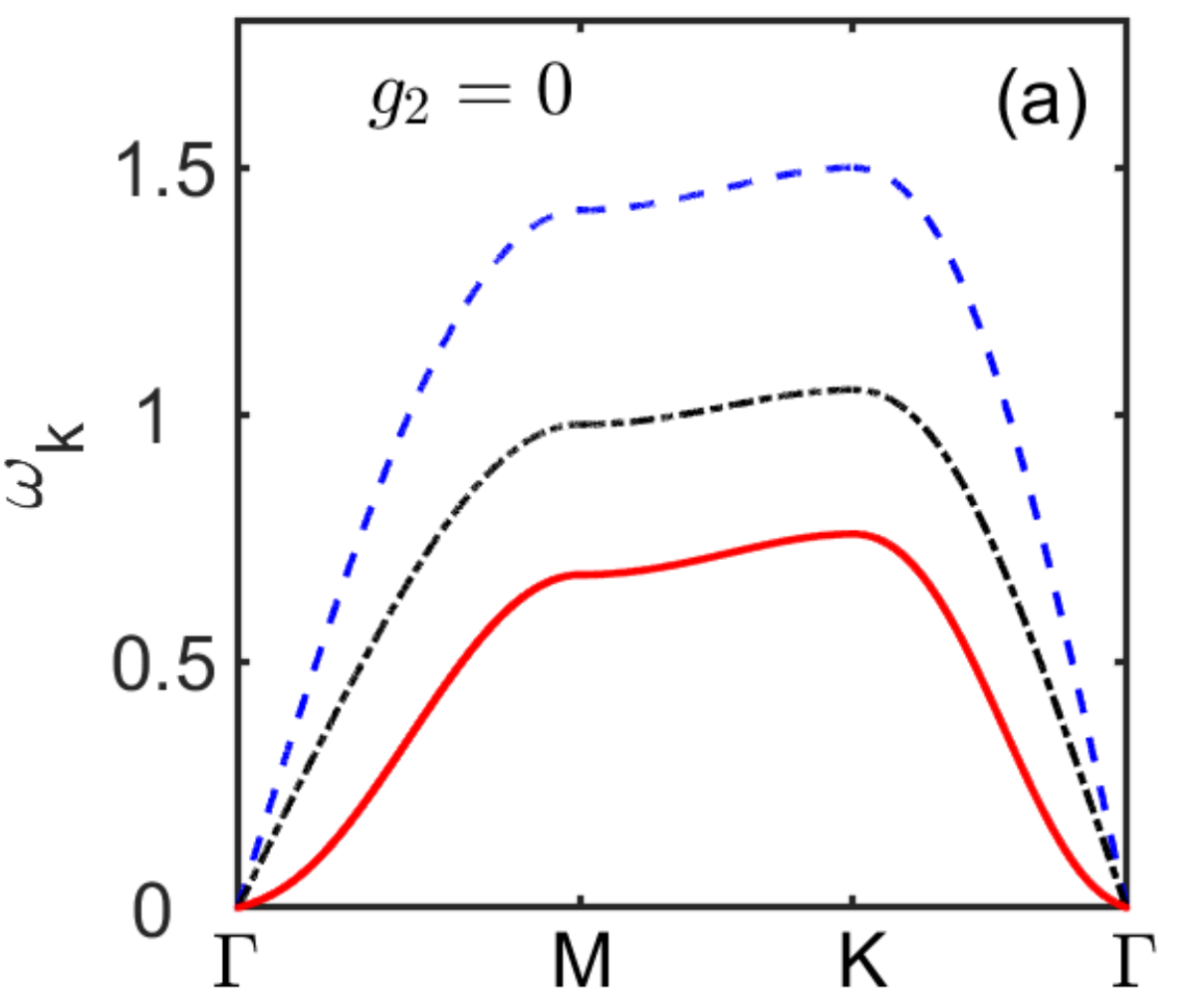}  \hspace{.0cm}
			\includegraphics[width=0.48\linewidth]{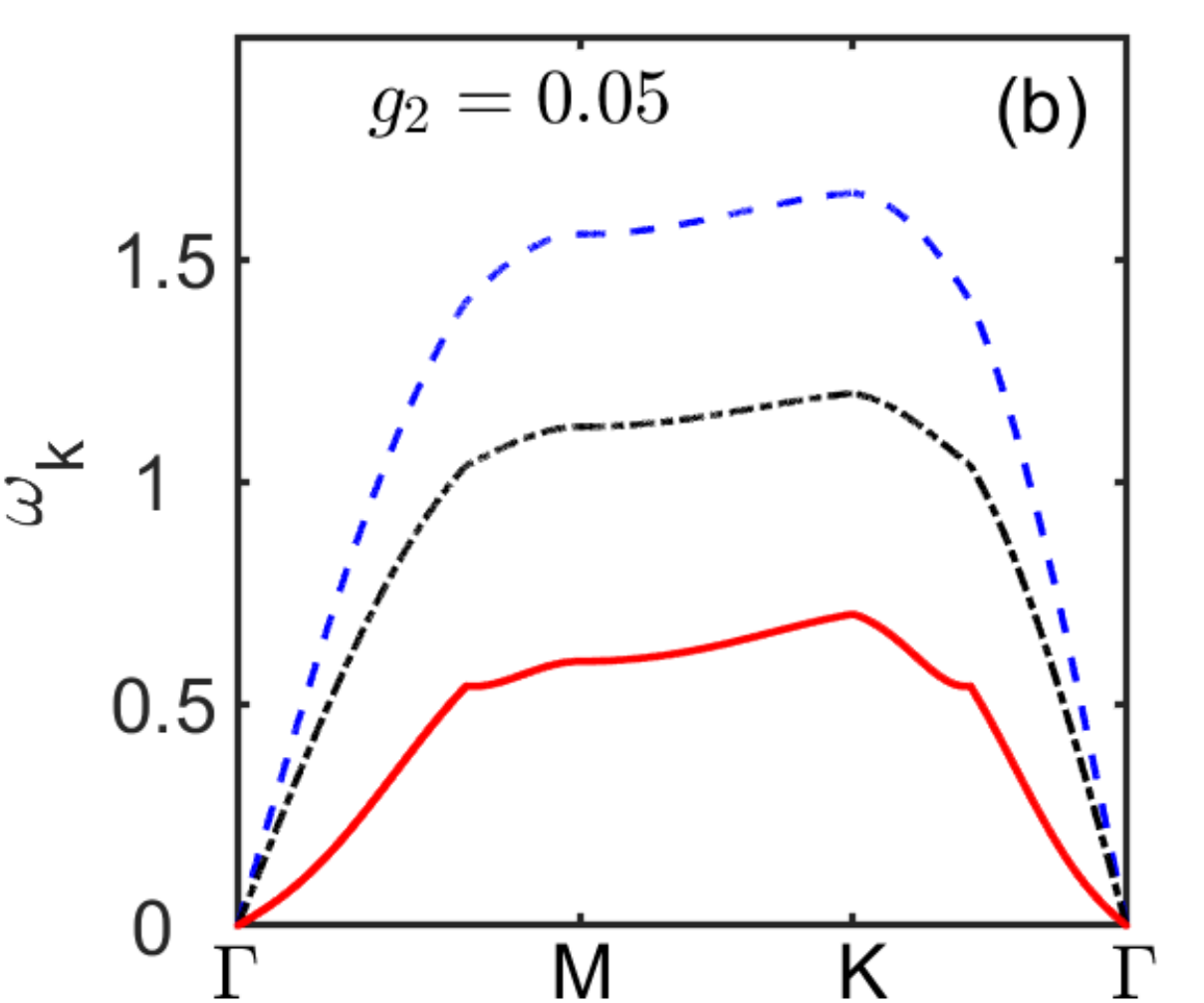} \hspace{.0cm}
		\end{center}
		\vspace{-0.2cm}
		\caption{(Color online) 
			The obtained magnon spectra, along high symmetry
			directions in the FBZ, via a LSW calculation in going from the AFM state to the spiral one for given $D=0.1$. The curves from top to bottom are associated with some values of  (a) $J_2$=0,~0.1,~0.165,  
			(b) $J_2$=0,~0.1,~0.215, respectively. The lower curve indicates the magnon instability boundary of the AFM phase.}
		\label{fig:6}
	\end{figure}
	
	Under the LSW approximation, we can quantize the spin operators in Eq.~(\ref{eq:22}) in terms of HP boson operators as follows:
	\begin{align}
		\begin{aligned}
			&
			{\tilde S}^{1}_{i,z}=S-a^{\dagger}_{i}a^{}_{i},~
			{\tilde S}^{1\dagger}_{i}=\sqrt{2S}a^{}_{i},~
			{\tilde S}^{1-}_{i}=\sqrt{2S}a^{\dagger}_{i},
			\\ &
			{\tilde S}^{2}_{i,z}=-S+b^{\dagger}_{i}b^{}_{i},~
			{\tilde S}^{2\dagger}_{i}=\sqrt{2S}b^{\dagger}_{i},~
			{\tilde S}^{2-}_{i}=\sqrt{2S}b^{}_{i}
			\label{eq:23}
		\end{aligned}
	\end{align}

	After performing FT,  
	the LSW Hamiltonian in momentum space can be written as ${\cal H}^{}_2= 2S\sum_{\mathbf{k}}\psi^{\dagger}_{\mathbf{k}}{\cal J}^{}_{\mathbf{k}}\psi_{\mathbf{k}}$
	with $\psi_{\mathbf{k}}=(a^{\dagger}_{\mathbf{k}},b^{\dagger}_{\mathbf{k}},a_{-{\mathbf{k}}},b_{-{\mathbf{k}}})$.
	The $4\times4$ matrix ${\cal J}^{}_{\mathbf{k}}$ takes the form
	\begin{align}
		{\cal J}_{\mathbf{k}} = \begin{pmatrix} {\cal J'}^{}_{\mathbf{k}}  & {\cal J''}^{}_{\mathbf{k}}  \\
			\\
			{\cal J''}^{\dagger}_{\mathbf{k}}  & {\cal J'}^{}_{\mathbf{k}} 
		\end{pmatrix},
		\label{eq:24}
	\end{align}
	%---------------------------------------------------------------------------------------------------
	%
	in which the $2\times2$ matrices ${\cal J'}^{}_{\mathbf{k}}$ and ${\cal J''}^{}_{\mathbf{k}}$ are given by
	
	\begin{align}
		\begin{aligned}
			&
			{\cal J'}^{}_{\mathbf{k}} = \begin{pmatrix} 
				{A}^{+}_{\mathbf{k}} & {B}^{}_{\mathbf{k}} \\ 
				{B}^{*}_{\mathbf{k}} & {A}^{-}_{\mathbf{k}} \\ 
			\end{pmatrix},
			~~~~~~
			{\cal J''}^{}_{\mathbf{k}} = \begin{pmatrix} 
				{C}^{+}_{\mathbf{k}} & {D}^{}_{\mathbf{k}} \\ 
				{D}^{*}_{\mathbf{k}} & {C}^{-}_{\mathbf{k}} \\ 
			\end{pmatrix},
			\label{eq:25}
		\end{aligned}    
	\end{align}
	with the expressions for the matrices elements given in 
	Appendix:~\ref{appB}. 

	To study the magnon instability boundaries of different magnetic structures in Fig.~\ref{fig:5},
	it is necessary to obtain the magnon spectra by diagonalization of the quadratic Hamiltonian (${\cal H}_2$) by means of a standard Bogoliubov transformation~\cite{Mucciolo_2004}.
	For the state with AFM order, with $\mathbf{Q}={\bm \Gamma}$ and $\phi^{\nu}_{i}=0$, the smallest positive magnon branch has been plotted in Fig.~\ref{fig:6} for given $D=0.1$  and some chosen values of the coupling constants $J_2$ and $g_2$ along high symmetry directions in the FBZ.
	The linear behavior of the magnon spectra close to the ${\bm \Gamma}$ point for the typical values of $J_2$ away from the instability boundary demonstrates the appearance of Goldstone modes due to spontaneously broken O$_2$ rotational symmetry.
	On further enhancing $J_2$, the magnon spectra in the vicinity of the ${\bm \Gamma}$ point exhibit a quadratic form. This result indicates that the soft modes are located around the symmetric point ${\bm \Gamma}$ inside the FBZ leading to the instability of the AFM order.
	It should be emphasized that the magnon instability boundary for the AFM phase in going to the classical spin liquid coincides with its classical boundary, and this result indicates a second-order phase transition between them.
	However, our results illustrate that the magnon instability boundary for both the AFM/120$^{\circ}$-AFM and AFM/NCL is far beyond the classical one; then, the first-order phase transition can be realized between these magnetic phases~\cite{Chubukov_1991}.
	As already discussed, the classical ground state energy associated with the spiral state (Eq.~\ref{eq:a2}) is minimized with the choice of the ordering vectors on the manifolds in Fig. \ref{fig:2}.    
	For this case with high classical degeneracy, we find no indication of quantum order-by-disorder using the LSW calculation.
	In contrast to the obtained results for the Kane-Mele Heisenberg model~\cite{Zare_2013}, no magnetic state with an ordering vector from the degenerate manifolds can be stabilized in the presence of the quantum fluctuations. 
	In the next section (in Sec.~\ref{secV}), we will numerically study the quantum nature of the classical spiral spin liquid in more detail.

	\subsection{$z$-AFM phase}
	For the $z$-AFM phase, the ordering vector is one of the $\mathbf{M}$ points of the FBZ.
	The magnetic unit cell of this state includes four spins, two in the $\hat z$ direction and two in the opposite direction [Fig.~\ref{fig:5}(g)].
	It clearly shows that the primitive transnational vectors of this magnetic order are 2$\mathbf{a}$ and $\mathbf{b}$~\cite{Zare_2013}.
	To study the magnon instability of the $z$-AFM state with ordering vector $\mathbf{Q}=\mathbf{M}=(\pi,\pi/\sqrt{3})$ and the phase difference $\pi$ within each unit cell, we define the HP transformations at the LSW approximation as follows:
	
	\begin{figure}[t]
	\includegraphics[  width=0.75\linewidth]{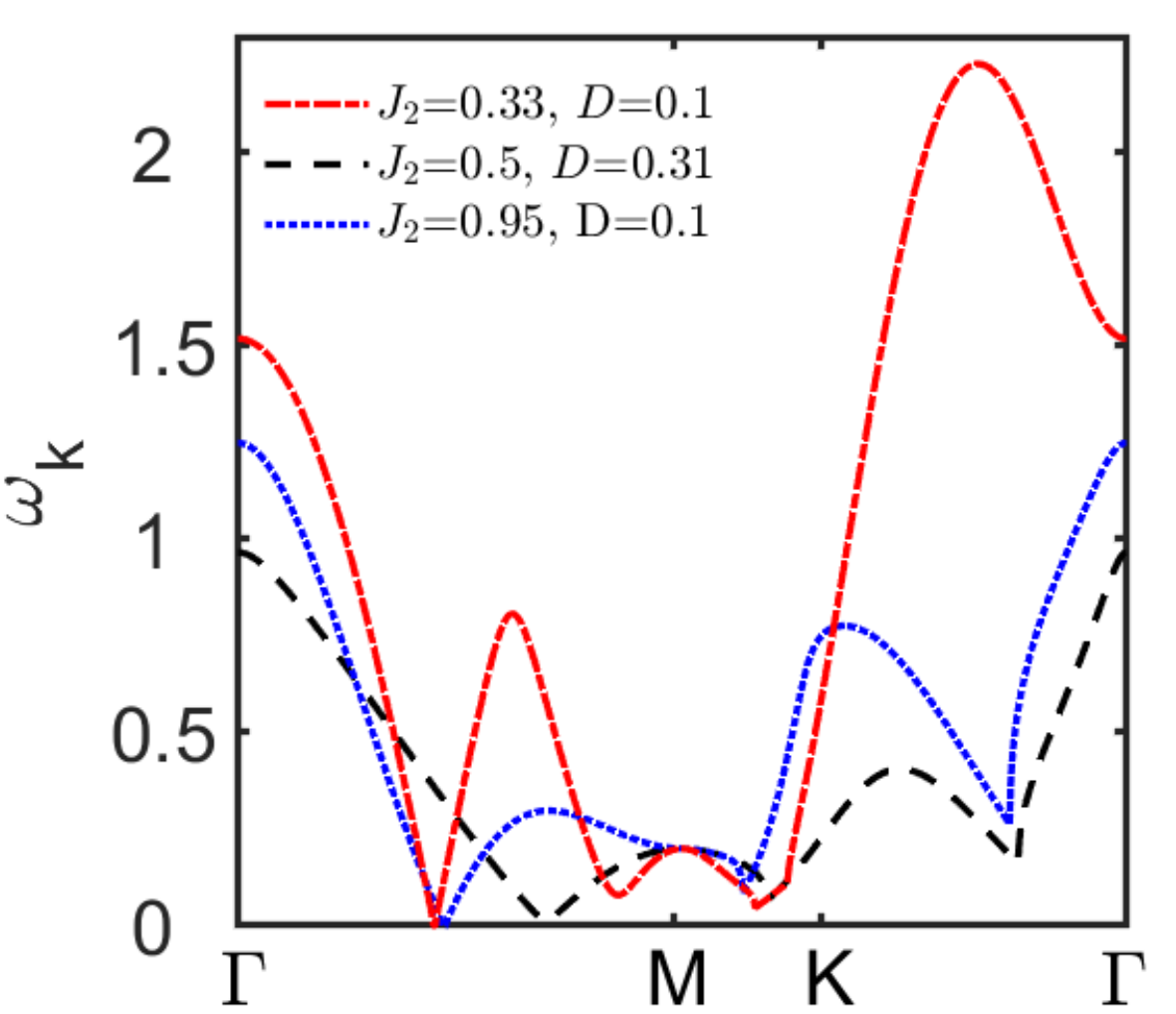}
	\caption{(Color online) 
		The obtained magnon spectra, along high symmetry
		directions in the FBZ, via a LSW calculation for some values of $J_2$ and $g_2$ in the border regions of the $z$-AFM state with its neighbor phases for $g_2=0.05$. 
	}	
	\label{fig:7}
\end{figure}
	
	\begin{align}
		\begin{aligned}
			&
			S^{z}_{i,\zeta}=S-a^{\dagger}_{i,\zeta}a^{}_{i,\zeta},~~~~ S^{z}_{i,\zeta}=-S+a^{\dagger}_{i,\zeta}a^{}_{i,\zeta},~
			\\ &
			S^{\dagger}_{i,\zeta}=\sqrt{2S}a^{}_{i,\zeta},~~~~~~~~
			S^{\dagger}_{i,\zeta}=\sqrt{2S}a^{\dagger}_{i,\zeta},~
			\\ &
			S^{-}_{i}=\sqrt{2S}a^{\dagger}_{i,\zeta},
			~~~~~~~~~ 
			S^{-}_{i,\zeta}=\sqrt{2S}a^{}_{i,\zeta}
			\\ &
			\zeta=1, 4,~~~~~~~~~~~~~~~~~ \zeta=2,3.
			\label{eq:26}
		\end{aligned}
	\end{align}

	To use these transformations,  
	the quadratic Hamiltonian in momentum space is given by
	
	\begin{align}
		{\cal H}_2= S \sum_{\mathbf{k}>0} \psi^{\dagger}_{\mathbf{k}} {\cal J}^{}_{\mathbf{k}} \Psi^{}_{\mathbf{k}},
		\label{eq:27} 
	\end{align} 
	with $\psi^{\dagger}_{\mathbf{k}}=(a^{\dagger}_{1\mathbf{k}},a^{\dagger}_{2\mathbf{k}},a^{\dagger}_{3\mathbf{k}},a^{\dagger}_{4\mathbf{k}},a^{}_{1-\mathbf{k}},a^{}_{2-\mathbf{k}},a^{}_{3-\mathbf{k}},a^{}_{4-\mathbf{k}})$
	and the $8\times8$ matrix ${\cal J}^{}_{\mathbf{k}}$ that is given by
	
	\begin{align}
		{\cal J}_{\mathbf{k}} = \begin{pmatrix} {\cal X}^{+}_{\mathbf{k}}  & 0 & 0 & J_1 & 0 & {\cal F}^{}_{\mathbf{k}} & {\cal G}^{}_{\mathbf{k}} & 0 \\
			0 & {\cal X}^{+}_{\mathbf{k}} & J_1 & 0 & {\cal F}^{}_{-{\mathbf{k}}} & 0 & 0 & {\cal L}^{}_{\mathbf{k}}
			\\
			0 & J_1 & {\cal X}^{-}_{\mathbf{k}} & 0 & {\cal G}^{*}_{\mathbf{k}} & 0 & 0 & {\cal F}^{*}_{\mathbf{k}}
			\\
			J_1 & 0 & 0 & {\cal X}^{-}_{\mathbf{k}} & 0 & {\cal L}^{*}_{\mathbf{k}} & {\cal F}^{*}_{-\mathbf{k}} & 0 
			\\
			0 & {\cal F}^{*}_{-{\mathbf{k}}} & {\cal G}^{}_{\mathbf{k}} & 0 & {\cal X}^{-}_{\mathbf{k}} &  0 & 0 & J_1
			\\
			{\cal F}^{*}_{\mathbf{k}} & 0 & 0 &{\cal L}^{}_{\mathbf{k}} & 0 & {\cal X}^{-}_{\mathbf{k}} & J_1 & 0
			\\
			{\cal G}^{*}_{\mathbf{k}} & 0 & 0 & {\cal F}^{}_{-\mathbf{k}} & 0 & J_1 & {\cal X}^{+}_{\mathbf{k}} & 0
			\\
			0 & {\cal L}^{*}_{\mathbf{k}} & {\cal F}^{}_{\mathbf{k}} & 0 & J_1 & 0 & 0 & {\cal X}^{+}_{\mathbf{k}} 
		\end{pmatrix},
		\label{eq:28}
	\end{align}
	%----------------------------------------------------
	%
	whose elements are written in Appendix: {\ref{appB}}.

	Here, we explore the magnon instability of the $z$-AFM phase in going to its neighbor phases.
	While not shown, the magnon spectra in the bulk phase of the $z$-AFM ordering are fully gaped owing to the spin rotational symmetry breaking~\cite{Zare_2013}.  
	Magnon instability for the $z$-AFM state in going to its neighbor phases occurs at the incommensurate wave vectors along the ${\bm \Gamma}-\mathbf{M}$ in the FBZ, as depicted in Fig.~\ref{fig:7}.
	Note that the magnon instability boundary of the $z$-AFM phase in going to the NCL and NCP phases is far beyond its classical boundaries resulting in a first-order transition. 
	While in the LSW approximation, the border region of the $z$-AFM phase with the classical spiral coincides with the result shown in Fig. \ref{fig:5}. Thus, this fact supports a second-order phase transition between them in the presence of the quantum fluctuations.

	\begin{figure}[t]
	%	\begin{center}
	\includegraphics[scale=.75]{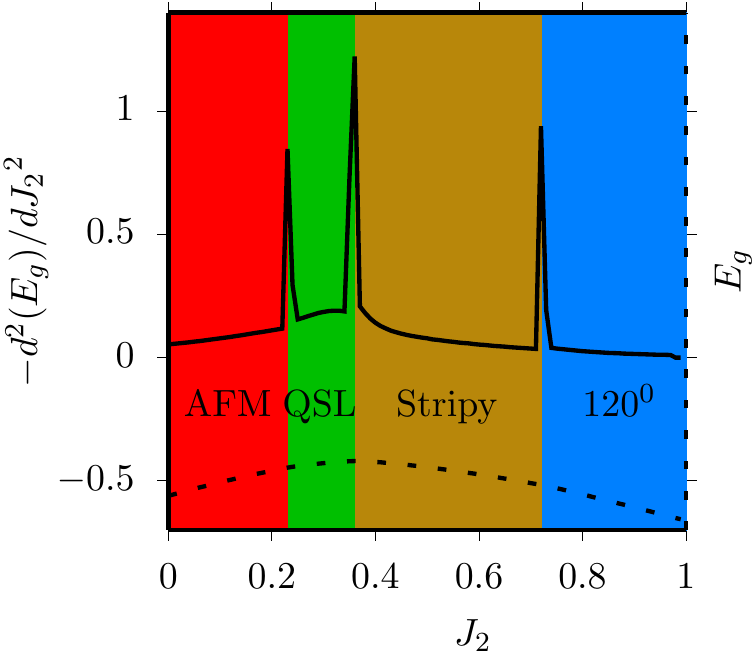}
	\vspace{-.2cm}
	\caption{ (Color online) The ground state energy per site $E_g$ (dashed line), and the second derivative of $E_g$ with respect to $J_2$, $-d^{2}E_{g}/dJ^{2}_{2}$, (solid line) as a function of $J_2$, for given $D=0.1$ and $g_2=0$.}
	\label{fig:8}
\end{figure}

	We know that the ordering vector according to 
	the classical energy minimization determines the spin configuration for the chosen set of the coupling constants and dictates the number of sublattices within each magnetic unit cell.
	In our spin-wave analysis, we shall neglect the magnon stability of the 120$^{\circ}$-AFM and NCL phases which have large magnetic unit cells. This assumption is reasonable by the fact that these large magnetic unit cells are destabilized by quantum fluctuations and therefore are expected to completely disappear in the small values of $S$, which we are mainly interested in.
	For the NCP phase with both incommensurate in-plane and out-of-plane ordering vectors, 
	the elements of the exchange matrix, ${\cal J}^{\alpha\beta}_{ij}$, are site-dependent and thus the LSW Hamiltonian (Eq.~\ref{eq:20}) cannot be diagonalized using FT~\cite{Zare_2013}.   
	Therefore, it is required to use a numerical method to figure out the stability of these classical magnetic 
	phases
	in contrast with the quantum fluctuations.
	This is the subject of the next section (in Sec.~\ref{secV}) that will utilize the DMRG method for this end. 

	\section{Quantum phase diagram}
	\label{secV}
	To identify the zero-temperature phases of the quantum $S=1/2$ case of the model (Eq.~\ref{eq:3}), here we perform numerical studies for small systems
	based on matrix product using the open-source ALPS libraries~\cite{Bauer_2011}. % 
	For the purpose of this work, we use honeycomb lattices under both periodic and twisted boundary conditions~\cite{Thesberg_2014}, containing  $2\times L \times W $ lattice sites.
	 We use clusters with $L$ and $W$ honeycomb unit cells along the ${\mathbf a}$ and ${\mathbf a}+{\mathbf b}$ directions, respectively [Fig.~\ref{fig:1}].
	Here, we take a cluster with $2\times3\times6$ sites, which has the full symmetry of the classical magnetic orders, thus allowing us to study the collinear, 120$^{\circ}$, and NCP phases. 
	To remove the finite size effects, we study the order parameters for several cluster sizes
	up to $N=2\times 5 \times 5$. In addition, the truncation error is decreased to $10^{-5}J_1$ or smaller by keeping 1000 density matrix eigenstates in the renormalization procedure and performing 10 sweeps. 
	This method may help us to gain further insight into the states with large magnetic unit cells and to realize a stable realm of the QSL behavior in the twisted TMDs where quantum fluctuations play a prominent role.

	\begin{figure}[t]
		\begin{center}
			\includegraphics[scale=.5]{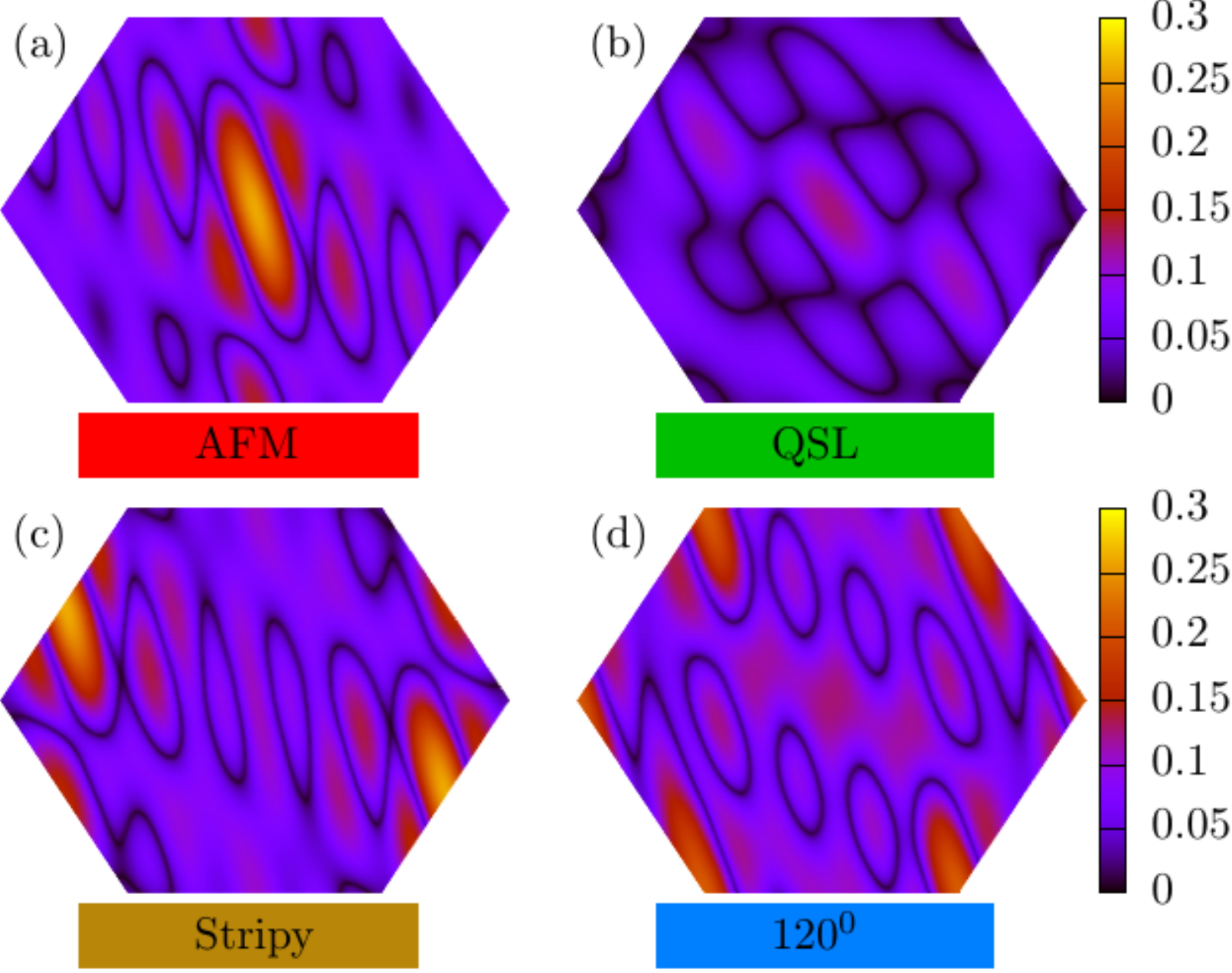}
			\vspace{.cm}
			\caption{ (Color online) Zero-temperature DMRG static magnetic structure factor, $\mathbf{S}({\mathbf{k}})$, with $D=0.1$ and $g_2=0$ for 
				(a) the AFM phase ($J_2=0.1$), (b) the QSL phase ($J_2=0.3$), (c) the stripy phase ($J_2=0.6$), and (d) the 120$^{\circ}$ ($J_2=0.9$).}
			\label{fig:9}
		\end{center}
	\end{figure}

	In order to construct a quantum phase diagram of the Hamiltonian (Eq.~\ref{eq:3}), we compute the ground state energy per site, $E_g$, its second derivative with respect to $J_2$, $-d^{2}E_{g}/dJ^{2}_{2}$, and the static spin structure factor defined as follows:

	\begin{equation}
		\mathbf{S}({\mathbf{k}})=N^{-1}_{}\sum_{ij}^{} \langle 0|\large(\mathbf{S}^{}_{i} \cdot \mathbf{S}^{}_{j}\large) |0\rangle e^{-i{\mathbf{k}}\cdot(\mathbf{r}_{i}-\mathbf{r}_{j})} 
	\end{equation}
	where $N$ is the number of sites and $\mathbf{r}_{i}$ is the lattice vector at site $i$.
	Here, $|0\rangle$ represents the ground state wave function.  They are frequently utilized as probes for the characterization of the phase transitions in a variety of models.

\begin{figure}[t]
	 	\includegraphics[scale=.8]{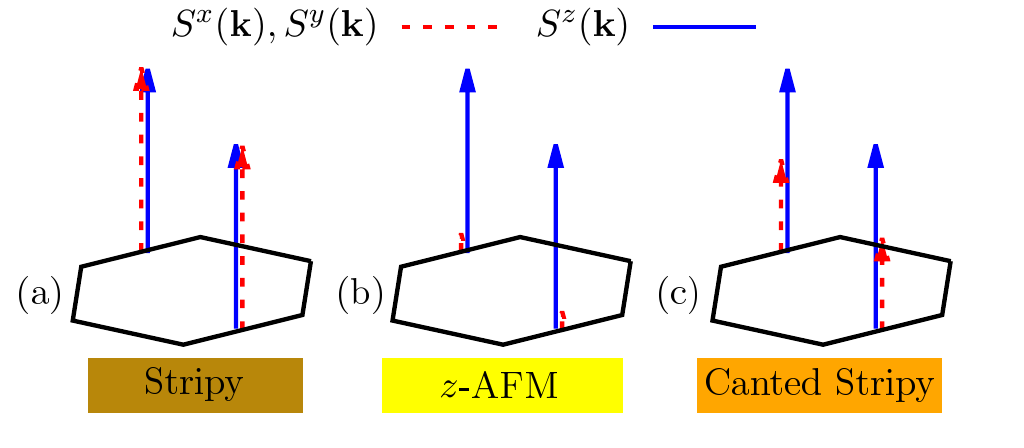}
	 	\caption{ (Color online) A comparison of scaled static magnetic structure factor for three spin components, $S^{x,y,z}({\mathbf k})$, in the momentum space
	 		for (a) the stripy phase, (b) the $z$-AFM phase, and (c) the canted stripy phase, which are characterized by a peak of ${\mathbf S}({\mathbf k})$ locating at the ${\bf M}$ point. 
	 		The dashed, and solid arrows represent the in-plane ($x$ and $y$), and out-of-plane ($z$) components of ${\mathbf S}({\mathbf k})$, respectively.}
	 	\label{fig:10}
	 \end{figure}

\begin{figure}[t]
	\begin{center}
		\vspace{0.10cm}
		\hspace{-0.1cm}
		\includegraphics[width=1.05 \linewidth]{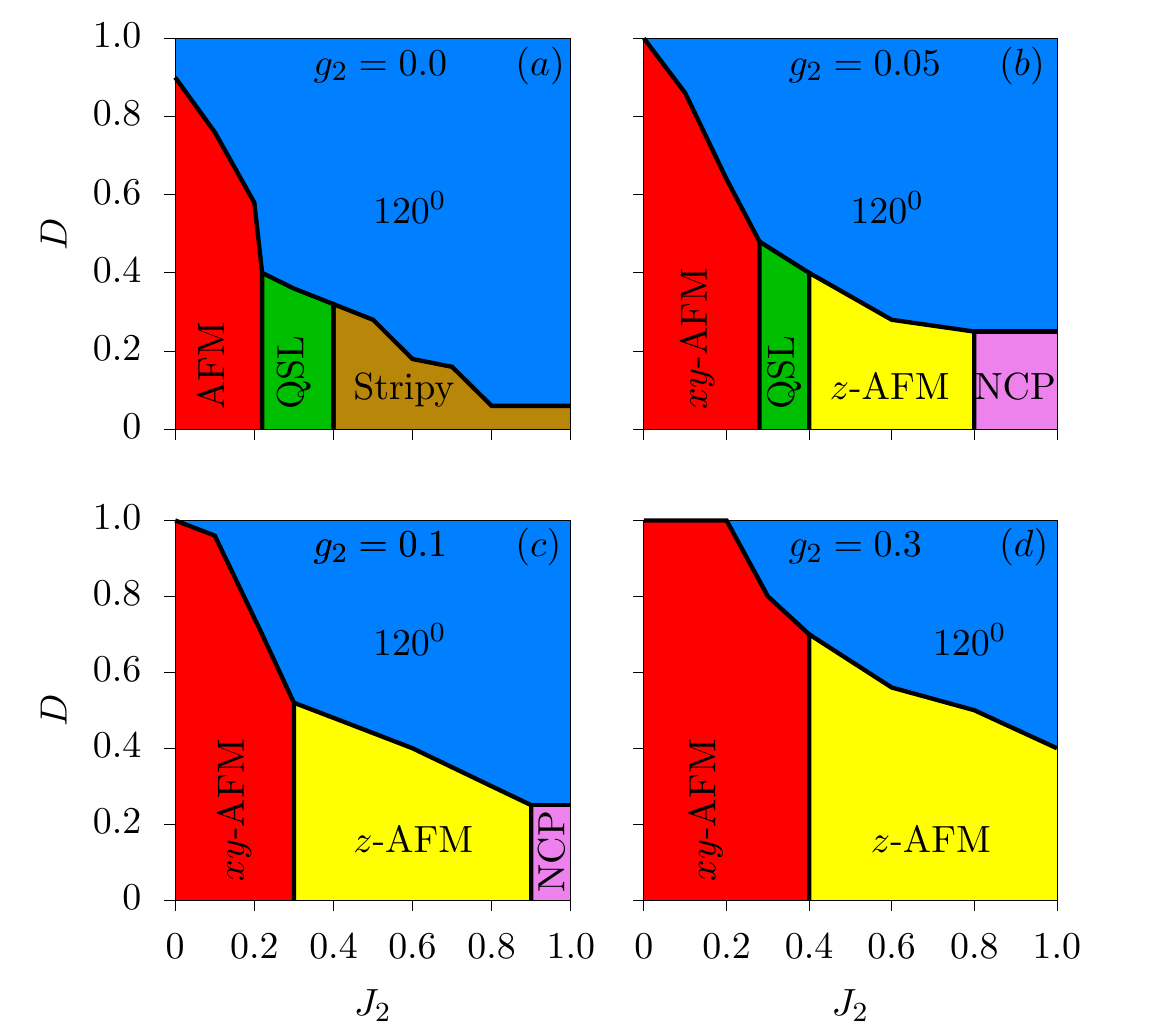}  \vspace{-.50cm}
		\includegraphics[width=0.85\linewidth]{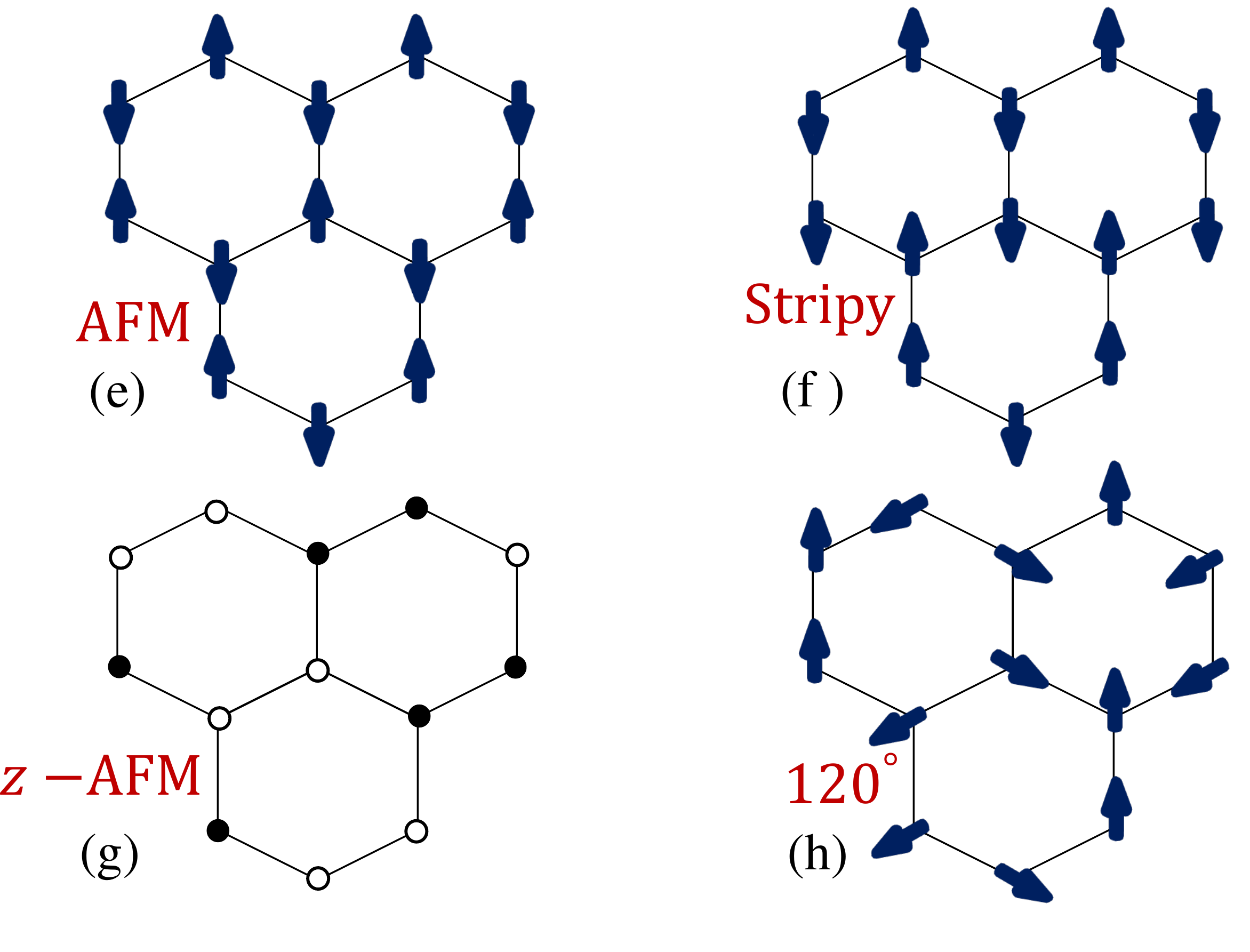} \hspace{.0cm}
	\end{center}
	\vspace{-0.2cm}
	\caption{ (Color online) 
	 Quantum ground state phase diagram obtained from the DMRG method in the plane of $J_2-D$ for some values of the KM term: (a) $g_2=0$, (b) $g_2=0.05$, (c) $g_2=0.1$, and (d) $g_2=0.3$. Real-space ordering for some spin configurations in (e) the N\'eel antiferromagnetic, (f) the stripy state, (g) the $z$-AFM ordered phase, and (h) the 120$^{\circ}$ phase.
}
	\label{fig:11}	 	%
\end{figure}

	Figure~\ref{fig:8} shows the ground state energy per site $E_g$ and the second derivative of $E_g$ as a function of $J_2$, which is obtained by the DMRG method, for given $D=0.1$ and $g_2=0$.
	As mentioned earlier, in analogy to their
	classical counterparts, the phase transitions are signaled by three singular behaviors in  
	$-d^{2}E_{g}/dJ^{2}_{2}$.
	In this section, we discriminate between the phases on either side of anomaly points using the static spin structure factor $\mathbf{S}({\mathbf{k}})$.
	Close to $J_2\approx0.22$,
	the system undergoes a continuous phase transition from the AFM phase to the QSL phase.
	 As shown in Fig.~\ref{fig:9}(a), the AFM Bragg peak appears only at the ${\bm \Gamma}$ point for $J_2<0.22$, while for the QSL phase ($0.22<J_2<0.38$), the spin structure factor is not necessarily in the form of Bragg peaks, and the absence of any peak in the reciprocal space confirms the QSL state [Fig.~\ref{fig:9}(b)].
	The other anomaly point at $J_2\approx0.38$ comes from the separation of the QSL state and the stripy phase.
	For $0.38<{J_2}<0.75$, an antiferromagnetic phase with a four-site magnetic unit cell (stripy) is stabilized with Bragg peaks at $\mathbf{M}$ point as shown in Fig.~\ref{fig:9}(c).
	To investigate in detail the stripy phase, we show the static structure factor for three spin components, ${S}^{\rho}({\mathbf k})$, in Fig.~\ref{fig:10}(a).
	It is indicated that the ${S}^{\rho}({\mathbf k})$ main peaks for the stripy phase are of the same magnitude at the $\mathbf{M}$ point.
	With an increase in $J_2$ up to $J_2\approx0.75$, the intensity of $\mathbf{S}({\mathbf{k}})$ increases at $\mathbf{K}$ point which represents a tendency toward the 120$^{\circ}$ order [Fig.~\ref{fig:9}(d)].
	Fixing $g_2=0$, there are four different phases in the plane of $J_2$ and $D$ [Fig.~\ref{fig:11}(a)]: a AFM phase, a spin liquid phase, a stripy-type antiferromagnetic phase (stripy), and a 120$^{\circ}$ phase.
	 Here, we reveal the intriguing role played by the DM interaction for the stability of the 120$^{\circ}$ phase. The unique aspect of this phase diagram is the stability of the 120$^{\circ}$ in a rather large extent region in the ${J_2}-D$ parameter space induced by the quantum fluctuations.
	 
%%%%%%%%%%%%%%%%%====== figure ======%%%%%%%%%%%%%%%%%

Finally, let us examine the case of $g_2\neq0$.
With an increase in $g_2$ up to $g_2=0.05$ for given $D=0.1$, phase changes from AFM to $xy$-AFM for $J_2<0.28$ in which in-plane spin-spin correlation has a finite value only at the ${\bm\Gamma}$ point, while $\langle S^{z}_{i}S^{z}_{j}\rangle$ is a very short range (not shown).
We next examine the frustrated case of $J_2=0.35$. Since no Bragg peaks exist in the FBZ, similar to the case of $g_2=0$, it can be used to deduce lack of ordering (not shown).
	As a result, the twisted homobilayer TMDs have been suggested as a possible QSL material with a frustrated honeycomb lattice containing DM magnetic interaction.
	With further increasing $J_2$ up to $J_2\approx0.4$, the system undergoes a quantum phase transition from the QSL phase to the $z$-AFM phase with the magnetic Bragg peak located at the $\mathbf{M}$ point of the FBZ in which the $x$ and $y$ components ${\mathbf S}({\mathbf k})$ are very small compared to the $ S^{z}({\mathbf k})$ one [Fig.~\ref{fig:10}(b)].
   	We then tune the next-nearest interaction $J_2$ for the range of  $0.8<J_2<1$. For these couplings, we identify that the Bragg peaks appear at incommensurate points in which both in-plane and out-of-plane components of the spin-spin correlation have finite values  (not shown).
   	Here, one observes an increase in the magnetic Bragg intensity of the in-plane components, $S^{x,y}({\mathbf k})$, in going from the $z$-AFM state to the NCP phase with further increasing $J_2$.
	In total, this analysis shows that
   	an infinitesimal KM coupling is already sufficient to generate the $z$-AFM and NCP phases, whereas the stripy phase is significantly suppressed as illustrated in Fig.~\ref{fig:11}(b).
	 However, for the small values of $D$ ($D<0.03$) in the phase diagram, a canted stripy phase with the Bragg peak at the ${\mathbf M}$ is observed (not shown), in which 
	 in-plane and out-of-plane components of the static spin structure factor are of different magnitudes [Fig.~\ref{fig:10} (c)]. 
	In addition, in a good agreement with the classical phase diagram, the numerical results from the DMRG method illustrate that the QSL phase is destabilized with an increase in $g_2$ up to $g_2\approx 0.1$  [Figs.~\ref{fig:11}(c) and \ref{fig:11}(d)].

	\section{conclusions}
	In our studies, inspired by the recent developments of the twisted bilayer materials, we constructed an effective spin Hamiltonian model (Eq.~\ref{eq:3}) on a honeycomb lattice for the twisted homobilayer TMDs in the strongly correlated limit. In the classical spin system, we obtained the ground state of the generic model Hamiltonian using the LT and VO methods.
	For the spin Hamiltonian (Eq.~\ref{eq:3}), in the context of classical calculations, we find that the second-neighbor DM coupling plays a crucial role in the stability of the magnetic phases which have large magnetic unit cells
	and extends the stability region of the spiral spin liquid in comparison to the $J_{1}-J_{2}$ Heisenberg model~\cite{Zare_2013,Mosadeq_2011}.
	Meanwhile, the quantum effects on the stability of the classical magnetic phases % 
	were examined using the LSW and DMRG methods. 
   To evaluate the effects of the quantum fluctuations on the ordered phases, we investigate the magnon stability using the LSW method. Our findings indicate that the classical phase boundaries were predicted to survive in contrast with the quantum fluctuations.  
   However, for the state with high classical degeneracy, we found no indication of quantum order-by-disorder 
   which can lift the degeneracy and induce a gap for these states due to the quantum zero-point fluctuations.
   In the quantum system, our numerical results show that there exist four distinct phases involving AFM, QSL, stripy, and 120$^{\circ}$ in the absence of the KM term, $g_2=0$: A region of the quantum spin liquid separates the AFM phase from the stripy phase for the finite values of $D$ up to $D\approx0.4$, while the 120$^{\circ}$ order can be stabilized by the DM interaction for all the values of $J_2$ with increasing $D$.    
   The main effect of $g_2$ is to reduce the frustration interaction for when it goes from a vanishing value to a finite one. As a result, the QSL phase, which is an extremely unusual magnetic state with highly correlated spins, vanishes for the value $g_2$ about $g_2\approx0.1$.
   With increasing $g_2$, we make the quantum phase diagram of the effective model (Eq.~\ref{eq:3}), whose phase diagrams are established. There are a $xy$-AFM phase, a $z$-AFM phase, a NCP phase, and a 120$^{\circ}$ phase.

  Finally, let us comment on the material trend in the twisted TMD bilayers.    
   The emergence of the strong correlation effect in the twisted materials can be provided by fine-tuning to a magic twist angle where the low energy bands become weakly dispersive.
   Therefore, these twisted van der Walls systems~\cite{Wu_2019,devakul_2021,zhang_2021} can be used for the realization of the Kane-Mele-Hubbard model for studying correlated insulating states, unconventional superconductors, fractional quantum Hall states, and quantum spin liquids.

   \section{ACKNOWLEDGMENTS}
	We thank S. Akbar Jafari and Mehdi Biderang for careful reading and constructive comments on the manuscript.	
	M. H. Z. was supported by grant No. G139163, research deputy of Qom University of Technology.
	The authors would like to appreciate the use of the computational clusters of the HPC center (School of Computer Science, IPM, Tehran, Iran), to complete this work.

	\begin{center}
		\textbf{\large Appendix}
	\end{center}
	\appendix{}

	\section{Functions Defining the Classical Energies}
	\label{appA}

	To substitute the parametrization of the spin by Eq.~(\ref{eq:14}) into Eq.~(\ref{eq:3}), the classical variational energy per spin for the states with commensurate wave vector (${\mathbf Q}^{'}_{C}$) find a general form as follows 
	
	\begin{widetext}   
		\begin{align}
			\begin{aligned}
				{\cal {E}} _C({\mathbf{Q}},{\mathbf{Q}'},\varphi,\gamma,\eta)
				= \frac{J_1}{2}\sum_{\bm{\delta}_1}[\cos({\mathbf{Q}}\cdot{\bm{ \delta}_1}-\varphi)\sin({\mathbf{Q}'}\cdot {\bm{ \delta}_1}+\gamma)\sin{\eta}+\cos({\mathbf{Q}}\cdot{\bm{ \delta}_1}+\varphi)\cos{\eta}]
				\\ &
				\hspace{-10 cm} + \frac{(J_2-g_2)}{4}\sum_{\bm{\delta}_2}\cos({\mathbf{Q}}\cdot{\bm{ \delta}_2})[2\cos({\mathbf{Q}'}\cdot {\bm{ \delta}_2})- \cos({\mathbf{Q}'}\cdot{\bm{ \delta}_2}+2\gamma)-\cos({\mathbf{Q}'}\cdot {\bm{ \delta}_2}+2{\eta})]
				\\ &
				\hspace{-10 cm} + \frac{(J_2+g_2)}{4}\sum_{\bm{\delta}_2}[2\cos({\mathbf{Q}'}\cdot {\bm{ \delta}_2})+ \cos({\mathbf{Q}'}\cdot{\bm{ \delta}_2}+2\gamma)+\cos({\mathbf{Q}'}\cdot {\bm{ \delta}_2}+2{\eta})]
				\\ &
				\hspace{-10 cm} -
				\frac{D}{2}\sum_{\bm{\delta}_2} \nu_{ij} [\sin({\mathbf{Q}'}\cdot {\bm{ \delta}_2}+\gamma)\sin({\gamma}) - \sin({\mathbf{Q}'}\cdot {\bm{ \delta}_2} + \eta)\sin({\eta})],
				\label{eq:a1}
			\end{aligned}    
		\end{align}
		while for the states with incommensurate wave vector (${\mathbf{Q}}^{'}_{IC}$), the variational energy read as
		\begin{align}
			\begin{aligned}
				{\cal {E}} _{IC}({\mathbf{Q}},{\mathbf{Q}'},\varphi,\gamma,\eta)
				= \frac{J_1}{4}\sum_{\bm{\delta}_1}[\cos({\mathbf{Q}}\cdot{\bm{ \delta}_1}-\varphi)+1]\cos({\mathbf{Q}'}\cdot {\bm{ \delta}_1}+\gamma-\eta)
				\\ &
				\hspace{-7 cm}
			 +  \frac{(J_2-g_2)}{2}\sum_{\bm{\delta}_2}\cos({\mathbf{Q}}\cdot{\bm{ \delta}_2})\cos({\mathbf{Q}'}\cdot {\bm{ \delta}_2})
				+ \frac{(J_2+g_2)}{2}\sum_{\bm{\delta}_2} \cos({\mathbf{Q}'}\cdot {\bm{ \delta}_2}).
				\label{eq:a2}
			\end{aligned}    
		\end{align}
	\end{widetext} 
	
	where
	${\bm\delta}_1=0, \mathbf{b}, \mathbf{a+b}$  and ${\bm\delta}_2=\mathbf{a}, \mathbf{b}, \mathbf{a+b}$ are the unit-cell position vectors of the nearest- and next-nearest neighbors
	of a given lattice point, respectively.
	
	\section{Matrices Elements of the LSW Hamiltonian  }
	\label{appB}
	
	The LSW Hamiltonian matrix elements related to in-plane magnetic structures
	in Eq.~(\ref{eq:25}) read as
	
	\begin{widetext}
		\begin{align}
			\begin{aligned}
				&
				A^{\pm}_{\mathbf{k}}=
				\frac{J_1}{2}\sum_{{\bm\delta}_1}\cos({\mathbf{Q}}\cdot{\bm\delta}_1-\varphi)-(J_2-g_2)\sum_{{\bm\delta}_2}\cos({\mathbf{Q}}\cdot{\bm\delta}_2)
				+\frac{1}{2} \sum_{{\bm\delta}_2} [J_2+g_2+(J_2-g_2)\cos({\mathbf{Q}}\cdot{\bm\delta}_2)]\cos({\mathbf{k}}\cdot{\bm\delta}_2)
				\\ & \hspace{1cm}
				\pm \frac{D}{2}[\sin{\mathbf{Q}}_a\cos{\mathbf{k}}_a
				+\sin{\mathbf{Q}}_b\cos{\mathbf{k}}_b
				-\sin({\mathbf{Q}}_a+{\mathbf{Q}}_b)\cos({\mathbf{k}}_a+{\mathbf{k}}_b)]
				\mp D [\sin{\mathbf{Q}}_a+\sin{\mathbf{Q}}_b -\sin({{\mathbf{Q}}_a}+{\mathbf{Q}}_b)]
				,
				\\ &
				B^{}_{\mathbf{k}}=\frac{J_1}{2}\sum_{{\bm\delta}_1} [1-\cos({\mathbf{Q}}\cdot{\bm\delta}_1-\varphi)]e^{-i{\mathbf{k}}\cdot{{\bm\delta}_1}}
				\\ &
				{C}^{\pm}_{\mathbf{k}}=
				\frac{1}{2}\sum_{{\bm\delta}_2} [J_2+g_2-(J_2-g_2)\cos({\mathbf{Q}}\cdot{\bm\delta}_2)]\cos({\mathbf{k}}\cdot{\bm\delta}_2)
				\pm \frac{D}{2}[\sin{\mathbf{Q}}_a\cos{\mathbf{k}}_a
				+\sin{\mathbf{Q}}_b\cos{\mathbf{k}}_b
				-\sin({\mathbf{Q}}_a+{\mathbf{Q}}_b)\cos({\mathbf{k}}_a+{\mathbf{k}}_b)],
				\\ &
				D^{}_{\mathbf{k}}=\frac{J_1}{2}\sum_{{\bm\delta}_1} [1+\cos({\mathbf{Q}}\cdot{\bm\delta}_1-\varphi)]e^{-i{\mathbf{k}}\cdot{\bm\delta}_1}
				\label{eq:b1}
			\end{aligned}
		\end{align}

		The explicit expressions of the Hamiltonian elements matrix associated with the $z$-AFM in Eq. (\ref{eq:28}) given by

		\begin{align}
			\begin{aligned}
				&
				{\cal X}^{\pm}_{\mathbf{k}}=J_1+2(J_2+g_2)+2(J_2-g_2)\cos{\mathbf{k}}_b \pm 2D \sin{\mathbf{k}}_b,
				\\ &
				{\cal F}_{\mathbf{k}}=(J_2-g_2)(1+e^{-i {\mathbf{k}}_a} + e^{i {\mathbf{k}}_b}+e^{-i ({\mathbf{k}}_a+{\mathbf{k}}_b)})+iD(1+e^{-i {\mathbf{k}}_a} - e^{i {\mathbf{k}}_b}-e^{-i ({\mathbf{k}}_a+{\mathbf{k}}_b)}),
				\\ &
				{\cal G}_{\mathbf{k}}=J_1(e^{-i {\mathbf{k}}_a}+e^{-i ({\mathbf{k}}_a+{\mathbf{k}}_b)}),
				\\ &
				{\cal L}_{\mathbf{k}}=J_1(1+e^{-i {\mathbf{k}}_b}).
				\label{b2}
			\end{aligned}
		\end{align}
	\end{widetext}

	%%%%%%%%%%%%%%%%%%%%%%%%%%%%%%%%%%%%%%%%%%%%%%%%%%%%%%%%%%%
	%%%%%%%%%%%%%%%%%%%%%%%%%%%%%%%%%%%%%%%%%%%%%%%%%%%%%%%%%%% 
	\bibliography{References}

\begin{thebibliography}{108}%
\makeatletter
\providecommand \@ifxundefined [1]{%
 \@ifx{#1\undefined}
}%
\providecommand \@ifnum [1]{%
 \ifnum #1\expandafter \@firstoftwo
 \else \expandafter \@secondoftwo
 \fi
}%
\providecommand \@ifx [1]{%
 \ifx #1\expandafter \@firstoftwo
 \else \expandafter \@secondoftwo
 \fi
}%
\providecommand \natexlab [1]{#1}%
\providecommand \enquote  [1]{``#1''}%
\providecommand \bibnamefont  [1]{#1}%
\providecommand \bibfnamefont [1]{#1}%
\providecommand \citenamefont [1]{#1}%
\providecommand \href@noop [0]{\@secondoftwo}%
\providecommand \href [0]{\begingroup \@sanitize@url \@href}%
\providecommand \@href[1]{\@@startlink{#1}\@@href}%
\providecommand \@@href[1]{\endgroup#1\@@endlink}%
\providecommand \@sanitize@url [0]{\catcode `\\12\catcode `\$12\catcode
  `\&12\catcode `\#12\catcode `\^12\catcode `\_12\catcode `\%12\relax}%
\providecommand \@@startlink[1]{}%
\providecommand \@@endlink[0]{}%
\providecommand \url  [0]{\begingroup\@sanitize@url \@url }%
\providecommand \@url [1]{\endgroup\@href {#1}{\urlprefix }}%
\providecommand \urlprefix  [0]{URL }%
\providecommand \Eprint [0]{\href }%
\providecommand \doibase [0]{http://dx.doi.org/}%
\providecommand \selectlanguage [0]{\@gobble}%
\providecommand \bibinfo  [0]{\@secondoftwo}%
\providecommand \bibfield  [0]{\@secondoftwo}%
\providecommand \translation [1]{[#1]}%
\providecommand \BibitemOpen [0]{}%
\providecommand \bibitemStop [0]{}%
\providecommand \bibitemNoStop [0]{.\EOS\space}%
\providecommand \EOS [0]{\spacefactor3000\relax}%
\providecommand \BibitemShut  [1]{\csname bibitem#1\endcsname}%
\let\auto@bib@innerbib\@empty
%</preamble>
\bibitem [{\citenamefont {Anderson}(1973)}]{ANDERSON_1973}%
  \BibitemOpen
  \bibfield  {author} {\bibinfo {author} {\bibfnamefont {P.}~\bibnamefont
  {Anderson}},\ }\href {\doibase https://doi.org/10.1016/0025-5408(73)90167-0}
  {\bibfield  {journal} {\bibinfo  {journal} {Materials Research Bulletin}\
  }\textbf {\bibinfo {volume} {8}},\ \bibinfo {pages} {153 } (\bibinfo {year}
  {1973})}\BibitemShut {NoStop}%
\bibitem [{\citenamefont {Lee}(2008)}]{Lee_2008}%
  \BibitemOpen
  \bibfield  {author} {\bibinfo {author} {\bibfnamefont {P.~A.}\ \bibnamefont
  {Lee}},\ }\href {\doibase 10.1126/science.1163196} {\bibfield  {journal}
  {\bibinfo  {journal} {Science}\ }\textbf {\bibinfo {volume} {321}},\ \bibinfo
  {pages} {1306} (\bibinfo {year} {2008})}\BibitemShut {NoStop}%
\bibitem [{\citenamefont {Balents}(2010)}]{Balents_2010}%
  \BibitemOpen
  \bibfield  {author} {\bibinfo {author} {\bibfnamefont {L.}~\bibnamefont
  {Balents}},\ }\href {https://doi.org/10.1038/nature08917} {\bibfield
  {journal} {\bibinfo  {journal} {Nature}\ }\textbf {\bibinfo {volume} {464}},\
  \bibinfo {pages} {199 EP } (\bibinfo {year} {2010})}\BibitemShut {NoStop}%
\bibitem [{\citenamefont {Savary}\ and\ \citenamefont
  {Balents}(2016)}]{Savary_2016}%
  \BibitemOpen
  \bibfield  {author} {\bibinfo {author} {\bibfnamefont {L.}~\bibnamefont
  {Savary}}\ and\ \bibinfo {author} {\bibfnamefont {L.}~\bibnamefont
  {Balents}},\ }\href {\doibase 10.1088/0034-4885/80/1/016502} {\bibfield
  {journal} {\bibinfo  {journal} {Reports on Progress in Physics}\ }\textbf
  {\bibinfo {volume} {80}},\ \bibinfo {pages} {016502} (\bibinfo {year}
  {2016})}\BibitemShut {NoStop}%
\bibitem [{\citenamefont {Zhou}\ \emph {et~al.}(2017)\citenamefont {Zhou},
  \citenamefont {Kanoda},\ and\ \citenamefont {Ng}}]{Zhou_2017}%
  \BibitemOpen
  \bibfield  {author} {\bibinfo {author} {\bibfnamefont {Y.}~\bibnamefont
  {Zhou}}, \bibinfo {author} {\bibfnamefont {K.}~\bibnamefont {Kanoda}}, \ and\
  \bibinfo {author} {\bibfnamefont {T.-K.}\ \bibnamefont {Ng}},\ }\href
  {\doibase 10.1103/RevModPhys.89.025003} {\bibfield  {journal} {\bibinfo
  {journal} {Rev. Mod. Phys.}\ }\textbf {\bibinfo {volume} {89}},\ \bibinfo
  {pages} {025003} (\bibinfo {year} {2017})}\BibitemShut {NoStop}%
\bibitem [{\citenamefont {Knolle}\ and\ \citenamefont
  {Moessner}(2019)}]{Knolle_2019}%
  \BibitemOpen
  \bibfield  {author} {\bibinfo {author} {\bibfnamefont {J.}~\bibnamefont
  {Knolle}}\ and\ \bibinfo {author} {\bibfnamefont {R.}~\bibnamefont
  {Moessner}},\ }\href {\doibase 10.1146/annurev-conmatphys-031218-013401}
  {\bibfield  {journal} {\bibinfo  {journal} {Annual Review of Condensed Matter
  Physics}\ }\textbf {\bibinfo {volume} {10}},\ \bibinfo {pages} {451}
  (\bibinfo {year} {2019})}\BibitemShut {NoStop}%
\bibitem [{\citenamefont {Takagi}\ \emph {et~al.}(2019)\citenamefont {Takagi},
  \citenamefont {Takayama}, \citenamefont {Jackeli}, \citenamefont
  {Khaliullin},\ and\ \citenamefont {Nagler}}]{Takagi_2019}%
  \BibitemOpen
  \bibfield  {author} {\bibinfo {author} {\bibfnamefont {H.}~\bibnamefont
  {Takagi}}, \bibinfo {author} {\bibfnamefont {T.}~\bibnamefont {Takayama}},
  \bibinfo {author} {\bibfnamefont {G.}~\bibnamefont {Jackeli}}, \bibinfo
  {author} {\bibfnamefont {G.}~\bibnamefont {Khaliullin}}, \ and\ \bibinfo
  {author} {\bibfnamefont {S.~E.}\ \bibnamefont {Nagler}},\ }\href {\doibase
  10.1038/s42254-019-0038-2} {\bibfield  {journal} {\bibinfo  {journal} {Nature
  Reviews Physics}\ }\textbf {\bibinfo {volume} {1}},\ \bibinfo {pages} {264}
  (\bibinfo {year} {2019})}\BibitemShut {NoStop}%
\bibitem [{\citenamefont {Kitaev}(2003)}]{Kitaev_2003}%
  \BibitemOpen
  \bibfield  {author} {\bibinfo {author} {\bibfnamefont {A.}~\bibnamefont
  {Kitaev}},\ }\href {\doibase https://doi.org/10.1016/S0003-4916(02)00018-0}
  {\bibfield  {journal} {\bibinfo  {journal} {Annals of Physics}\ }\textbf
  {\bibinfo {volume} {303}},\ \bibinfo {pages} {2 } (\bibinfo {year}
  {2003})}\BibitemShut {NoStop}%
\bibitem [{\citenamefont {Kitaev}\ and\ \citenamefont
  {Preskill}(2006)}]{Kitaev_2006}%
  \BibitemOpen
  \bibfield  {author} {\bibinfo {author} {\bibfnamefont {A.}~\bibnamefont
  {Kitaev}}\ and\ \bibinfo {author} {\bibfnamefont {J.}~\bibnamefont
  {Preskill}},\ }\href {\doibase 10.1103/PhysRevLett.96.110404} {\bibfield
  {journal} {\bibinfo  {journal} {Phys. Rev. Lett.}\ }\textbf {\bibinfo
  {volume} {96}},\ \bibinfo {pages} {110404} (\bibinfo {year}
  {2006})}\BibitemShut {NoStop}%
\bibitem [{\citenamefont {Jiang}\ \emph {et~al.}(2012)\citenamefont {Jiang},
  \citenamefont {Wang},\ and\ \citenamefont {Balents}}]{Jiang_2012}%
  \BibitemOpen
  \bibfield  {author} {\bibinfo {author} {\bibfnamefont {H.-C.}\ \bibnamefont
  {Jiang}}, \bibinfo {author} {\bibfnamefont {Z.}~\bibnamefont {Wang}}, \ and\
  \bibinfo {author} {\bibfnamefont {L.}~\bibnamefont {Balents}},\ }\href
  {https://doi.org/10.1038/nphys2465} {\bibfield  {journal} {\bibinfo
  {journal} {Nature Physics}\ }\textbf {\bibinfo {volume} {8}},\ \bibinfo
  {pages} {902} (\bibinfo {year} {2012})}\BibitemShut {NoStop}%
\bibitem [{\citenamefont {ANDERSON}(1987)}]{Anderson_1987}%
  \BibitemOpen
  \bibfield  {author} {\bibinfo {author} {\bibfnamefont {P.~W.}\ \bibnamefont
  {ANDERSON}},\ }\href {\doibase 10.1126/science.235.4793.1196} {\bibfield
  {journal} {\bibinfo  {journal} {Science}\ }\textbf {\bibinfo {volume}
  {235}},\ \bibinfo {pages} {1196} (\bibinfo {year} {1987})}\BibitemShut
  {NoStop}%
\bibitem [{\citenamefont {Yamashita}\ \emph
  {et~al.}(2008{\natexlab{a}})\citenamefont {Yamashita}, \citenamefont
  {Nakata}, \citenamefont {Kasahara}, \citenamefont {Sasaki}, \citenamefont
  {Yoneyama}, \citenamefont {Kobayashi}, \citenamefont {Fujimoto},
  \citenamefont {Shibauchi},\ and\ \citenamefont {Matsuda}}]{Yamashita_2008a}%
  \BibitemOpen
  \bibfield  {author} {\bibinfo {author} {\bibfnamefont {M.}~\bibnamefont
  {Yamashita}}, \bibinfo {author} {\bibfnamefont {N.}~\bibnamefont {Nakata}},
  \bibinfo {author} {\bibfnamefont {Y.}~\bibnamefont {Kasahara}}, \bibinfo
  {author} {\bibfnamefont {T.}~\bibnamefont {Sasaki}}, \bibinfo {author}
  {\bibfnamefont {N.}~\bibnamefont {Yoneyama}}, \bibinfo {author}
  {\bibfnamefont {N.}~\bibnamefont {Kobayashi}}, \bibinfo {author}
  {\bibfnamefont {S.}~\bibnamefont {Fujimoto}}, \bibinfo {author}
  {\bibfnamefont {T.}~\bibnamefont {Shibauchi}}, \ and\ \bibinfo {author}
  {\bibfnamefont {Y.}~\bibnamefont {Matsuda}},\ }\href
  {https://doi.org/10.1038/nphys1134} {\bibfield  {journal} {\bibinfo
  {journal} {Nature Physics}\ }\textbf {\bibinfo {volume} {5}},\ \bibinfo
  {pages} {44} (\bibinfo {year} {2008}{\natexlab{a}})}\BibitemShut {NoStop}%
\bibitem [{\citenamefont {Yamashita}\ \emph
  {et~al.}(2008{\natexlab{b}})\citenamefont {Yamashita}, \citenamefont
  {Nakazawa}, \citenamefont {Oguni}, \citenamefont {Oshima}, \citenamefont
  {Nojiri}, \citenamefont {Shimizu}, \citenamefont {Miyagawa},\ and\
  \citenamefont {Kanoda}}]{Yamashita_2008}%
  \BibitemOpen
  \bibfield  {author} {\bibinfo {author} {\bibfnamefont {S.}~\bibnamefont
  {Yamashita}}, \bibinfo {author} {\bibfnamefont {Y.}~\bibnamefont {Nakazawa}},
  \bibinfo {author} {\bibfnamefont {M.}~\bibnamefont {Oguni}}, \bibinfo
  {author} {\bibfnamefont {Y.}~\bibnamefont {Oshima}}, \bibinfo {author}
  {\bibfnamefont {H.}~\bibnamefont {Nojiri}}, \bibinfo {author} {\bibfnamefont
  {Y.}~\bibnamefont {Shimizu}}, \bibinfo {author} {\bibfnamefont
  {K.}~\bibnamefont {Miyagawa}}, \ and\ \bibinfo {author} {\bibfnamefont
  {K.}~\bibnamefont {Kanoda}},\ }\href {https://doi.org/10.1038/nphys942}
  {\bibfield  {journal} {\bibinfo  {journal} {Nature Physics}\ }\textbf
  {\bibinfo {volume} {4}},\ \bibinfo {pages} {459} (\bibinfo {year}
  {2008}{\natexlab{b}})}\BibitemShut {NoStop}%
\bibitem [{\citenamefont {Yamashita}\ \emph {et~al.}(2010)\citenamefont
  {Yamashita}, \citenamefont {Nakata}, \citenamefont {Senshu}, \citenamefont
  {Nagata}, \citenamefont {Yamamoto}, \citenamefont {Kato}, \citenamefont
  {Shibauchi},\ and\ \citenamefont {Matsuda}}]{Yamashita_2010}%
  \BibitemOpen
  \bibfield  {author} {\bibinfo {author} {\bibfnamefont {M.}~\bibnamefont
  {Yamashita}}, \bibinfo {author} {\bibfnamefont {N.}~\bibnamefont {Nakata}},
  \bibinfo {author} {\bibfnamefont {Y.}~\bibnamefont {Senshu}}, \bibinfo
  {author} {\bibfnamefont {M.}~\bibnamefont {Nagata}}, \bibinfo {author}
  {\bibfnamefont {H.~M.}\ \bibnamefont {Yamamoto}}, \bibinfo {author}
  {\bibfnamefont {R.}~\bibnamefont {Kato}}, \bibinfo {author} {\bibfnamefont
  {T.}~\bibnamefont {Shibauchi}}, \ and\ \bibinfo {author} {\bibfnamefont
  {Y.}~\bibnamefont {Matsuda}},\ }\href {\doibase 10.1126/science.1188200}
  {\bibfield  {journal} {\bibinfo  {journal} {Science}\ }\textbf {\bibinfo
  {volume} {328}},\ \bibinfo {pages} {1246} (\bibinfo {year}
  {2010})}\BibitemShut {NoStop}%
\bibitem [{\citenamefont {Yamashita}\ \emph {et~al.}(2011)\citenamefont
  {Yamashita}, \citenamefont {Yamamoto}, \citenamefont {Nakazawa},
  \citenamefont {Tamura},\ and\ \citenamefont {Kato}}]{Yamashita_2011}%
  \BibitemOpen
  \bibfield  {author} {\bibinfo {author} {\bibfnamefont {S.}~\bibnamefont
  {Yamashita}}, \bibinfo {author} {\bibfnamefont {T.}~\bibnamefont {Yamamoto}},
  \bibinfo {author} {\bibfnamefont {Y.}~\bibnamefont {Nakazawa}}, \bibinfo
  {author} {\bibfnamefont {M.}~\bibnamefont {Tamura}}, \ and\ \bibinfo {author}
  {\bibfnamefont {R.}~\bibnamefont {Kato}},\ }\href
  {https://doi.org/10.1038/ncomms1274} {\bibfield  {journal} {\bibinfo
  {journal} {Nature Communications}\ }\textbf {\bibinfo {volume} {2}},\
  \bibinfo {pages} {275} (\bibinfo {year} {2011})}\BibitemShut {NoStop}%
\bibitem [{\citenamefont {Powell}\ and\ \citenamefont
  {McKenzie}(2011)}]{Powell_2011}%
  \BibitemOpen
  \bibfield  {author} {\bibinfo {author} {\bibfnamefont {B.~J.}\ \bibnamefont
  {Powell}}\ and\ \bibinfo {author} {\bibfnamefont {R.~H.}\ \bibnamefont
  {McKenzie}},\ }\href {\doibase 10.1088/0034-4885/74/5/056501} {\bibfield
  {journal} {\bibinfo  {journal} {Reports on Progress in Physics}\ }\textbf
  {\bibinfo {volume} {74}},\ \bibinfo {pages} {056501} (\bibinfo {year}
  {2011})}\BibitemShut {NoStop}%
\bibitem [{\citenamefont {Li}\ \emph {et~al.}(2015{\natexlab{a}})\citenamefont
  {Li}, \citenamefont {Liao}, \citenamefont {Zhang}, \citenamefont {Li},
  \citenamefont {Jin}, \citenamefont {Ling}, \citenamefont {Zhang},
  \citenamefont {Zou}, \citenamefont {Pi}, \citenamefont {Yang}, \citenamefont
  {Wang}, \citenamefont {Wu},\ and\ \citenamefont {Zhang}}]{Li_2015}%
  \BibitemOpen
  \bibfield  {author} {\bibinfo {author} {\bibfnamefont {Y.}~\bibnamefont
  {Li}}, \bibinfo {author} {\bibfnamefont {H.}~\bibnamefont {Liao}}, \bibinfo
  {author} {\bibfnamefont {Z.}~\bibnamefont {Zhang}}, \bibinfo {author}
  {\bibfnamefont {S.}~\bibnamefont {Li}}, \bibinfo {author} {\bibfnamefont
  {F.}~\bibnamefont {Jin}}, \bibinfo {author} {\bibfnamefont {L.}~\bibnamefont
  {Ling}}, \bibinfo {author} {\bibfnamefont {L.}~\bibnamefont {Zhang}},
  \bibinfo {author} {\bibfnamefont {Y.}~\bibnamefont {Zou}}, \bibinfo {author}
  {\bibfnamefont {L.}~\bibnamefont {Pi}}, \bibinfo {author} {\bibfnamefont
  {Z.}~\bibnamefont {Yang}}, \bibinfo {author} {\bibfnamefont {J.}~\bibnamefont
  {Wang}}, \bibinfo {author} {\bibfnamefont {Z.}~\bibnamefont {Wu}}, \ and\
  \bibinfo {author} {\bibfnamefont {Q.}~\bibnamefont {Zhang}},\ }\href
  {https://doi.org/10.1038/srep16419} {\bibfield  {journal} {\bibinfo
  {journal} {Scientific Reports}\ }\textbf {\bibinfo {volume} {5}},\ \bibinfo
  {pages} {16419} (\bibinfo {year} {2015}{\natexlab{a}})}\BibitemShut {NoStop}%
\bibitem [{\citenamefont {Li}\ \emph {et~al.}(2015{\natexlab{b}})\citenamefont
  {Li}, \citenamefont {Chen}, \citenamefont {Tong}, \citenamefont {Pi},
  \citenamefont {Liu}, \citenamefont {Yang}, \citenamefont {Wang},\ and\
  \citenamefont {Zhang}}]{Li_2015a}%
  \BibitemOpen
  \bibfield  {author} {\bibinfo {author} {\bibfnamefont {Y.}~\bibnamefont
  {Li}}, \bibinfo {author} {\bibfnamefont {G.}~\bibnamefont {Chen}}, \bibinfo
  {author} {\bibfnamefont {W.}~\bibnamefont {Tong}}, \bibinfo {author}
  {\bibfnamefont {L.}~\bibnamefont {Pi}}, \bibinfo {author} {\bibfnamefont
  {J.}~\bibnamefont {Liu}}, \bibinfo {author} {\bibfnamefont {Z.}~\bibnamefont
  {Yang}}, \bibinfo {author} {\bibfnamefont {X.}~\bibnamefont {Wang}}, \ and\
  \bibinfo {author} {\bibfnamefont {Q.}~\bibnamefont {Zhang}},\ }\href
  {\doibase 10.1103/PhysRevLett.115.167203} {\bibfield  {journal} {\bibinfo
  {journal} {Phys. Rev. Lett.}\ }\textbf {\bibinfo {volume} {115}},\ \bibinfo
  {pages} {167203} (\bibinfo {year} {2015}{\natexlab{b}})}\BibitemShut
  {NoStop}%
\bibitem [{\citenamefont {Paddison}\ \emph {et~al.}(2016)\citenamefont
  {Paddison}, \citenamefont {Daum}, \citenamefont {Dun}, \citenamefont
  {Ehlers}, \citenamefont {Liu}, \citenamefont {Stone}, \citenamefont {Zhou},\
  and\ \citenamefont {Mourigal}}]{Paddison_2016}%
  \BibitemOpen
  \bibfield  {author} {\bibinfo {author} {\bibfnamefont {J.~A.~M.}\
  \bibnamefont {Paddison}}, \bibinfo {author} {\bibfnamefont {M.}~\bibnamefont
  {Daum}}, \bibinfo {author} {\bibfnamefont {Z.}~\bibnamefont {Dun}}, \bibinfo
  {author} {\bibfnamefont {G.}~\bibnamefont {Ehlers}}, \bibinfo {author}
  {\bibfnamefont {Y.}~\bibnamefont {Liu}}, \bibinfo {author} {\bibfnamefont
  {M.}~\bibnamefont {Stone}}, \bibinfo {author} {\bibfnamefont
  {H.}~\bibnamefont {Zhou}}, \ and\ \bibinfo {author} {\bibfnamefont
  {M.}~\bibnamefont {Mourigal}},\ }\href {https://doi.org/10.1038/nphys3971}
  {\bibfield  {journal} {\bibinfo  {journal} {Nature Physics}\ }\textbf
  {\bibinfo {volume} {13}},\ \bibinfo {pages} {117} (\bibinfo {year}
  {2016})}\BibitemShut {NoStop}%
\bibitem [{\citenamefont {Shen}\ \emph {et~al.}(2016)\citenamefont {Shen},
  \citenamefont {Li}, \citenamefont {Wo}, \citenamefont {Li}, \citenamefont
  {Shen}, \citenamefont {Pan}, \citenamefont {Wang}, \citenamefont {Walker},
  \citenamefont {Steffens}, \citenamefont {Boehm}, \citenamefont {Hao},
  \citenamefont {Quintero-Castro}, \citenamefont {Harriger}, \citenamefont
  {Frontzek}, \citenamefont {Hao}, \citenamefont {Meng}, \citenamefont {Zhang},
  \citenamefont {Chen},\ and\ \citenamefont {Zhao}}]{Shen_2016}%
  \BibitemOpen
  \bibfield  {author} {\bibinfo {author} {\bibfnamefont {Y.}~\bibnamefont
  {Shen}}, \bibinfo {author} {\bibfnamefont {Y.-D.}\ \bibnamefont {Li}},
  \bibinfo {author} {\bibfnamefont {H.}~\bibnamefont {Wo}}, \bibinfo {author}
  {\bibfnamefont {Y.}~\bibnamefont {Li}}, \bibinfo {author} {\bibfnamefont
  {S.}~\bibnamefont {Shen}}, \bibinfo {author} {\bibfnamefont {B.}~\bibnamefont
  {Pan}}, \bibinfo {author} {\bibfnamefont {Q.}~\bibnamefont {Wang}}, \bibinfo
  {author} {\bibfnamefont {H.~C.}\ \bibnamefont {Walker}}, \bibinfo {author}
  {\bibfnamefont {P.}~\bibnamefont {Steffens}}, \bibinfo {author}
  {\bibfnamefont {M.}~\bibnamefont {Boehm}}, \bibinfo {author} {\bibfnamefont
  {Y.}~\bibnamefont {Hao}}, \bibinfo {author} {\bibfnamefont {D.~L.}\
  \bibnamefont {Quintero-Castro}}, \bibinfo {author} {\bibfnamefont {L.~W.}\
  \bibnamefont {Harriger}}, \bibinfo {author} {\bibfnamefont {M.~D.}\
  \bibnamefont {Frontzek}}, \bibinfo {author} {\bibfnamefont {L.}~\bibnamefont
  {Hao}}, \bibinfo {author} {\bibfnamefont {S.}~\bibnamefont {Meng}}, \bibinfo
  {author} {\bibfnamefont {Q.}~\bibnamefont {Zhang}}, \bibinfo {author}
  {\bibfnamefont {G.}~\bibnamefont {Chen}}, \ and\ \bibinfo {author}
  {\bibfnamefont {J.}~\bibnamefont {Zhao}},\ }\href
  {https://doi.org/10.1038/nature20614} {\bibfield  {journal} {\bibinfo
  {journal} {Nature}\ }\textbf {\bibinfo {volume} {540}},\ \bibinfo {pages}
  {559} (\bibinfo {year} {2016})}\BibitemShut {NoStop}%
\bibitem [{\citenamefont {Li}\ \emph {et~al.}(2016)\citenamefont {Li},
  \citenamefont {Adroja}, \citenamefont {Biswas}, \citenamefont {Baker},
  \citenamefont {Zhang}, \citenamefont {Liu}, \citenamefont {Tsirlin},
  \citenamefont {Gegenwart},\ and\ \citenamefont {Zhang}}]{Li_2016}%
  \BibitemOpen
  \bibfield  {author} {\bibinfo {author} {\bibfnamefont {Y.}~\bibnamefont
  {Li}}, \bibinfo {author} {\bibfnamefont {D.}~\bibnamefont {Adroja}}, \bibinfo
  {author} {\bibfnamefont {P.~K.}\ \bibnamefont {Biswas}}, \bibinfo {author}
  {\bibfnamefont {P.~J.}\ \bibnamefont {Baker}}, \bibinfo {author}
  {\bibfnamefont {Q.}~\bibnamefont {Zhang}}, \bibinfo {author} {\bibfnamefont
  {J.}~\bibnamefont {Liu}}, \bibinfo {author} {\bibfnamefont {A.~A.}\
  \bibnamefont {Tsirlin}}, \bibinfo {author} {\bibfnamefont {P.}~\bibnamefont
  {Gegenwart}}, \ and\ \bibinfo {author} {\bibfnamefont {Q.}~\bibnamefont
  {Zhang}},\ }\href {\doibase 10.1103/PhysRevLett.117.097201} {\bibfield
  {journal} {\bibinfo  {journal} {Phys. Rev. Lett.}\ }\textbf {\bibinfo
  {volume} {117}},\ \bibinfo {pages} {097201} (\bibinfo {year}
  {2016})}\BibitemShut {NoStop}%
\bibitem [{\citenamefont {Li}\ \emph {et~al.}(2017{\natexlab{a}})\citenamefont
  {Li}, \citenamefont {Adroja}, \citenamefont {Bewley}, \citenamefont
  {Voneshen}, \citenamefont {Tsirlin}, \citenamefont {Gegenwart},\ and\
  \citenamefont {Zhang}}]{Li_2017}%
  \BibitemOpen
  \bibfield  {author} {\bibinfo {author} {\bibfnamefont {Y.}~\bibnamefont
  {Li}}, \bibinfo {author} {\bibfnamefont {D.}~\bibnamefont {Adroja}}, \bibinfo
  {author} {\bibfnamefont {R.~I.}\ \bibnamefont {Bewley}}, \bibinfo {author}
  {\bibfnamefont {D.}~\bibnamefont {Voneshen}}, \bibinfo {author}
  {\bibfnamefont {A.~A.}\ \bibnamefont {Tsirlin}}, \bibinfo {author}
  {\bibfnamefont {P.}~\bibnamefont {Gegenwart}}, \ and\ \bibinfo {author}
  {\bibfnamefont {Q.}~\bibnamefont {Zhang}},\ }\href {\doibase
  10.1103/PhysRevLett.118.107202} {\bibfield  {journal} {\bibinfo  {journal}
  {Phys. Rev. Lett.}\ }\textbf {\bibinfo {volume} {118}},\ \bibinfo {pages}
  {107202} (\bibinfo {year} {2017}{\natexlab{a}})}\BibitemShut {NoStop}%
\bibitem [{\citenamefont {Li}\ \emph {et~al.}(2017{\natexlab{b}})\citenamefont
  {Li}, \citenamefont {Adroja}, \citenamefont {Voneshen}, \citenamefont
  {Bewley}, \citenamefont {Zhang}, \citenamefont {Tsirlin},\ and\ \citenamefont
  {Gegenwart}}]{Li_2017a}%
  \BibitemOpen
  \bibfield  {author} {\bibinfo {author} {\bibfnamefont {Y.}~\bibnamefont
  {Li}}, \bibinfo {author} {\bibfnamefont {D.}~\bibnamefont {Adroja}}, \bibinfo
  {author} {\bibfnamefont {D.}~\bibnamefont {Voneshen}}, \bibinfo {author}
  {\bibfnamefont {R.~I.}\ \bibnamefont {Bewley}}, \bibinfo {author}
  {\bibfnamefont {Q.}~\bibnamefont {Zhang}}, \bibinfo {author} {\bibfnamefont
  {A.~A.}\ \bibnamefont {Tsirlin}}, \ and\ \bibinfo {author} {\bibfnamefont
  {P.}~\bibnamefont {Gegenwart}},\ }\href {https://doi.org/10.1038/ncomms15814}
  {\bibfield  {journal} {\bibinfo  {journal} {Nature Communications}\ }\textbf
  {\bibinfo {volume} {8}},\ \bibinfo {pages} {15814} (\bibinfo {year}
  {2017}{\natexlab{b}})}\BibitemShut {NoStop}%
\bibitem [{\citenamefont {Shores}\ \emph {et~al.}(2005)\citenamefont {Shores},
  \citenamefont {Nytko}, \citenamefont {Bartlett},\ and\ \citenamefont
  {Nocera}}]{Shores_2005}%
  \BibitemOpen
  \bibfield  {author} {\bibinfo {author} {\bibfnamefont {M.~P.}\ \bibnamefont
  {Shores}}, \bibinfo {author} {\bibfnamefont {E.~A.}\ \bibnamefont {Nytko}},
  \bibinfo {author} {\bibfnamefont {B.~M.}\ \bibnamefont {Bartlett}}, \ and\
  \bibinfo {author} {\bibfnamefont {D.~G.}\ \bibnamefont {Nocera}},\ }\href
  {\doibase 10.1021/ja053891p} {\bibfield  {journal} {\bibinfo  {journal}
  {Journal of the American Chemical Society}\ }\textbf {\bibinfo {volume}
  {127}},\ \bibinfo {pages} {13462} (\bibinfo {year} {2005})}\BibitemShut
  {NoStop}%
\bibitem [{\citenamefont {Bert}\ \emph {et~al.}(2007)\citenamefont {Bert},
  \citenamefont {Nakamae}, \citenamefont {Ladieu}, \citenamefont {L'H\^ote},
  \citenamefont {Bonville}, \citenamefont {Duc}, \citenamefont {Trombe},\ and\
  \citenamefont {Mendels}}]{Bert_2007}%
  \BibitemOpen
  \bibfield  {author} {\bibinfo {author} {\bibfnamefont {F.}~\bibnamefont
  {Bert}}, \bibinfo {author} {\bibfnamefont {S.}~\bibnamefont {Nakamae}},
  \bibinfo {author} {\bibfnamefont {F.}~\bibnamefont {Ladieu}}, \bibinfo
  {author} {\bibfnamefont {D.}~\bibnamefont {L'H\^ote}}, \bibinfo {author}
  {\bibfnamefont {P.}~\bibnamefont {Bonville}}, \bibinfo {author}
  {\bibfnamefont {F.}~\bibnamefont {Duc}}, \bibinfo {author} {\bibfnamefont
  {J.-C.}\ \bibnamefont {Trombe}}, \ and\ \bibinfo {author} {\bibfnamefont
  {P.}~\bibnamefont {Mendels}},\ }\href {\doibase 10.1103/PhysRevB.76.132411}
  {\bibfield  {journal} {\bibinfo  {journal} {Phys. Rev. B}\ }\textbf {\bibinfo
  {volume} {76}},\ \bibinfo {pages} {132411} (\bibinfo {year}
  {2007})}\BibitemShut {NoStop}%
\bibitem [{\citenamefont {Olariu}\ \emph {et~al.}(2008)\citenamefont {Olariu},
  \citenamefont {Mendels}, \citenamefont {Bert}, \citenamefont {Duc},
  \citenamefont {Trombe}, \citenamefont {de~Vries},\ and\ \citenamefont
  {Harrison}}]{Olariu_2008}%
  \BibitemOpen
  \bibfield  {author} {\bibinfo {author} {\bibfnamefont {A.}~\bibnamefont
  {Olariu}}, \bibinfo {author} {\bibfnamefont {P.}~\bibnamefont {Mendels}},
  \bibinfo {author} {\bibfnamefont {F.}~\bibnamefont {Bert}}, \bibinfo {author}
  {\bibfnamefont {F.}~\bibnamefont {Duc}}, \bibinfo {author} {\bibfnamefont
  {J.~C.}\ \bibnamefont {Trombe}}, \bibinfo {author} {\bibfnamefont {M.~A.}\
  \bibnamefont {de~Vries}}, \ and\ \bibinfo {author} {\bibfnamefont
  {A.}~\bibnamefont {Harrison}},\ }\href {\doibase
  10.1103/PhysRevLett.100.087202} {\bibfield  {journal} {\bibinfo  {journal}
  {Phys. Rev. Lett.}\ }\textbf {\bibinfo {volume} {100}},\ \bibinfo {pages}
  {087202} (\bibinfo {year} {2008})}\BibitemShut {NoStop}%
\bibitem [{\citenamefont {de~Vries}\ \emph {et~al.}(2009)\citenamefont
  {de~Vries}, \citenamefont {Stewart}, \citenamefont {Deen}, \citenamefont
  {Piatek}, \citenamefont {Nilsen}, \citenamefont {R\o{}nnow},\ and\
  \citenamefont {Harrison}}]{Vries_2009}%
  \BibitemOpen
  \bibfield  {author} {\bibinfo {author} {\bibfnamefont {M.~A.}\ \bibnamefont
  {de~Vries}}, \bibinfo {author} {\bibfnamefont {J.~R.}\ \bibnamefont
  {Stewart}}, \bibinfo {author} {\bibfnamefont {P.~P.}\ \bibnamefont {Deen}},
  \bibinfo {author} {\bibfnamefont {J.~O.}\ \bibnamefont {Piatek}}, \bibinfo
  {author} {\bibfnamefont {G.~J.}\ \bibnamefont {Nilsen}}, \bibinfo {author}
  {\bibfnamefont {H.~M.}\ \bibnamefont {R\o{}nnow}}, \ and\ \bibinfo {author}
  {\bibfnamefont {A.}~\bibnamefont {Harrison}},\ }\href {\doibase
  10.1103/PhysRevLett.103.237201} {\bibfield  {journal} {\bibinfo  {journal}
  {Phys. Rev. Lett.}\ }\textbf {\bibinfo {volume} {103}},\ \bibinfo {pages}
  {237201} (\bibinfo {year} {2009})}\BibitemShut {NoStop}%
\bibitem [{\citenamefont {Mendels}\ and\ \citenamefont
  {Bert}(2010)}]{Mendels_2010}%
  \BibitemOpen
  \bibfield  {author} {\bibinfo {author} {\bibfnamefont {P.}~\bibnamefont
  {Mendels}}\ and\ \bibinfo {author} {\bibfnamefont {F.}~\bibnamefont {Bert}},\
  }\href {\doibase 10.1143/JPSJ.79.011001} {\bibfield  {journal} {\bibinfo
  {journal} {Journal of the Physical Society of Japan}\ }\textbf {\bibinfo
  {volume} {79}},\ \bibinfo {pages} {011001} (\bibinfo {year}
  {2010})}\BibitemShut {NoStop}%
\bibitem [{\citenamefont {Helton}\ \emph {et~al.}(2010)\citenamefont {Helton},
  \citenamefont {Matan}, \citenamefont {Shores}, \citenamefont {Nytko},
  \citenamefont {Bartlett}, \citenamefont {Qiu}, \citenamefont {Nocera},\ and\
  \citenamefont {Lee}}]{Helton_2010}%
  \BibitemOpen
  \bibfield  {author} {\bibinfo {author} {\bibfnamefont {J.~S.}\ \bibnamefont
  {Helton}}, \bibinfo {author} {\bibfnamefont {K.}~\bibnamefont {Matan}},
  \bibinfo {author} {\bibfnamefont {M.~P.}\ \bibnamefont {Shores}}, \bibinfo
  {author} {\bibfnamefont {E.~A.}\ \bibnamefont {Nytko}}, \bibinfo {author}
  {\bibfnamefont {B.~M.}\ \bibnamefont {Bartlett}}, \bibinfo {author}
  {\bibfnamefont {Y.}~\bibnamefont {Qiu}}, \bibinfo {author} {\bibfnamefont
  {D.~G.}\ \bibnamefont {Nocera}}, \ and\ \bibinfo {author} {\bibfnamefont
  {Y.~S.}\ \bibnamefont {Lee}},\ }\href {\doibase
  10.1103/PhysRevLett.104.147201} {\bibfield  {journal} {\bibinfo  {journal}
  {Phys. Rev. Lett.}\ }\textbf {\bibinfo {volume} {104}},\ \bibinfo {pages}
  {147201} (\bibinfo {year} {2010})}\BibitemShut {NoStop}%
\bibitem [{\citenamefont {Han}\ \emph {et~al.}(2012)\citenamefont {Han},
  \citenamefont {Helton}, \citenamefont {Chu}, \citenamefont {Nocera},
  \citenamefont {Rodriguez-Rivera}, \citenamefont {Broholm},\ and\
  \citenamefont {Lee}}]{Han_2012}%
  \BibitemOpen
  \bibfield  {author} {\bibinfo {author} {\bibfnamefont {T.-H.}\ \bibnamefont
  {Han}}, \bibinfo {author} {\bibfnamefont {J.~S.}\ \bibnamefont {Helton}},
  \bibinfo {author} {\bibfnamefont {S.}~\bibnamefont {Chu}}, \bibinfo {author}
  {\bibfnamefont {D.~G.}\ \bibnamefont {Nocera}}, \bibinfo {author}
  {\bibfnamefont {J.~A.}\ \bibnamefont {Rodriguez-Rivera}}, \bibinfo {author}
  {\bibfnamefont {C.}~\bibnamefont {Broholm}}, \ and\ \bibinfo {author}
  {\bibfnamefont {Y.~S.}\ \bibnamefont {Lee}},\ }\href
  {https://doi.org/10.1038/nature11659} {\bibfield  {journal} {\bibinfo
  {journal} {Nature}\ }\textbf {\bibinfo {volume} {492}},\ \bibinfo {pages}
  {406} (\bibinfo {year} {2012})}\BibitemShut {NoStop}%
\bibitem [{\citenamefont {Jeschke}\ \emph {et~al.}(2013)\citenamefont
  {Jeschke}, \citenamefont {Salvat-Pujol},\ and\ \citenamefont
  {Valent\'{\i}}}]{Jeschke_2013}%
  \BibitemOpen
  \bibfield  {author} {\bibinfo {author} {\bibfnamefont {H.~O.}\ \bibnamefont
  {Jeschke}}, \bibinfo {author} {\bibfnamefont {F.}~\bibnamefont
  {Salvat-Pujol}}, \ and\ \bibinfo {author} {\bibfnamefont {R.}~\bibnamefont
  {Valent\'{\i}}},\ }\href {\doibase 10.1103/PhysRevB.88.075106} {\bibfield
  {journal} {\bibinfo  {journal} {Phys. Rev. B}\ }\textbf {\bibinfo {volume}
  {88}},\ \bibinfo {pages} {075106} (\bibinfo {year} {2013})}\BibitemShut
  {NoStop}%
\bibitem [{\citenamefont {Pilon}\ \emph {et~al.}(2013)\citenamefont {Pilon},
  \citenamefont {Lui}, \citenamefont {Han}, \citenamefont {Shrekenhamer},
  \citenamefont {Frenzel}, \citenamefont {Padilla}, \citenamefont {Lee},\ and\
  \citenamefont {Gedik}}]{Pilon_2013}%
  \BibitemOpen
  \bibfield  {author} {\bibinfo {author} {\bibfnamefont {D.~V.}\ \bibnamefont
  {Pilon}}, \bibinfo {author} {\bibfnamefont {C.~H.}\ \bibnamefont {Lui}},
  \bibinfo {author} {\bibfnamefont {T.~H.}\ \bibnamefont {Han}}, \bibinfo
  {author} {\bibfnamefont {D.}~\bibnamefont {Shrekenhamer}}, \bibinfo {author}
  {\bibfnamefont {A.~J.}\ \bibnamefont {Frenzel}}, \bibinfo {author}
  {\bibfnamefont {W.~J.}\ \bibnamefont {Padilla}}, \bibinfo {author}
  {\bibfnamefont {Y.~S.}\ \bibnamefont {Lee}}, \ and\ \bibinfo {author}
  {\bibfnamefont {N.}~\bibnamefont {Gedik}},\ }\href {\doibase
  10.1103/PhysRevLett.111.127401} {\bibfield  {journal} {\bibinfo  {journal}
  {Phys. Rev. Lett.}\ }\textbf {\bibinfo {volume} {111}},\ \bibinfo {pages}
  {127401} (\bibinfo {year} {2013})}\BibitemShut {NoStop}%
\bibitem [{\citenamefont {Witczak-Krempa}\ \emph {et~al.}(2014)\citenamefont
  {Witczak-Krempa}, \citenamefont {Chen}, \citenamefont {Kim},\ and\
  \citenamefont {Balents}}]{Witczak_2014}%
  \BibitemOpen
  \bibfield  {author} {\bibinfo {author} {\bibfnamefont {W.}~\bibnamefont
  {Witczak-Krempa}}, \bibinfo {author} {\bibfnamefont {G.}~\bibnamefont
  {Chen}}, \bibinfo {author} {\bibfnamefont {Y.~B.}\ \bibnamefont {Kim}}, \
  and\ \bibinfo {author} {\bibfnamefont {L.}~\bibnamefont {Balents}},\ }\href
  {\doibase 10.1146/annurev-conmatphys-020911-125138} {\bibfield  {journal}
  {\bibinfo  {journal} {Annual Review of Condensed Matter Physics}\ }\textbf
  {\bibinfo {volume} {5}},\ \bibinfo {pages} {57} (\bibinfo {year}
  {2014})}\BibitemShut {NoStop}%
\bibitem [{\citenamefont {Rau}\ \emph {et~al.}(2016)\citenamefont {Rau},
  \citenamefont {Lee},\ and\ \citenamefont {Kee}}]{Rau_2016}%
  \BibitemOpen
  \bibfield  {author} {\bibinfo {author} {\bibfnamefont {J.~G.}\ \bibnamefont
  {Rau}}, \bibinfo {author} {\bibfnamefont {E.~K.-H.}\ \bibnamefont {Lee}}, \
  and\ \bibinfo {author} {\bibfnamefont {H.-Y.}\ \bibnamefont {Kee}},\ }\href
  {\doibase 10.1146/annurev-conmatphys-031115-011319} {\bibfield  {journal}
  {\bibinfo  {journal} {Annual Review of Condensed Matter Physics}\ }\textbf
  {\bibinfo {volume} {7}},\ \bibinfo {pages} {195} (\bibinfo {year}
  {2016})}\BibitemShut {NoStop}%
\bibitem [{\citenamefont {Winter}\ \emph {et~al.}(2017)\citenamefont {Winter},
  \citenamefont {Tsirlin}, \citenamefont {Daghofer}, \citenamefont {van~den
  Brink}, \citenamefont {Singh}, \citenamefont {Gegenwart},\ and\ \citenamefont
  {Valent{\'{\i}}}}]{Winter_2017}%
  \BibitemOpen
  \bibfield  {author} {\bibinfo {author} {\bibfnamefont {S.~M.}\ \bibnamefont
  {Winter}}, \bibinfo {author} {\bibfnamefont {A.~A.}\ \bibnamefont {Tsirlin}},
  \bibinfo {author} {\bibfnamefont {M.}~\bibnamefont {Daghofer}}, \bibinfo
  {author} {\bibfnamefont {J.}~\bibnamefont {van~den Brink}}, \bibinfo {author}
  {\bibfnamefont {Y.}~\bibnamefont {Singh}}, \bibinfo {author} {\bibfnamefont
  {P.}~\bibnamefont {Gegenwart}}, \ and\ \bibinfo {author} {\bibfnamefont
  {R.}~\bibnamefont {Valent{\'{\i}}}},\ }\href {\doibase
  10.1088/1361-648x/aa8cf5} {\bibfield  {journal} {\bibinfo  {journal} {Journal
  of Physics: Condensed Matter}\ }\textbf {\bibinfo {volume} {29}},\ \bibinfo
  {pages} {493002} (\bibinfo {year} {2017})}\BibitemShut {NoStop}%
\bibitem [{\citenamefont {{Trebst}}(2017)}]{Trebst_2017}%
  \BibitemOpen
  \bibfield  {author} {\bibinfo {author} {\bibfnamefont {S.}~\bibnamefont
  {{Trebst}}},\ }\href@noop {} {\bibfield  {journal} {\bibinfo  {journal}
  {arXiv e-prints}\ } (\bibinfo {year} {2017})},\ \Eprint
  {http://arxiv.org/abs/1701.07056} {arXiv:1701.07056} \BibitemShut {NoStop}%
\bibitem [{\citenamefont {Kitagawa}\ \emph {et~al.}(2018)\citenamefont
  {Kitagawa}, \citenamefont {Takayama}, \citenamefont {Matsumoto},
  \citenamefont {Kato}, \citenamefont {Takano}, \citenamefont {Kishimoto},
  \citenamefont {Bette}, \citenamefont {Dinnebier}, \citenamefont {Jackeli},\
  and\ \citenamefont {Takagi}}]{Kitagawa_2018}%
  \BibitemOpen
  \bibfield  {author} {\bibinfo {author} {\bibfnamefont {K.}~\bibnamefont
  {Kitagawa}}, \bibinfo {author} {\bibfnamefont {T.}~\bibnamefont {Takayama}},
  \bibinfo {author} {\bibfnamefont {Y.}~\bibnamefont {Matsumoto}}, \bibinfo
  {author} {\bibfnamefont {A.}~\bibnamefont {Kato}}, \bibinfo {author}
  {\bibfnamefont {R.}~\bibnamefont {Takano}}, \bibinfo {author} {\bibfnamefont
  {Y.}~\bibnamefont {Kishimoto}}, \bibinfo {author} {\bibfnamefont
  {S.}~\bibnamefont {Bette}}, \bibinfo {author} {\bibfnamefont
  {R.}~\bibnamefont {Dinnebier}}, \bibinfo {author} {\bibfnamefont
  {G.}~\bibnamefont {Jackeli}}, \ and\ \bibinfo {author} {\bibfnamefont
  {H.}~\bibnamefont {Takagi}},\ }\href {https://doi.org/10.1038/nature25482}
  {\bibfield  {journal} {\bibinfo  {journal} {Nature}\ }\textbf {\bibinfo
  {volume} {554}},\ \bibinfo {pages} {341} (\bibinfo {year}
  {2018})}\BibitemShut {NoStop}%
\bibitem [{\citenamefont {Yadav}\ \emph
  {et~al.}(2018{\natexlab{a}})\citenamefont {Yadav}, \citenamefont {Ray},
  \citenamefont {Eldeeb}, \citenamefont {Nishimoto}, \citenamefont {Hozoi},\
  and\ \citenamefont {van~den Brink}}]{Yadav_2018}%
  \BibitemOpen
  \bibfield  {author} {\bibinfo {author} {\bibfnamefont {R.}~\bibnamefont
  {Yadav}}, \bibinfo {author} {\bibfnamefont {R.}~\bibnamefont {Ray}}, \bibinfo
  {author} {\bibfnamefont {M.~S.}\ \bibnamefont {Eldeeb}}, \bibinfo {author}
  {\bibfnamefont {S.}~\bibnamefont {Nishimoto}}, \bibinfo {author}
  {\bibfnamefont {L.}~\bibnamefont {Hozoi}}, \ and\ \bibinfo {author}
  {\bibfnamefont {J.}~\bibnamefont {van~den Brink}},\ }\href {\doibase
  10.1103/PhysRevLett.121.197203} {\bibfield  {journal} {\bibinfo  {journal}
  {Phys. Rev. Lett.}\ }\textbf {\bibinfo {volume} {121}},\ \bibinfo {pages}
  {197203} (\bibinfo {year} {2018}{\natexlab{a}})}\BibitemShut {NoStop}%
\bibitem [{\citenamefont {Li}\ \emph {et~al.}(2018)\citenamefont {Li},
  \citenamefont {Winter},\ and\ \citenamefont {Valent\'{\i}}}]{Li_2018}%
  \BibitemOpen
  \bibfield  {author} {\bibinfo {author} {\bibfnamefont {Y.}~\bibnamefont
  {Li}}, \bibinfo {author} {\bibfnamefont {S.~M.}\ \bibnamefont {Winter}}, \
  and\ \bibinfo {author} {\bibfnamefont {R.}~\bibnamefont {Valent\'{\i}}},\
  }\href {\doibase 10.1103/PhysRevLett.121.247202} {\bibfield  {journal}
  {\bibinfo  {journal} {Phys. Rev. Lett.}\ }\textbf {\bibinfo {volume} {121}},\
  \bibinfo {pages} {247202} (\bibinfo {year} {2018})}\BibitemShut {NoStop}%
\bibitem [{\citenamefont {Knolle}\ \emph {et~al.}(2019)\citenamefont {Knolle},
  \citenamefont {Moessner},\ and\ \citenamefont {Perkins}}]{Knolle_2019a}%
  \BibitemOpen
  \bibfield  {author} {\bibinfo {author} {\bibfnamefont {J.}~\bibnamefont
  {Knolle}}, \bibinfo {author} {\bibfnamefont {R.}~\bibnamefont {Moessner}}, \
  and\ \bibinfo {author} {\bibfnamefont {N.~B.}\ \bibnamefont {Perkins}},\
  }\href {\doibase 10.1103/PhysRevLett.122.047202} {\bibfield  {journal}
  {\bibinfo  {journal} {Phys. Rev. Lett.}\ }\textbf {\bibinfo {volume} {122}},\
  \bibinfo {pages} {047202} (\bibinfo {year} {2019})}\BibitemShut {NoStop}%
\bibitem [{\citenamefont {Hwan~Chun}\ \emph {et~al.}(2015)\citenamefont
  {Hwan~Chun}, \citenamefont {Kim}, \citenamefont {Kim}, \citenamefont {Zheng},
  \citenamefont {Stoumpos}, \citenamefont {Malliakas}, \citenamefont
  {Mitchell}, \citenamefont {Mehlawat}, \citenamefont {Singh}, \citenamefont
  {Choi}, \citenamefont {Gog}, \citenamefont {Al-Zein}, \citenamefont {Sala},
  \citenamefont {Krisch}, \citenamefont {Chaloupka}, \citenamefont {Jackeli},
  \citenamefont {Khaliullin},\ and\ \citenamefont {Kim}}]{HwanChun_2015}%
  \BibitemOpen
  \bibfield  {author} {\bibinfo {author} {\bibfnamefont {S.}~\bibnamefont
  {Hwan~Chun}}, \bibinfo {author} {\bibfnamefont {J.-W.}\ \bibnamefont {Kim}},
  \bibinfo {author} {\bibfnamefont {J.}~\bibnamefont {Kim}}, \bibinfo {author}
  {\bibfnamefont {H.}~\bibnamefont {Zheng}}, \bibinfo {author} {\bibfnamefont
  {C.}~\bibnamefont {Stoumpos}}, \bibinfo {author} {\bibfnamefont {C.~.~D.}\
  \bibnamefont {Malliakas}}, \bibinfo {author} {\bibfnamefont {J.~.~F.}\
  \bibnamefont {Mitchell}}, \bibinfo {author} {\bibfnamefont {K.}~\bibnamefont
  {Mehlawat}}, \bibinfo {author} {\bibfnamefont {Y.}~\bibnamefont {Singh}},
  \bibinfo {author} {\bibfnamefont {Y.}~\bibnamefont {Choi}}, \bibinfo {author}
  {\bibfnamefont {T.}~\bibnamefont {Gog}}, \bibinfo {author} {\bibfnamefont
  {A.}~\bibnamefont {Al-Zein}}, \bibinfo {author} {\bibfnamefont {M.~M.}\
  \bibnamefont {Sala}}, \bibinfo {author} {\bibfnamefont {M.}~\bibnamefont
  {Krisch}}, \bibinfo {author} {\bibfnamefont {J.}~\bibnamefont {Chaloupka}},
  \bibinfo {author} {\bibfnamefont {G.}~\bibnamefont {Jackeli}}, \bibinfo
  {author} {\bibfnamefont {G.}~\bibnamefont {Khaliullin}}, \ and\ \bibinfo
  {author} {\bibfnamefont {B.~J.}\ \bibnamefont {Kim}},\ }\href
  {https://doi.org/10.1038/nphys3322} {\bibfield  {journal} {\bibinfo
  {journal} {Nature Physics}\ }\textbf {\bibinfo {volume} {11}},\ \bibinfo
  {pages} {462} (\bibinfo {year} {2015})}\BibitemShut {NoStop}%
\bibitem [{\citenamefont {Takayama}\ \emph {et~al.}(2015)\citenamefont
  {Takayama}, \citenamefont {Kato}, \citenamefont {Dinnebier}, \citenamefont
  {Nuss}, \citenamefont {Kono}, \citenamefont {Veiga}, \citenamefont {Fabbris},
  \citenamefont {Haskel},\ and\ \citenamefont {Takagi}}]{Takayama_2015}%
  \BibitemOpen
  \bibfield  {author} {\bibinfo {author} {\bibfnamefont {T.}~\bibnamefont
  {Takayama}}, \bibinfo {author} {\bibfnamefont {A.}~\bibnamefont {Kato}},
  \bibinfo {author} {\bibfnamefont {R.}~\bibnamefont {Dinnebier}}, \bibinfo
  {author} {\bibfnamefont {J.}~\bibnamefont {Nuss}}, \bibinfo {author}
  {\bibfnamefont {H.}~\bibnamefont {Kono}}, \bibinfo {author} {\bibfnamefont
  {L.~S.~I.}\ \bibnamefont {Veiga}}, \bibinfo {author} {\bibfnamefont
  {G.}~\bibnamefont {Fabbris}}, \bibinfo {author} {\bibfnamefont
  {D.}~\bibnamefont {Haskel}}, \ and\ \bibinfo {author} {\bibfnamefont
  {H.}~\bibnamefont {Takagi}},\ }\href {\doibase
  10.1103/PhysRevLett.114.077202} {\bibfield  {journal} {\bibinfo  {journal}
  {Phys. Rev. Lett.}\ }\textbf {\bibinfo {volume} {114}},\ \bibinfo {pages}
  {077202} (\bibinfo {year} {2015})}\BibitemShut {NoStop}%
\bibitem [{\citenamefont {Sandilands}\ \emph {et~al.}(2015)\citenamefont
  {Sandilands}, \citenamefont {Tian}, \citenamefont {Plumb}, \citenamefont
  {Kim},\ and\ \citenamefont {Burch}}]{Sandilands_2015}%
  \BibitemOpen
  \bibfield  {author} {\bibinfo {author} {\bibfnamefont {L.~J.}\ \bibnamefont
  {Sandilands}}, \bibinfo {author} {\bibfnamefont {Y.}~\bibnamefont {Tian}},
  \bibinfo {author} {\bibfnamefont {K.~W.}\ \bibnamefont {Plumb}}, \bibinfo
  {author} {\bibfnamefont {Y.-J.}\ \bibnamefont {Kim}}, \ and\ \bibinfo
  {author} {\bibfnamefont {K.~S.}\ \bibnamefont {Burch}},\ }\href {\doibase
  10.1103/PhysRevLett.114.147201} {\bibfield  {journal} {\bibinfo  {journal}
  {Phys. Rev. Lett.}\ }\textbf {\bibinfo {volume} {114}},\ \bibinfo {pages}
  {147201} (\bibinfo {year} {2015})}\BibitemShut {NoStop}%
\bibitem [{\citenamefont {Banerjee}\ \emph {et~al.}(2016)\citenamefont
  {Banerjee}, \citenamefont {Bridges}, \citenamefont {Yan}, \citenamefont
  {Aczel}, \citenamefont {Li}, \citenamefont {Stone}, \citenamefont {Granroth},
  \citenamefont {Lumsden}, \citenamefont {Yiu}, \citenamefont {Knolle},
  \citenamefont {Bhattacharjee}, \citenamefont {Kovrizhin}, \citenamefont
  {Moessner}, \citenamefont {Tennant}, \citenamefont {Mandrus},\ and\
  \citenamefont {Nagler}}]{Banerjee_2016}%
  \BibitemOpen
  \bibfield  {author} {\bibinfo {author} {\bibfnamefont {A.}~\bibnamefont
  {Banerjee}}, \bibinfo {author} {\bibfnamefont {C.~A.}\ \bibnamefont
  {Bridges}}, \bibinfo {author} {\bibfnamefont {J.-Q.}\ \bibnamefont {Yan}},
  \bibinfo {author} {\bibfnamefont {A.~A.}\ \bibnamefont {Aczel}}, \bibinfo
  {author} {\bibfnamefont {L.}~\bibnamefont {Li}}, \bibinfo {author}
  {\bibfnamefont {M.~B.}\ \bibnamefont {Stone}}, \bibinfo {author}
  {\bibfnamefont {G.~E.}\ \bibnamefont {Granroth}}, \bibinfo {author}
  {\bibfnamefont {M.~D.}\ \bibnamefont {Lumsden}}, \bibinfo {author}
  {\bibfnamefont {Y.}~\bibnamefont {Yiu}}, \bibinfo {author} {\bibfnamefont
  {J.}~\bibnamefont {Knolle}}, \bibinfo {author} {\bibfnamefont
  {S.}~\bibnamefont {Bhattacharjee}}, \bibinfo {author} {\bibfnamefont {D.~L.}\
  \bibnamefont {Kovrizhin}}, \bibinfo {author} {\bibfnamefont {R.}~\bibnamefont
  {Moessner}}, \bibinfo {author} {\bibfnamefont {D.~A.}\ \bibnamefont
  {Tennant}}, \bibinfo {author} {\bibfnamefont {D.~G.}\ \bibnamefont
  {Mandrus}}, \ and\ \bibinfo {author} {\bibfnamefont {S.~E.}\ \bibnamefont
  {Nagler}},\ }\href {https://doi.org/10.1038/nmat4604} {\bibfield  {journal}
  {\bibinfo  {journal} {Nature Materials}\ }\textbf {\bibinfo {volume} {15}},\
  \bibinfo {pages} {733 EP } (\bibinfo {year} {2016})},\ \bibinfo {note}
  {article}\BibitemShut {NoStop}%
\bibitem [{\citenamefont {Cao}\ \emph {et~al.}(2016)\citenamefont {Cao},
  \citenamefont {Banerjee}, \citenamefont {Yan}, \citenamefont {Bridges},
  \citenamefont {Lumsden}, \citenamefont {Mandrus}, \citenamefont {Tennant},
  \citenamefont {Chakoumakos},\ and\ \citenamefont {Nagler}}]{Cao_2016}%
  \BibitemOpen
  \bibfield  {author} {\bibinfo {author} {\bibfnamefont {H.~B.}\ \bibnamefont
  {Cao}}, \bibinfo {author} {\bibfnamefont {A.}~\bibnamefont {Banerjee}},
  \bibinfo {author} {\bibfnamefont {J.-Q.}\ \bibnamefont {Yan}}, \bibinfo
  {author} {\bibfnamefont {C.~A.}\ \bibnamefont {Bridges}}, \bibinfo {author}
  {\bibfnamefont {M.~D.}\ \bibnamefont {Lumsden}}, \bibinfo {author}
  {\bibfnamefont {D.~G.}\ \bibnamefont {Mandrus}}, \bibinfo {author}
  {\bibfnamefont {D.~A.}\ \bibnamefont {Tennant}}, \bibinfo {author}
  {\bibfnamefont {B.~C.}\ \bibnamefont {Chakoumakos}}, \ and\ \bibinfo {author}
  {\bibfnamefont {S.~E.}\ \bibnamefont {Nagler}},\ }\href {\doibase
  10.1103/PhysRevB.93.134423} {\bibfield  {journal} {\bibinfo  {journal} {Phys.
  Rev. B}\ }\textbf {\bibinfo {volume} {93}},\ \bibinfo {pages} {134423}
  (\bibinfo {year} {2016})}\BibitemShut {NoStop}%
\bibitem [{\citenamefont {Banerjee}\ \emph {et~al.}(2017)\citenamefont
  {Banerjee}, \citenamefont {Yan}, \citenamefont {Knolle}, \citenamefont
  {Bridges}, \citenamefont {Stone}, \citenamefont {Lumsden}, \citenamefont
  {Mandrus}, \citenamefont {Tennant}, \citenamefont {Moessner},\ and\
  \citenamefont {Nagler}}]{Banerjee_2017}%
  \BibitemOpen
  \bibfield  {author} {\bibinfo {author} {\bibfnamefont {A.}~\bibnamefont
  {Banerjee}}, \bibinfo {author} {\bibfnamefont {J.}~\bibnamefont {Yan}},
  \bibinfo {author} {\bibfnamefont {J.}~\bibnamefont {Knolle}}, \bibinfo
  {author} {\bibfnamefont {C.~A.}\ \bibnamefont {Bridges}}, \bibinfo {author}
  {\bibfnamefont {M.~B.}\ \bibnamefont {Stone}}, \bibinfo {author}
  {\bibfnamefont {M.~D.}\ \bibnamefont {Lumsden}}, \bibinfo {author}
  {\bibfnamefont {D.~G.}\ \bibnamefont {Mandrus}}, \bibinfo {author}
  {\bibfnamefont {D.~A.}\ \bibnamefont {Tennant}}, \bibinfo {author}
  {\bibfnamefont {R.}~\bibnamefont {Moessner}}, \ and\ \bibinfo {author}
  {\bibfnamefont {S.~E.}\ \bibnamefont {Nagler}},\ }\href {\doibase
  10.1126/science.aah6015} {\bibfield  {journal} {\bibinfo  {journal}
  {Science}\ }\textbf {\bibinfo {volume} {356}},\ \bibinfo {pages} {1055}
  (\bibinfo {year} {2017})}\BibitemShut {NoStop}%
\bibitem [{\citenamefont {{Revelli}}\ \emph {et~al.}(2019)\citenamefont
  {{Revelli}}, \citenamefont {{Moretti Sala}}, \citenamefont {{Monaco}},
  \citenamefont {{Hickey}}, \citenamefont {{Becker}}, \citenamefont {{Freund}},
  \citenamefont {{Jesche}}, \citenamefont {{Gegenwart}}, \citenamefont
  {{Eschmann}}, \citenamefont {{Buessen}}, \citenamefont {{Trebst}},
  \citenamefont {{van Loosdrecht}}, \citenamefont {{van den Brink}},\ and\
  \citenamefont {{Gr{\"u}ninger}}}]{Revelli_2019}%
  \BibitemOpen
  \bibfield  {author} {\bibinfo {author} {\bibfnamefont {A.}~\bibnamefont
  {{Revelli}}}, \bibinfo {author} {\bibfnamefont {M.}~\bibnamefont {{Moretti
  Sala}}}, \bibinfo {author} {\bibfnamefont {G.}~\bibnamefont {{Monaco}}},
  \bibinfo {author} {\bibfnamefont {C.}~\bibnamefont {{Hickey}}}, \bibinfo
  {author} {\bibfnamefont {P.}~\bibnamefont {{Becker}}}, \bibinfo {author}
  {\bibfnamefont {F.}~\bibnamefont {{Freund}}}, \bibinfo {author}
  {\bibfnamefont {A.}~\bibnamefont {{Jesche}}}, \bibinfo {author}
  {\bibfnamefont {P.}~\bibnamefont {{Gegenwart}}}, \bibinfo {author}
  {\bibfnamefont {T.}~\bibnamefont {{Eschmann}}}, \bibinfo {author}
  {\bibfnamefont {F.~L.}\ \bibnamefont {{Buessen}}}, \bibinfo {author}
  {\bibfnamefont {S.}~\bibnamefont {{Trebst}}}, \bibinfo {author}
  {\bibfnamefont {P.~H.~M.}\ \bibnamefont {{van Loosdrecht}}}, \bibinfo
  {author} {\bibfnamefont {J.}~\bibnamefont {{van den Brink}}}, \ and\ \bibinfo
  {author} {\bibfnamefont {M.}~\bibnamefont {{Gr{\"u}ninger}}},\ }\href@noop {}
  {\bibfield  {journal} {\bibinfo  {journal} {arXiv e-prints}\ } (\bibinfo
  {year} {2019})},\ \Eprint {http://arxiv.org/abs/1905.13590}
  {arXiv:1905.13590} \BibitemShut {NoStop}%
\bibitem [{\citenamefont {Breznay}\ \emph {et~al.}(2017)\citenamefont
  {Breznay}, \citenamefont {Ruiz}, \citenamefont {Frano}, \citenamefont {Bi},
  \citenamefont {Birgeneau}, \citenamefont {Haskel},\ and\ \citenamefont
  {Analytis}}]{Breznay_2017}%
  \BibitemOpen
  \bibfield  {author} {\bibinfo {author} {\bibfnamefont {N.~P.}\ \bibnamefont
  {Breznay}}, \bibinfo {author} {\bibfnamefont {A.}~\bibnamefont {Ruiz}},
  \bibinfo {author} {\bibfnamefont {A.}~\bibnamefont {Frano}}, \bibinfo
  {author} {\bibfnamefont {W.}~\bibnamefont {Bi}}, \bibinfo {author}
  {\bibfnamefont {R.~J.}\ \bibnamefont {Birgeneau}}, \bibinfo {author}
  {\bibfnamefont {D.}~\bibnamefont {Haskel}}, \ and\ \bibinfo {author}
  {\bibfnamefont {J.~G.}\ \bibnamefont {Analytis}},\ }\href {\doibase
  10.1103/PhysRevB.96.020402} {\bibfield  {journal} {\bibinfo  {journal} {Phys.
  Rev. B}\ }\textbf {\bibinfo {volume} {96}},\ \bibinfo {pages} {020402}
  (\bibinfo {year} {2017})}\BibitemShut {NoStop}%
\bibitem [{\citenamefont {Wang}\ \emph {et~al.}(2018)\citenamefont {Wang},
  \citenamefont {Guo}, \citenamefont {Tafti}, \citenamefont {Hegg},
  \citenamefont {Sen}, \citenamefont {Sidorov}, \citenamefont {Wang},
  \citenamefont {Cai}, \citenamefont {Yi}, \citenamefont {Zhou}, \citenamefont
  {Wang}, \citenamefont {Zhang}, \citenamefont {Yang}, \citenamefont {Li},
  \citenamefont {Li}, \citenamefont {Li}, \citenamefont {Liu}, \citenamefont
  {Shi}, \citenamefont {Ku}, \citenamefont {Wu}, \citenamefont {Cava},\ and\
  \citenamefont {Sun}}]{Wang_2018}%
  \BibitemOpen
  \bibfield  {author} {\bibinfo {author} {\bibfnamefont {Z.}~\bibnamefont
  {Wang}}, \bibinfo {author} {\bibfnamefont {J.}~\bibnamefont {Guo}}, \bibinfo
  {author} {\bibfnamefont {F.~F.}\ \bibnamefont {Tafti}}, \bibinfo {author}
  {\bibfnamefont {A.}~\bibnamefont {Hegg}}, \bibinfo {author} {\bibfnamefont
  {S.}~\bibnamefont {Sen}}, \bibinfo {author} {\bibfnamefont {V.~A.}\
  \bibnamefont {Sidorov}}, \bibinfo {author} {\bibfnamefont {L.}~\bibnamefont
  {Wang}}, \bibinfo {author} {\bibfnamefont {S.}~\bibnamefont {Cai}}, \bibinfo
  {author} {\bibfnamefont {W.}~\bibnamefont {Yi}}, \bibinfo {author}
  {\bibfnamefont {Y.}~\bibnamefont {Zhou}}, \bibinfo {author} {\bibfnamefont
  {H.}~\bibnamefont {Wang}}, \bibinfo {author} {\bibfnamefont {S.}~\bibnamefont
  {Zhang}}, \bibinfo {author} {\bibfnamefont {K.}~\bibnamefont {Yang}},
  \bibinfo {author} {\bibfnamefont {A.}~\bibnamefont {Li}}, \bibinfo {author}
  {\bibfnamefont {X.}~\bibnamefont {Li}}, \bibinfo {author} {\bibfnamefont
  {Y.}~\bibnamefont {Li}}, \bibinfo {author} {\bibfnamefont {J.}~\bibnamefont
  {Liu}}, \bibinfo {author} {\bibfnamefont {Y.}~\bibnamefont {Shi}}, \bibinfo
  {author} {\bibfnamefont {W.}~\bibnamefont {Ku}}, \bibinfo {author}
  {\bibfnamefont {Q.}~\bibnamefont {Wu}}, \bibinfo {author} {\bibfnamefont
  {R.~J.}\ \bibnamefont {Cava}}, \ and\ \bibinfo {author} {\bibfnamefont
  {L.}~\bibnamefont {Sun}},\ }\href {\doibase 10.1103/PhysRevB.97.245149}
  {\bibfield  {journal} {\bibinfo  {journal} {Phys. Rev. B}\ }\textbf {\bibinfo
  {volume} {97}},\ \bibinfo {pages} {245149} (\bibinfo {year}
  {2018})}\BibitemShut {NoStop}%
\bibitem [{\citenamefont {Clancy}\ \emph {et~al.}(2018)\citenamefont {Clancy},
  \citenamefont {Gretarsson}, \citenamefont {Sears}, \citenamefont {Singh},
  \citenamefont {Desgreniers}, \citenamefont {Mehlawat}, \citenamefont {Layek},
  \citenamefont {Rozenberg}, \citenamefont {Ding}, \citenamefont {Upton},
  \citenamefont {Casa}, \citenamefont {Chen}, \citenamefont {Im}, \citenamefont
  {Lee}, \citenamefont {Yadav}, \citenamefont {Hozoi}, \citenamefont {Efremov},
  \citenamefont {van~den Brink},\ and\ \citenamefont {Kim}}]{Clancy_2018}%
  \BibitemOpen
  \bibfield  {author} {\bibinfo {author} {\bibfnamefont {J.~P.}\ \bibnamefont
  {Clancy}}, \bibinfo {author} {\bibfnamefont {H.}~\bibnamefont {Gretarsson}},
  \bibinfo {author} {\bibfnamefont {J.~A.}\ \bibnamefont {Sears}}, \bibinfo
  {author} {\bibfnamefont {Y.}~\bibnamefont {Singh}}, \bibinfo {author}
  {\bibfnamefont {S.}~\bibnamefont {Desgreniers}}, \bibinfo {author}
  {\bibfnamefont {K.}~\bibnamefont {Mehlawat}}, \bibinfo {author}
  {\bibfnamefont {S.}~\bibnamefont {Layek}}, \bibinfo {author} {\bibfnamefont
  {G.~K.}\ \bibnamefont {Rozenberg}}, \bibinfo {author} {\bibfnamefont
  {Y.}~\bibnamefont {Ding}}, \bibinfo {author} {\bibfnamefont {M.~H.}\
  \bibnamefont {Upton}}, \bibinfo {author} {\bibfnamefont {D.}~\bibnamefont
  {Casa}}, \bibinfo {author} {\bibfnamefont {N.}~\bibnamefont {Chen}}, \bibinfo
  {author} {\bibfnamefont {J.}~\bibnamefont {Im}}, \bibinfo {author}
  {\bibfnamefont {Y.}~\bibnamefont {Lee}}, \bibinfo {author} {\bibfnamefont
  {R.}~\bibnamefont {Yadav}}, \bibinfo {author} {\bibfnamefont
  {L.}~\bibnamefont {Hozoi}}, \bibinfo {author} {\bibfnamefont
  {D.}~\bibnamefont {Efremov}}, \bibinfo {author} {\bibfnamefont
  {J.}~\bibnamefont {van~den Brink}}, \ and\ \bibinfo {author} {\bibfnamefont
  {Y.-J.}\ \bibnamefont {Kim}},\ }\href {\doibase 10.1038/s41535-018-0109-0}
  {\bibfield  {journal} {\bibinfo  {journal} {npj Quantum Materials}\ }\textbf
  {\bibinfo {volume} {3}},\ \bibinfo {pages} {35} (\bibinfo {year}
  {2018})}\BibitemShut {NoStop}%
\bibitem [{\citenamefont {Yadav}\ \emph
  {et~al.}(2018{\natexlab{b}})\citenamefont {Yadav}, \citenamefont {Rachel},
  \citenamefont {Hozoi}, \citenamefont {van~den Brink},\ and\ \citenamefont
  {Jackeli}}]{Yadav_2018a}%
  \BibitemOpen
  \bibfield  {author} {\bibinfo {author} {\bibfnamefont {R.}~\bibnamefont
  {Yadav}}, \bibinfo {author} {\bibfnamefont {S.}~\bibnamefont {Rachel}},
  \bibinfo {author} {\bibfnamefont {L.}~\bibnamefont {Hozoi}}, \bibinfo
  {author} {\bibfnamefont {J.}~\bibnamefont {van~den Brink}}, \ and\ \bibinfo
  {author} {\bibfnamefont {G.}~\bibnamefont {Jackeli}},\ }\href {\doibase
  10.1103/PhysRevB.98.121107} {\bibfield  {journal} {\bibinfo  {journal} {Phys.
  Rev. B}\ }\textbf {\bibinfo {volume} {98}},\ \bibinfo {pages} {121107}
  (\bibinfo {year} {2018}{\natexlab{b}})}\BibitemShut {NoStop}%
\bibitem [{\citenamefont {Simutis}\ \emph {et~al.}(2018)\citenamefont
  {Simutis}, \citenamefont {Barbero}, \citenamefont {Rolfs}, \citenamefont
  {Leroy-Calatayud}, \citenamefont {Mehlawat}, \citenamefont {Khasanov},
  \citenamefont {Luetkens}, \citenamefont {Pomjakushina}, \citenamefont
  {Singh}, \citenamefont {Ott}, \citenamefont {Mesot}, \citenamefont {Amato},\
  and\ \citenamefont {Shiroka}}]{Simutis_2018}%
  \BibitemOpen
  \bibfield  {author} {\bibinfo {author} {\bibfnamefont {G.}~\bibnamefont
  {Simutis}}, \bibinfo {author} {\bibfnamefont {N.}~\bibnamefont {Barbero}},
  \bibinfo {author} {\bibfnamefont {K.}~\bibnamefont {Rolfs}}, \bibinfo
  {author} {\bibfnamefont {P.}~\bibnamefont {Leroy-Calatayud}}, \bibinfo
  {author} {\bibfnamefont {K.}~\bibnamefont {Mehlawat}}, \bibinfo {author}
  {\bibfnamefont {R.}~\bibnamefont {Khasanov}}, \bibinfo {author}
  {\bibfnamefont {H.}~\bibnamefont {Luetkens}}, \bibinfo {author}
  {\bibfnamefont {E.}~\bibnamefont {Pomjakushina}}, \bibinfo {author}
  {\bibfnamefont {Y.}~\bibnamefont {Singh}}, \bibinfo {author} {\bibfnamefont
  {H.-R.}\ \bibnamefont {Ott}}, \bibinfo {author} {\bibfnamefont
  {J.}~\bibnamefont {Mesot}}, \bibinfo {author} {\bibfnamefont
  {A.}~\bibnamefont {Amato}}, \ and\ \bibinfo {author} {\bibfnamefont
  {T.}~\bibnamefont {Shiroka}},\ }\href {\doibase 10.1103/PhysRevB.98.104421}
  {\bibfield  {journal} {\bibinfo  {journal} {Phys. Rev. B}\ }\textbf {\bibinfo
  {volume} {98}},\ \bibinfo {pages} {104421} (\bibinfo {year}
  {2018})}\BibitemShut {NoStop}%
\bibitem [{\citenamefont {Yadav}\ \emph {et~al.}(2016)\citenamefont {Yadav},
  \citenamefont {Bogdanov}, \citenamefont {Katukuri}, \citenamefont
  {Nishimoto}, \citenamefont {van~den Brink},\ and\ \citenamefont
  {Hozoi}}]{Yadav_2016}%
  \BibitemOpen
  \bibfield  {author} {\bibinfo {author} {\bibfnamefont {R.}~\bibnamefont
  {Yadav}}, \bibinfo {author} {\bibfnamefont {N.~A.}\ \bibnamefont {Bogdanov}},
  \bibinfo {author} {\bibfnamefont {V.~M.}\ \bibnamefont {Katukuri}}, \bibinfo
  {author} {\bibfnamefont {S.}~\bibnamefont {Nishimoto}}, \bibinfo {author}
  {\bibfnamefont {J.}~\bibnamefont {van~den Brink}}, \ and\ \bibinfo {author}
  {\bibfnamefont {L.}~\bibnamefont {Hozoi}},\ }\href
  {https://doi.org/10.1038/srep37925} {\bibfield  {journal} {\bibinfo
  {journal} {Scientific Reports}\ }\textbf {\bibinfo {volume} {6}},\ \bibinfo
  {pages} {37925} (\bibinfo {year} {2016})}\BibitemShut {NoStop}%
\bibitem [{\citenamefont {Sears}\ \emph {et~al.}(2017)\citenamefont {Sears},
  \citenamefont {Zhao}, \citenamefont {Xu}, \citenamefont {Lynn},\ and\
  \citenamefont {Kim}}]{Sears_2017}%
  \BibitemOpen
  \bibfield  {author} {\bibinfo {author} {\bibfnamefont {J.~A.}\ \bibnamefont
  {Sears}}, \bibinfo {author} {\bibfnamefont {Y.}~\bibnamefont {Zhao}},
  \bibinfo {author} {\bibfnamefont {Z.}~\bibnamefont {Xu}}, \bibinfo {author}
  {\bibfnamefont {J.~W.}\ \bibnamefont {Lynn}}, \ and\ \bibinfo {author}
  {\bibfnamefont {Y.-J.}\ \bibnamefont {Kim}},\ }\href {\doibase
  10.1103/PhysRevB.95.180411} {\bibfield  {journal} {\bibinfo  {journal} {Phys.
  Rev. B}\ }\textbf {\bibinfo {volume} {95}},\ \bibinfo {pages} {180411}
  (\bibinfo {year} {2017})}\BibitemShut {NoStop}%
\bibitem [{\citenamefont {Kasahara}\ \emph {et~al.}(2018)\citenamefont
  {Kasahara}, \citenamefont {Ohnishi}, \citenamefont {Mizukami}, \citenamefont
  {Tanaka}, \citenamefont {Ma}, \citenamefont {Sugii}, \citenamefont {Kurita},
  \citenamefont {Tanaka}, \citenamefont {Nasu}, \citenamefont {Motome},
  \citenamefont {Shibauchi},\ and\ \citenamefont {Matsuda}}]{Kasahara_2018}%
  \BibitemOpen
  \bibfield  {author} {\bibinfo {author} {\bibfnamefont {Y.}~\bibnamefont
  {Kasahara}}, \bibinfo {author} {\bibfnamefont {T.}~\bibnamefont {Ohnishi}},
  \bibinfo {author} {\bibfnamefont {Y.}~\bibnamefont {Mizukami}}, \bibinfo
  {author} {\bibfnamefont {O.}~\bibnamefont {Tanaka}}, \bibinfo {author}
  {\bibfnamefont {S.}~\bibnamefont {Ma}}, \bibinfo {author} {\bibfnamefont
  {K.}~\bibnamefont {Sugii}}, \bibinfo {author} {\bibfnamefont
  {N.}~\bibnamefont {Kurita}}, \bibinfo {author} {\bibfnamefont
  {H.}~\bibnamefont {Tanaka}}, \bibinfo {author} {\bibfnamefont
  {J.}~\bibnamefont {Nasu}}, \bibinfo {author} {\bibfnamefont {Y.}~\bibnamefont
  {Motome}}, \bibinfo {author} {\bibfnamefont {T.}~\bibnamefont {Shibauchi}}, \
  and\ \bibinfo {author} {\bibfnamefont {Y.}~\bibnamefont {Matsuda}},\ }\href
  {\doibase 10.1038/s41586-018-0274-0} {\bibfield  {journal} {\bibinfo
  {journal} {Nature}\ }\textbf {\bibinfo {volume} {559}},\ \bibinfo {pages}
  {227} (\bibinfo {year} {2018})}\BibitemShut {NoStop}%
\bibitem [{\citenamefont {Balz}\ \emph {et~al.}(2019)\citenamefont {Balz},
  \citenamefont {Lampen-Kelley}, \citenamefont {Banerjee}, \citenamefont {Yan},
  \citenamefont {Lu}, \citenamefont {Hu}, \citenamefont {Yadav}, \citenamefont
  {Takano}, \citenamefont {Liu}, \citenamefont {Tennant}, \citenamefont
  {Lumsden}, \citenamefont {Mandrus},\ and\ \citenamefont
  {Nagler}}]{Balz_2019}%
  \BibitemOpen
  \bibfield  {author} {\bibinfo {author} {\bibfnamefont {C.}~\bibnamefont
  {Balz}}, \bibinfo {author} {\bibfnamefont {P.}~\bibnamefont {Lampen-Kelley}},
  \bibinfo {author} {\bibfnamefont {A.}~\bibnamefont {Banerjee}}, \bibinfo
  {author} {\bibfnamefont {J.}~\bibnamefont {Yan}}, \bibinfo {author}
  {\bibfnamefont {Z.}~\bibnamefont {Lu}}, \bibinfo {author} {\bibfnamefont
  {X.}~\bibnamefont {Hu}}, \bibinfo {author} {\bibfnamefont {S.~M.}\
  \bibnamefont {Yadav}}, \bibinfo {author} {\bibfnamefont {Y.}~\bibnamefont
  {Takano}}, \bibinfo {author} {\bibfnamefont {Y.}~\bibnamefont {Liu}},
  \bibinfo {author} {\bibfnamefont {D.~A.}\ \bibnamefont {Tennant}}, \bibinfo
  {author} {\bibfnamefont {M.~D.}\ \bibnamefont {Lumsden}}, \bibinfo {author}
  {\bibfnamefont {D.}~\bibnamefont {Mandrus}}, \ and\ \bibinfo {author}
  {\bibfnamefont {S.~E.}\ \bibnamefont {Nagler}},\ }\href {\doibase
  10.1103/PhysRevB.100.060405} {\bibfield  {journal} {\bibinfo  {journal}
  {Phys. Rev. B}\ }\textbf {\bibinfo {volume} {100}},\ \bibinfo {pages}
  {060405} (\bibinfo {year} {2019})}\BibitemShut {NoStop}%
\bibitem [{\citenamefont {Hunt}\ \emph {et~al.}(2013)\citenamefont {Hunt},
  \citenamefont {Sanchez-Yamagishi}, \citenamefont {Young}, \citenamefont
  {Yankowitz}, \citenamefont {LeRoy}, \citenamefont {Watanabe}, \citenamefont
  {Taniguchi}, \citenamefont {Moon}, \citenamefont {Koshino}, \citenamefont
  {Jarillo-Herrero},\ and\ \citenamefont {Ashoori}}]{Hunt_2013}%
  \BibitemOpen
  \bibfield  {author} {\bibinfo {author} {\bibfnamefont {B.}~\bibnamefont
  {Hunt}}, \bibinfo {author} {\bibfnamefont {J.~D.}\ \bibnamefont
  {Sanchez-Yamagishi}}, \bibinfo {author} {\bibfnamefont {A.~F.}\ \bibnamefont
  {Young}}, \bibinfo {author} {\bibfnamefont {M.}~\bibnamefont {Yankowitz}},
  \bibinfo {author} {\bibfnamefont {B.~J.}\ \bibnamefont {LeRoy}}, \bibinfo
  {author} {\bibfnamefont {K.}~\bibnamefont {Watanabe}}, \bibinfo {author}
  {\bibfnamefont {T.}~\bibnamefont {Taniguchi}}, \bibinfo {author}
  {\bibfnamefont {P.}~\bibnamefont {Moon}}, \bibinfo {author} {\bibfnamefont
  {M.}~\bibnamefont {Koshino}}, \bibinfo {author} {\bibfnamefont
  {P.}~\bibnamefont {Jarillo-Herrero}}, \ and\ \bibinfo {author} {\bibfnamefont
  {R.~C.}\ \bibnamefont {Ashoori}},\ }\href {\doibase 10.1126/science.1237240}
  {\bibfield  {journal} {\bibinfo  {journal} {Science}\ }\textbf {\bibinfo
  {volume} {340}},\ \bibinfo {pages} {1427} (\bibinfo {year}
  {2013})}\BibitemShut {NoStop}%
\bibitem [{\citenamefont {Dean}\ \emph {et~al.}(2013)\citenamefont {Dean},
  \citenamefont {Wang}, \citenamefont {Maher}, \citenamefont {Forsythe},
  \citenamefont {Ghahari}, \citenamefont {Gao}, \citenamefont {Katoch},
  \citenamefont {Ishigami}, \citenamefont {Moon}, \citenamefont {Koshino},
  \citenamefont {Taniguchi}, \citenamefont {Watanabe}, \citenamefont {Shepard},
  \citenamefont {Hone},\ and\ \citenamefont {Kim}}]{Dean_2013}%
  \BibitemOpen
  \bibfield  {author} {\bibinfo {author} {\bibfnamefont {C.~R.}\ \bibnamefont
  {Dean}}, \bibinfo {author} {\bibfnamefont {L.}~\bibnamefont {Wang}}, \bibinfo
  {author} {\bibfnamefont {P.}~\bibnamefont {Maher}}, \bibinfo {author}
  {\bibfnamefont {C.}~\bibnamefont {Forsythe}}, \bibinfo {author}
  {\bibfnamefont {F.}~\bibnamefont {Ghahari}}, \bibinfo {author} {\bibfnamefont
  {Y.}~\bibnamefont {Gao}}, \bibinfo {author} {\bibfnamefont {J.}~\bibnamefont
  {Katoch}}, \bibinfo {author} {\bibfnamefont {M.}~\bibnamefont {Ishigami}},
  \bibinfo {author} {\bibfnamefont {P.}~\bibnamefont {Moon}}, \bibinfo {author}
  {\bibfnamefont {M.}~\bibnamefont {Koshino}}, \bibinfo {author} {\bibfnamefont
  {T.}~\bibnamefont {Taniguchi}}, \bibinfo {author} {\bibfnamefont
  {K.}~\bibnamefont {Watanabe}}, \bibinfo {author} {\bibfnamefont {K.~L.}\
  \bibnamefont {Shepard}}, \bibinfo {author} {\bibfnamefont {J.}~\bibnamefont
  {Hone}}, \ and\ \bibinfo {author} {\bibfnamefont {P.}~\bibnamefont {Kim}},\
  }\href {http://dx.doi.org/10.1038/nature12186} {\bibfield  {journal}
  {\bibinfo  {journal} {Nature}\ }\textbf {\bibinfo {volume} {497}} (\bibinfo
  {year} {2013})}\BibitemShut {NoStop}%
\bibitem [{\citenamefont {Wang}\ \emph {et~al.}(2015)\citenamefont {Wang},
  \citenamefont {Gao}, \citenamefont {Wen}, \citenamefont {Han}, \citenamefont
  {Taniguchi}, \citenamefont {Watanabe}, \citenamefont {Koshino}, \citenamefont
  {Hone},\ and\ \citenamefont {Dean}}]{Wang_2015}%
  \BibitemOpen
  \bibfield  {author} {\bibinfo {author} {\bibfnamefont {L.}~\bibnamefont
  {Wang}}, \bibinfo {author} {\bibfnamefont {Y.}~\bibnamefont {Gao}}, \bibinfo
  {author} {\bibfnamefont {B.}~\bibnamefont {Wen}}, \bibinfo {author}
  {\bibfnamefont {Z.}~\bibnamefont {Han}}, \bibinfo {author} {\bibfnamefont
  {T.}~\bibnamefont {Taniguchi}}, \bibinfo {author} {\bibfnamefont
  {K.}~\bibnamefont {Watanabe}}, \bibinfo {author} {\bibfnamefont
  {M.}~\bibnamefont {Koshino}}, \bibinfo {author} {\bibfnamefont
  {J.}~\bibnamefont {Hone}}, \ and\ \bibinfo {author} {\bibfnamefont {C.~R.}\
  \bibnamefont {Dean}},\ }\href {\doibase 10.1126/science.aad2102} {\bibfield
  {journal} {\bibinfo  {journal} {Science}\ }\textbf {\bibinfo {volume}
  {350}},\ \bibinfo {pages} {1231} (\bibinfo {year} {2015})}\BibitemShut
  {NoStop}%
\bibitem [{\citenamefont {Spanton}\ \emph {et~al.}(2018)\citenamefont
  {Spanton}, \citenamefont {Zibrov}, \citenamefont {Zhou}, \citenamefont
  {Taniguchi}, \citenamefont {Watanabe}, \citenamefont {Zaletel},\ and\
  \citenamefont {Young}}]{Spanton_2018}%
  \BibitemOpen
  \bibfield  {author} {\bibinfo {author} {\bibfnamefont {E.~M.}\ \bibnamefont
  {Spanton}}, \bibinfo {author} {\bibfnamefont {A.~A.}\ \bibnamefont {Zibrov}},
  \bibinfo {author} {\bibfnamefont {H.}~\bibnamefont {Zhou}}, \bibinfo {author}
  {\bibfnamefont {T.}~\bibnamefont {Taniguchi}}, \bibinfo {author}
  {\bibfnamefont {K.}~\bibnamefont {Watanabe}}, \bibinfo {author}
  {\bibfnamefont {M.~P.}\ \bibnamefont {Zaletel}}, \ and\ \bibinfo {author}
  {\bibfnamefont {A.~F.}\ \bibnamefont {Young}},\ }\href {\doibase
  10.1126/science.aan8458} {\bibfield  {journal} {\bibinfo  {journal}
  {Science}\ }\textbf {\bibinfo {volume} {360}},\ \bibinfo {pages} {62}
  (\bibinfo {year} {2018})}\BibitemShut {NoStop}%
\bibitem [{\citenamefont {Lopes~dos Santos}\ \emph {et~al.}(2007)\citenamefont
  {Lopes~dos Santos}, \citenamefont {Peres},\ and\ \citenamefont
  {Castro~Neto}}]{Lopes_2007}%
  \BibitemOpen
  \bibfield  {author} {\bibinfo {author} {\bibfnamefont {J.~M.~B.}\
  \bibnamefont {Lopes~dos Santos}}, \bibinfo {author} {\bibfnamefont
  {N.~M.~R.}\ \bibnamefont {Peres}}, \ and\ \bibinfo {author} {\bibfnamefont
  {A.~H.}\ \bibnamefont {Castro~Neto}},\ }\href {\doibase
  10.1103/PhysRevLett.99.256802} {\bibfield  {journal} {\bibinfo  {journal}
  {Phys. Rev. Lett.}\ }\textbf {\bibinfo {volume} {99}},\ \bibinfo {pages}
  {256802} (\bibinfo {year} {2007})}\BibitemShut {NoStop}%
\bibitem [{\citenamefont {Li}\ \emph {et~al.}(2009)\citenamefont {Li},
  \citenamefont {Luican}, \citenamefont {Lopes~dos Santos}, \citenamefont
  {Castro~Neto}, \citenamefont {Reina}, \citenamefont {Kong},\ and\
  \citenamefont {Andrei}}]{Li_2009}%
  \BibitemOpen
  \bibfield  {author} {\bibinfo {author} {\bibfnamefont {G.}~\bibnamefont
  {Li}}, \bibinfo {author} {\bibfnamefont {A.}~\bibnamefont {Luican}}, \bibinfo
  {author} {\bibfnamefont {J.~M.~B.}\ \bibnamefont {Lopes~dos Santos}},
  \bibinfo {author} {\bibfnamefont {A.~H.}\ \bibnamefont {Castro~Neto}},
  \bibinfo {author} {\bibfnamefont {A.}~\bibnamefont {Reina}}, \bibinfo
  {author} {\bibfnamefont {J.}~\bibnamefont {Kong}}, \ and\ \bibinfo {author}
  {\bibfnamefont {E.~Y.}\ \bibnamefont {Andrei}},\ }\href
  {http://dx.doi.org/10.1038/nphys1463} {\bibfield  {journal} {\bibinfo
  {journal} {Nature Physics}\ }\textbf {\bibinfo {volume} {6}},\ \bibinfo
  {pages} {109} (\bibinfo {year} {2009})}\BibitemShut {NoStop}%
\bibitem [{\citenamefont {Bistritzer}\ and\ \citenamefont
  {MacDonald}(2011)}]{Bistritzer_2010}%
  \BibitemOpen
  \bibfield  {author} {\bibinfo {author} {\bibfnamefont {R.}~\bibnamefont
  {Bistritzer}}\ and\ \bibinfo {author} {\bibfnamefont {A.~H.}\ \bibnamefont
  {MacDonald}},\ }\href {\doibase 10.1073/pnas.1108174108} {\bibfield
  {journal} {\bibinfo  {journal} {Proceedings of the National Academy of
  Sciences}\ }\textbf {\bibinfo {volume} {108}},\ \bibinfo {pages} {12233}
  (\bibinfo {year} {2011})}\BibitemShut {NoStop}%
\bibitem [{\citenamefont {Mele}(2010)}]{Mele_2010}%
  \BibitemOpen
  \bibfield  {author} {\bibinfo {author} {\bibfnamefont {E.~J.}\ \bibnamefont
  {Mele}},\ }\href {\doibase 10.1103/PhysRevB.81.161405} {\bibfield  {journal}
  {\bibinfo  {journal} {Phys. Rev. B}\ }\textbf {\bibinfo {volume} {81}},\
  \bibinfo {pages} {161405} (\bibinfo {year} {2010})}\BibitemShut {NoStop}%
\bibitem [{\citenamefont {Luican}\ \emph {et~al.}(2011)\citenamefont {Luican},
  \citenamefont {Li}, \citenamefont {Reina}, \citenamefont {Kong},
  \citenamefont {Nair}, \citenamefont {Novoselov}, \citenamefont {Geim},\ and\
  \citenamefont {Andrei}}]{Luican_2011}%
  \BibitemOpen
  \bibfield  {author} {\bibinfo {author} {\bibfnamefont {A.}~\bibnamefont
  {Luican}}, \bibinfo {author} {\bibfnamefont {G.}~\bibnamefont {Li}}, \bibinfo
  {author} {\bibfnamefont {A.}~\bibnamefont {Reina}}, \bibinfo {author}
  {\bibfnamefont {J.}~\bibnamefont {Kong}}, \bibinfo {author} {\bibfnamefont
  {R.~R.}\ \bibnamefont {Nair}}, \bibinfo {author} {\bibfnamefont {K.~S.}\
  \bibnamefont {Novoselov}}, \bibinfo {author} {\bibfnamefont {A.~K.}\
  \bibnamefont {Geim}}, \ and\ \bibinfo {author} {\bibfnamefont {E.~Y.}\
  \bibnamefont {Andrei}},\ }\href {\doibase 10.1103/PhysRevLett.106.126802}
  {\bibfield  {journal} {\bibinfo  {journal} {Phys. Rev. Lett.}\ }\textbf
  {\bibinfo {volume} {106}},\ \bibinfo {pages} {126802} (\bibinfo {year}
  {2011})}\BibitemShut {NoStop}%
\bibitem [{\citenamefont {Lopes~dos Santos}\ \emph {et~al.}(2012)\citenamefont
  {Lopes~dos Santos}, \citenamefont {Peres},\ and\ \citenamefont
  {Castro~Neto}}]{Lopes_2012}%
  \BibitemOpen
  \bibfield  {author} {\bibinfo {author} {\bibfnamefont {J.~M.~B.}\
  \bibnamefont {Lopes~dos Santos}}, \bibinfo {author} {\bibfnamefont
  {N.~M.~R.}\ \bibnamefont {Peres}}, \ and\ \bibinfo {author} {\bibfnamefont
  {A.~H.}\ \bibnamefont {Castro~Neto}},\ }\href {\doibase
  10.1103/PhysRevB.86.155449} {\bibfield  {journal} {\bibinfo  {journal} {Phys.
  Rev. B}\ }\textbf {\bibinfo {volume} {86}},\ \bibinfo {pages} {155449}
  (\bibinfo {year} {2012})}\BibitemShut {NoStop}%
\bibitem [{\citenamefont {Jung}\ \emph {et~al.}(2014)\citenamefont {Jung},
  \citenamefont {Raoux}, \citenamefont {Qiao},\ and\ \citenamefont
  {MacDonald}}]{Jung_2014}%
  \BibitemOpen
  \bibfield  {author} {\bibinfo {author} {\bibfnamefont {J.}~\bibnamefont
  {Jung}}, \bibinfo {author} {\bibfnamefont {A.}~\bibnamefont {Raoux}},
  \bibinfo {author} {\bibfnamefont {Z.}~\bibnamefont {Qiao}}, \ and\ \bibinfo
  {author} {\bibfnamefont {A.~H.}\ \bibnamefont {MacDonald}},\ }\href {\doibase
  10.1103/PhysRevB.89.205414} {\bibfield  {journal} {\bibinfo  {journal} {Phys.
  Rev. B}\ }\textbf {\bibinfo {volume} {89}},\ \bibinfo {pages} {205414}
  (\bibinfo {year} {2014})}\BibitemShut {NoStop}%
\bibitem [{\citenamefont {Wong}\ \emph {et~al.}(2015)\citenamefont {Wong},
  \citenamefont {Wang}, \citenamefont {Jung}, \citenamefont {Pezzini},
  \citenamefont {DaSilva}, \citenamefont {Tsai}, \citenamefont {Jung},
  \citenamefont {Khajeh}, \citenamefont {Kim}, \citenamefont {Lee},
  \citenamefont {Kahn}, \citenamefont {Tollabimazraehno}, \citenamefont
  {Rasool}, \citenamefont {Watanabe}, \citenamefont {Taniguchi}, \citenamefont
  {Zettl}, \citenamefont {Adam}, \citenamefont {MacDonald},\ and\ \citenamefont
  {Crommie}}]{Wong_2015}%
  \BibitemOpen
  \bibfield  {author} {\bibinfo {author} {\bibfnamefont {D.}~\bibnamefont
  {Wong}}, \bibinfo {author} {\bibfnamefont {Y.}~\bibnamefont {Wang}}, \bibinfo
  {author} {\bibfnamefont {J.}~\bibnamefont {Jung}}, \bibinfo {author}
  {\bibfnamefont {S.}~\bibnamefont {Pezzini}}, \bibinfo {author} {\bibfnamefont
  {A.~M.}\ \bibnamefont {DaSilva}}, \bibinfo {author} {\bibfnamefont {H.-Z.}\
  \bibnamefont {Tsai}}, \bibinfo {author} {\bibfnamefont {H.~S.}\ \bibnamefont
  {Jung}}, \bibinfo {author} {\bibfnamefont {R.}~\bibnamefont {Khajeh}},
  \bibinfo {author} {\bibfnamefont {Y.}~\bibnamefont {Kim}}, \bibinfo {author}
  {\bibfnamefont {J.}~\bibnamefont {Lee}}, \bibinfo {author} {\bibfnamefont
  {S.}~\bibnamefont {Kahn}}, \bibinfo {author} {\bibfnamefont {S.}~\bibnamefont
  {Tollabimazraehno}}, \bibinfo {author} {\bibfnamefont {H.}~\bibnamefont
  {Rasool}}, \bibinfo {author} {\bibfnamefont {K.}~\bibnamefont {Watanabe}},
  \bibinfo {author} {\bibfnamefont {T.}~\bibnamefont {Taniguchi}}, \bibinfo
  {author} {\bibfnamefont {A.}~\bibnamefont {Zettl}}, \bibinfo {author}
  {\bibfnamefont {S.}~\bibnamefont {Adam}}, \bibinfo {author} {\bibfnamefont
  {A.~H.}\ \bibnamefont {MacDonald}}, \ and\ \bibinfo {author} {\bibfnamefont
  {M.~F.}\ \bibnamefont {Crommie}},\ }\href {\doibase
  10.1103/PhysRevB.92.155409} {\bibfield  {journal} {\bibinfo  {journal} {Phys.
  Rev. B}\ }\textbf {\bibinfo {volume} {92}},\ \bibinfo {pages} {155409}
  (\bibinfo {year} {2015})}\BibitemShut {NoStop}%
\bibitem [{\citenamefont {Kim}\ \emph {et~al.}(2017)\citenamefont {Kim},
  \citenamefont {DaSilva}, \citenamefont {Huang}, \citenamefont {Fallahazad},
  \citenamefont {Larentis}, \citenamefont {Taniguchi}, \citenamefont
  {Watanabe}, \citenamefont {LeRoy}, \citenamefont {MacDonald},\ and\
  \citenamefont {Tutuc}}]{Kim_2017}%
  \BibitemOpen
  \bibfield  {author} {\bibinfo {author} {\bibfnamefont {K.}~\bibnamefont
  {Kim}}, \bibinfo {author} {\bibfnamefont {A.}~\bibnamefont {DaSilva}},
  \bibinfo {author} {\bibfnamefont {S.}~\bibnamefont {Huang}}, \bibinfo
  {author} {\bibfnamefont {B.}~\bibnamefont {Fallahazad}}, \bibinfo {author}
  {\bibfnamefont {S.}~\bibnamefont {Larentis}}, \bibinfo {author}
  {\bibfnamefont {T.}~\bibnamefont {Taniguchi}}, \bibinfo {author}
  {\bibfnamefont {K.}~\bibnamefont {Watanabe}}, \bibinfo {author}
  {\bibfnamefont {B.~J.}\ \bibnamefont {LeRoy}}, \bibinfo {author}
  {\bibfnamefont {A.~H.}\ \bibnamefont {MacDonald}}, \ and\ \bibinfo {author}
  {\bibfnamefont {E.}~\bibnamefont {Tutuc}},\ }\href {\doibase
  10.1073/pnas.1620140114} {\bibfield  {journal} {\bibinfo  {journal}
  {Proceedings of the National Academy of Sciences}\ }\textbf {\bibinfo
  {volume} {114}},\ \bibinfo {pages} {3364} (\bibinfo {year}
  {2017})}\BibitemShut {NoStop}%
\bibitem [{\citenamefont {Nam}\ and\ \citenamefont {Koshino}(2017)}]{Nam_2017}%
  \BibitemOpen
  \bibfield  {author} {\bibinfo {author} {\bibfnamefont {N.~N.~T.}\
  \bibnamefont {Nam}}\ and\ \bibinfo {author} {\bibfnamefont {M.}~\bibnamefont
  {Koshino}},\ }\href {\doibase 10.1103/PhysRevB.96.075311} {\bibfield
  {journal} {\bibinfo  {journal} {Phys. Rev. B}\ }\textbf {\bibinfo {volume}
  {96}},\ \bibinfo {pages} {075311} (\bibinfo {year} {2017})}\BibitemShut
  {NoStop}%
\bibitem [{\citenamefont {Efimkin}\ and\ \citenamefont
  {MacDonald}(2018)}]{Efimkin_2018}%
  \BibitemOpen
  \bibfield  {author} {\bibinfo {author} {\bibfnamefont {D.~K.}\ \bibnamefont
  {Efimkin}}\ and\ \bibinfo {author} {\bibfnamefont {A.~H.}\ \bibnamefont
  {MacDonald}},\ }\href {\doibase 10.1103/PhysRevB.98.035404} {\bibfield
  {journal} {\bibinfo  {journal} {Phys. Rev. B}\ }\textbf {\bibinfo {volume}
  {98}},\ \bibinfo {pages} {035404} (\bibinfo {year} {2018})}\BibitemShut
  {NoStop}%
\bibitem [{\citenamefont {Cao}\ \emph {et~al.}(2018{\natexlab{a}})\citenamefont
  {Cao}, \citenamefont {Fatemi}, \citenamefont {Demir}, \citenamefont {Fang},
  \citenamefont {Tomarken}, \citenamefont {Luo}, \citenamefont
  {Sanchez-Yamagishi}, \citenamefont {Watanabe}, \citenamefont {Taniguchi},
  \citenamefont {Kaxiras}, \citenamefont {Ashoori},\ and\ \citenamefont
  {Jarillo-Herrero}}]{Cao_2018}%
  \BibitemOpen
  \bibfield  {author} {\bibinfo {author} {\bibfnamefont {Y.}~\bibnamefont
  {Cao}}, \bibinfo {author} {\bibfnamefont {V.}~\bibnamefont {Fatemi}},
  \bibinfo {author} {\bibfnamefont {A.}~\bibnamefont {Demir}}, \bibinfo
  {author} {\bibfnamefont {S.}~\bibnamefont {Fang}}, \bibinfo {author}
  {\bibfnamefont {S.~L.}\ \bibnamefont {Tomarken}}, \bibinfo {author}
  {\bibfnamefont {J.~Y.}\ \bibnamefont {Luo}}, \bibinfo {author} {\bibfnamefont
  {J.~D.}\ \bibnamefont {Sanchez-Yamagishi}}, \bibinfo {author} {\bibfnamefont
  {K.}~\bibnamefont {Watanabe}}, \bibinfo {author} {\bibfnamefont
  {T.}~\bibnamefont {Taniguchi}}, \bibinfo {author} {\bibfnamefont
  {E.}~\bibnamefont {Kaxiras}}, \bibinfo {author} {\bibfnamefont {R.~C.}\
  \bibnamefont {Ashoori}}, \ and\ \bibinfo {author} {\bibfnamefont
  {P.}~\bibnamefont {Jarillo-Herrero}},\ }\href
  {http://dx.doi.org/10.1038/nature26154} {\bibfield  {journal} {\bibinfo
  {journal} {Nature}\ }\textbf {\bibinfo {volume} {556}},\ \bibinfo {pages}
  {80} (\bibinfo {year} {2018}{\natexlab{a}})}\BibitemShut {NoStop}%
\bibitem [{\citenamefont {Cao}\ \emph {et~al.}(2018{\natexlab{b}})\citenamefont
  {Cao}, \citenamefont {Fatemi}, \citenamefont {Fang}, \citenamefont
  {Watanabe}, \citenamefont {Taniguchi}, \citenamefont {Kaxiras},\ and\
  \citenamefont {Jarillo-Herrero}}]{Cao_2018a}%
  \BibitemOpen
  \bibfield  {author} {\bibinfo {author} {\bibfnamefont {Y.}~\bibnamefont
  {Cao}}, \bibinfo {author} {\bibfnamefont {V.}~\bibnamefont {Fatemi}},
  \bibinfo {author} {\bibfnamefont {S.}~\bibnamefont {Fang}}, \bibinfo {author}
  {\bibfnamefont {K.}~\bibnamefont {Watanabe}}, \bibinfo {author}
  {\bibfnamefont {T.}~\bibnamefont {Taniguchi}}, \bibinfo {author}
  {\bibfnamefont {E.}~\bibnamefont {Kaxiras}}, \ and\ \bibinfo {author}
  {\bibfnamefont {P.}~\bibnamefont {Jarillo-Herrero}},\ }\href
  {http://dx.doi.org/10.1038/nature26160} {\bibfield  {journal} {\bibinfo
  {journal} {Nature}\ }\textbf {\bibinfo {volume} {556}},\ \bibinfo {pages}
  {43} (\bibinfo {year} {2018}{\natexlab{b}})}\BibitemShut {NoStop}%
\bibitem [{\citenamefont {Xu}\ and\ \citenamefont {Balents}(2018)}]{Xu_2018}%
  \BibitemOpen
  \bibfield  {author} {\bibinfo {author} {\bibfnamefont {C.}~\bibnamefont
  {Xu}}\ and\ \bibinfo {author} {\bibfnamefont {L.}~\bibnamefont {Balents}},\
  }\href {\doibase 10.1103/PhysRevLett.121.087001} {\bibfield  {journal}
  {\bibinfo  {journal} {Phys. Rev. Lett.}\ }\textbf {\bibinfo {volume} {121}},\
  \bibinfo {pages} {087001} (\bibinfo {year} {2018})}\BibitemShut {NoStop}%
\bibitem [{\citenamefont {Isobe}\ \emph {et~al.}(2018)\citenamefont {Isobe},
  \citenamefont {Yuan},\ and\ \citenamefont {Fu}}]{Isobe_2018}%
  \BibitemOpen
  \bibfield  {author} {\bibinfo {author} {\bibfnamefont {H.}~\bibnamefont
  {Isobe}}, \bibinfo {author} {\bibfnamefont {N.~F.~Q.}\ \bibnamefont {Yuan}},
  \ and\ \bibinfo {author} {\bibfnamefont {L.}~\bibnamefont {Fu}},\ }\href
  {\doibase 10.1103/PhysRevX.8.041041} {\bibfield  {journal} {\bibinfo
  {journal} {Phys. Rev. X}\ }\textbf {\bibinfo {volume} {8}},\ \bibinfo {pages}
  {041041} (\bibinfo {year} {2018})}\BibitemShut {NoStop}%
\bibitem [{\citenamefont {Wu}\ \emph {et~al.}(2018{\natexlab{a}})\citenamefont
  {Wu}, \citenamefont {Lovorn}, \citenamefont {Tutuc},\ and\ \citenamefont
  {MacDonald}}]{Wu_2018}%
  \BibitemOpen
  \bibfield  {author} {\bibinfo {author} {\bibfnamefont {F.}~\bibnamefont
  {Wu}}, \bibinfo {author} {\bibfnamefont {T.}~\bibnamefont {Lovorn}}, \bibinfo
  {author} {\bibfnamefont {E.}~\bibnamefont {Tutuc}}, \ and\ \bibinfo {author}
  {\bibfnamefont {A.~H.}\ \bibnamefont {MacDonald}},\ }\href {\doibase
  10.1103/PhysRevLett.121.026402} {\bibfield  {journal} {\bibinfo  {journal}
  {Phys. Rev. Lett.}\ }\textbf {\bibinfo {volume} {121}},\ \bibinfo {pages}
  {026402} (\bibinfo {year} {2018}{\natexlab{a}})}\BibitemShut {NoStop}%
\bibitem [{\citenamefont {You}\ and\ \citenamefont
  {Vishwanath}(2019)}]{You_2019}%
  \BibitemOpen
  \bibfield  {author} {\bibinfo {author} {\bibfnamefont {Y.-Z.}\ \bibnamefont
  {You}}\ and\ \bibinfo {author} {\bibfnamefont {A.}~\bibnamefont
  {Vishwanath}},\ }\href {\doibase 10.1038/s41535-019-0153-4} {\bibfield
  {journal} {\bibinfo  {journal} {npj Quantum Materials}\ }\textbf {\bibinfo
  {volume} {4}},\ \bibinfo {pages} {16} (\bibinfo {year} {2019})}\BibitemShut
  {NoStop}%
\bibitem [{\citenamefont {Roy}\ and\ \citenamefont {Juri\ifmmode \check{c}\else
  \v{c}\fi{}i\ifmmode~\acute{c}\else \'{c}\fi{}}(2019)}]{Roy_2019}%
  \BibitemOpen
  \bibfield  {author} {\bibinfo {author} {\bibfnamefont {B.}~\bibnamefont
  {Roy}}\ and\ \bibinfo {author} {\bibfnamefont {V.}~\bibnamefont {Juri\ifmmode
  \check{c}\else \v{c}\fi{}i\ifmmode~\acute{c}\else \'{c}\fi{}}},\ }\href
  {\doibase 10.1103/PhysRevB.99.121407} {\bibfield  {journal} {\bibinfo
  {journal} {Phys. Rev. B}\ }\textbf {\bibinfo {volume} {99}},\ \bibinfo
  {pages} {121407} (\bibinfo {year} {2019})}\BibitemShut {NoStop}%
\bibitem [{\citenamefont {Ruiz-Tijerina}\ and\ \citenamefont
  {Fal'ko}(2019)}]{Ruiz_2019}%
  \BibitemOpen
  \bibfield  {author} {\bibinfo {author} {\bibfnamefont {D.~A.}\ \bibnamefont
  {Ruiz-Tijerina}}\ and\ \bibinfo {author} {\bibfnamefont {V.~I.}\ \bibnamefont
  {Fal'ko}},\ }\href {\doibase 10.1103/PhysRevB.99.125424} {\bibfield
  {journal} {\bibinfo  {journal} {Phys. Rev. B}\ }\textbf {\bibinfo {volume}
  {99}},\ \bibinfo {pages} {125424} (\bibinfo {year} {2019})}\BibitemShut
  {NoStop}%
\bibitem [{\citenamefont {Wu}\ \emph {et~al.}(2018{\natexlab{b}})\citenamefont
  {Wu}, \citenamefont {MacDonald},\ and\ \citenamefont {Martin}}]{Wu_2019a}%
  \BibitemOpen
  \bibfield  {author} {\bibinfo {author} {\bibfnamefont {F.}~\bibnamefont
  {Wu}}, \bibinfo {author} {\bibfnamefont {A.~H.}\ \bibnamefont {MacDonald}}, \
  and\ \bibinfo {author} {\bibfnamefont {I.}~\bibnamefont {Martin}},\ }\href
  {\doibase 10.1103/PhysRevLett.121.257001} {\bibfield  {journal} {\bibinfo
  {journal} {Phys. Rev. Lett.}\ }\textbf {\bibinfo {volume} {121}},\ \bibinfo
  {pages} {257001} (\bibinfo {year} {2018}{\natexlab{b}})}\BibitemShut
  {NoStop}%
\bibitem [{\citenamefont {Wu}\ \emph {et~al.}(2019)\citenamefont {Wu},
  \citenamefont {Lovorn}, \citenamefont {Tutuc}, \citenamefont {Martin},\ and\
  \citenamefont {MacDonald}}]{Wu_2019}%
  \BibitemOpen
  \bibfield  {author} {\bibinfo {author} {\bibfnamefont {F.}~\bibnamefont
  {Wu}}, \bibinfo {author} {\bibfnamefont {T.}~\bibnamefont {Lovorn}}, \bibinfo
  {author} {\bibfnamefont {E.}~\bibnamefont {Tutuc}}, \bibinfo {author}
  {\bibfnamefont {I.}~\bibnamefont {Martin}}, \ and\ \bibinfo {author}
  {\bibfnamefont {A.~H.}\ \bibnamefont {MacDonald}},\ }\href {\doibase
  10.1103/PhysRevLett.122.086402} {\bibfield  {journal} {\bibinfo  {journal}
  {Phys. Rev. Lett.}\ }\textbf {\bibinfo {volume} {122}},\ \bibinfo {pages}
  {086402} (\bibinfo {year} {2019})}\BibitemShut {NoStop}%
\bibitem [{\citenamefont {Naik}\ and\ \citenamefont {Jain}(2018)}]{Naik_2018}%
  \BibitemOpen
  \bibfield  {author} {\bibinfo {author} {\bibfnamefont {M.~H.}\ \bibnamefont
  {Naik}}\ and\ \bibinfo {author} {\bibfnamefont {M.}~\bibnamefont {Jain}},\
  }\href {\doibase 10.1103/PhysRevLett.121.266401} {\bibfield  {journal}
  {\bibinfo  {journal} {Phys. Rev. Lett.}\ }\textbf {\bibinfo {volume} {121}},\
  \bibinfo {pages} {266401} (\bibinfo {year} {2018})}\BibitemShut {NoStop}%
\bibitem [{\citenamefont {Tang}\ \emph {et~al.}(2020)\citenamefont {Tang},
  \citenamefont {Li}, \citenamefont {Li}, \citenamefont {Xu}, \citenamefont
  {Liu}, \citenamefont {Barmak}, \citenamefont {Watanabe}, \citenamefont
  {Taniguchi}, \citenamefont {MacDonald}, \citenamefont {Shan},\ and\
  \citenamefont {Mak}}]{Tang_2020}%
  \BibitemOpen
  \bibfield  {author} {\bibinfo {author} {\bibfnamefont {Y.}~\bibnamefont
  {Tang}}, \bibinfo {author} {\bibfnamefont {L.}~\bibnamefont {Li}}, \bibinfo
  {author} {\bibfnamefont {T.}~\bibnamefont {Li}}, \bibinfo {author}
  {\bibfnamefont {Y.}~\bibnamefont {Xu}}, \bibinfo {author} {\bibfnamefont
  {S.}~\bibnamefont {Liu}}, \bibinfo {author} {\bibfnamefont {K.}~\bibnamefont
  {Barmak}}, \bibinfo {author} {\bibfnamefont {K.}~\bibnamefont {Watanabe}},
  \bibinfo {author} {\bibfnamefont {T.}~\bibnamefont {Taniguchi}}, \bibinfo
  {author} {\bibfnamefont {A.~H.}\ \bibnamefont {MacDonald}}, \bibinfo {author}
  {\bibfnamefont {J.}~\bibnamefont {Shan}}, \ and\ \bibinfo {author}
  {\bibfnamefont {K.~F.}\ \bibnamefont {Mak}},\ }\href {\doibase
  10.1038/s41586-020-2085-3} {\bibfield  {journal} {\bibinfo  {journal}
  {Nature}\ }\textbf {\bibinfo {volume} {579}},\ \bibinfo {pages} {353}
  (\bibinfo {year} {2020})}\BibitemShut {NoStop}%
\bibitem [{\citenamefont {Regan}\ \emph {et~al.}(2020)\citenamefont {Regan},
  \citenamefont {Wang}, \citenamefont {Jin}, \citenamefont {Bakti~Utama},
  \citenamefont {Gao}, \citenamefont {Wei}, \citenamefont {Zhao}, \citenamefont
  {Zhao}, \citenamefont {Zhang}, \citenamefont {Yumigeta}, \citenamefont
  {Blei}, \citenamefont {Carlstr{\"o}m}, \citenamefont {Watanabe},
  \citenamefont {Taniguchi}, \citenamefont {Tongay}, \citenamefont {Crommie},
  \citenamefont {Zettl},\ and\ \citenamefont {Wang}}]{Regan_2020}%
  \BibitemOpen
  \bibfield  {author} {\bibinfo {author} {\bibfnamefont {E.~C.}\ \bibnamefont
  {Regan}}, \bibinfo {author} {\bibfnamefont {D.}~\bibnamefont {Wang}},
  \bibinfo {author} {\bibfnamefont {C.}~\bibnamefont {Jin}}, \bibinfo {author}
  {\bibfnamefont {M.~I.}\ \bibnamefont {Bakti~Utama}}, \bibinfo {author}
  {\bibfnamefont {B.}~\bibnamefont {Gao}}, \bibinfo {author} {\bibfnamefont
  {X.}~\bibnamefont {Wei}}, \bibinfo {author} {\bibfnamefont {S.}~\bibnamefont
  {Zhao}}, \bibinfo {author} {\bibfnamefont {W.}~\bibnamefont {Zhao}}, \bibinfo
  {author} {\bibfnamefont {Z.}~\bibnamefont {Zhang}}, \bibinfo {author}
  {\bibfnamefont {K.}~\bibnamefont {Yumigeta}}, \bibinfo {author}
  {\bibfnamefont {M.}~\bibnamefont {Blei}}, \bibinfo {author} {\bibfnamefont
  {J.~D.}\ \bibnamefont {Carlstr{\"o}m}}, \bibinfo {author} {\bibfnamefont
  {K.}~\bibnamefont {Watanabe}}, \bibinfo {author} {\bibfnamefont
  {T.}~\bibnamefont {Taniguchi}}, \bibinfo {author} {\bibfnamefont
  {S.}~\bibnamefont {Tongay}}, \bibinfo {author} {\bibfnamefont
  {M.}~\bibnamefont {Crommie}}, \bibinfo {author} {\bibfnamefont
  {A.}~\bibnamefont {Zettl}}, \ and\ \bibinfo {author} {\bibfnamefont
  {F.}~\bibnamefont {Wang}},\ }\href {\doibase 10.1038/s41586-020-2092-4}
  {\bibfield  {journal} {\bibinfo  {journal} {Nature}\ }\textbf {\bibinfo
  {volume} {579}},\ \bibinfo {pages} {359} (\bibinfo {year}
  {2020})}\BibitemShut {NoStop}%
\bibitem [{\citenamefont {Zhang}\ \emph {et~al.}(2020)\citenamefont {Zhang},
  \citenamefont {Wang}, \citenamefont {Watanabe}, \citenamefont {Taniguchi},
  \citenamefont {Ueno}, \citenamefont {Tutuc},\ and\ \citenamefont
  {LeRoy}}]{Zhang_2020}%
  \BibitemOpen
  \bibfield  {author} {\bibinfo {author} {\bibfnamefont {Z.}~\bibnamefont
  {Zhang}}, \bibinfo {author} {\bibfnamefont {Y.}~\bibnamefont {Wang}},
  \bibinfo {author} {\bibfnamefont {K.}~\bibnamefont {Watanabe}}, \bibinfo
  {author} {\bibfnamefont {T.}~\bibnamefont {Taniguchi}}, \bibinfo {author}
  {\bibfnamefont {K.}~\bibnamefont {Ueno}}, \bibinfo {author} {\bibfnamefont
  {E.}~\bibnamefont {Tutuc}}, \ and\ \bibinfo {author} {\bibfnamefont {B.~J.}\
  \bibnamefont {LeRoy}},\ }\href {\doibase 10.1038/s41567-020-0958-x}
  {\bibfield  {journal} {\bibinfo  {journal} {Nature Physics}\ }\textbf
  {\bibinfo {volume} {16}},\ \bibinfo {pages} {1093} (\bibinfo {year}
  {2020})}\BibitemShut {NoStop}%
\bibitem [{\citenamefont {Wang}\ \emph {et~al.}(2020)\citenamefont {Wang},
  \citenamefont {Shih}, \citenamefont {Ghiotto}, \citenamefont {Xian},
  \citenamefont {Rhodes}, \citenamefont {Tan}, \citenamefont {Claassen},
  \citenamefont {Kennes}, \citenamefont {Bai}, \citenamefont {Kim},
  \citenamefont {Watanabe}, \citenamefont {Taniguchi}, \citenamefont {Zhu},
  \citenamefont {Hone}, \citenamefont {Rubio}, \citenamefont {Pasupathy},\ and\
  \citenamefont {Dean}}]{Wang_2020}%
  \BibitemOpen
  \bibfield  {author} {\bibinfo {author} {\bibfnamefont {L.}~\bibnamefont
  {Wang}}, \bibinfo {author} {\bibfnamefont {E.-M.}\ \bibnamefont {Shih}},
  \bibinfo {author} {\bibfnamefont {A.}~\bibnamefont {Ghiotto}}, \bibinfo
  {author} {\bibfnamefont {L.}~\bibnamefont {Xian}}, \bibinfo {author}
  {\bibfnamefont {D.~A.}\ \bibnamefont {Rhodes}}, \bibinfo {author}
  {\bibfnamefont {C.}~\bibnamefont {Tan}}, \bibinfo {author} {\bibfnamefont
  {M.}~\bibnamefont {Claassen}}, \bibinfo {author} {\bibfnamefont {D.~M.}\
  \bibnamefont {Kennes}}, \bibinfo {author} {\bibfnamefont {Y.}~\bibnamefont
  {Bai}}, \bibinfo {author} {\bibfnamefont {B.}~\bibnamefont {Kim}}, \bibinfo
  {author} {\bibfnamefont {K.}~\bibnamefont {Watanabe}}, \bibinfo {author}
  {\bibfnamefont {T.}~\bibnamefont {Taniguchi}}, \bibinfo {author}
  {\bibfnamefont {X.}~\bibnamefont {Zhu}}, \bibinfo {author} {\bibfnamefont
  {J.}~\bibnamefont {Hone}}, \bibinfo {author} {\bibfnamefont {A.}~\bibnamefont
  {Rubio}}, \bibinfo {author} {\bibfnamefont {A.~N.}\ \bibnamefont
  {Pasupathy}}, \ and\ \bibinfo {author} {\bibfnamefont {C.~R.}\ \bibnamefont
  {Dean}},\ }\href {\doibase 10.1038/s41563-020-0708-6} {\bibfield  {journal}
  {\bibinfo  {journal} {Nature Materials}\ }\textbf {\bibinfo {volume} {19}},\
  \bibinfo {pages} {861} (\bibinfo {year} {2020})}\BibitemShut {NoStop}%
\bibitem [{\citenamefont {Pan}\ \emph {et~al.}(2020)\citenamefont {Pan},
  \citenamefont {Wu},\ and\ \citenamefont {Das~Sarma}}]{Pan_2020}%
  \BibitemOpen
  \bibfield  {author} {\bibinfo {author} {\bibfnamefont {H.}~\bibnamefont
  {Pan}}, \bibinfo {author} {\bibfnamefont {F.}~\bibnamefont {Wu}}, \ and\
  \bibinfo {author} {\bibfnamefont {S.}~\bibnamefont {Das~Sarma}},\ }\href
  {\doibase 10.1103/PhysRevResearch.2.033087} {\bibfield  {journal} {\bibinfo
  {journal} {Phys. Rev. Research}\ }\textbf {\bibinfo {volume} {2}},\ \bibinfo
  {pages} {033087} (\bibinfo {year} {2020})}\BibitemShut {NoStop}%
\bibitem [{\citenamefont {Kane}\ and\ \citenamefont {Mele}(2005)}]{Kane_2005}%
  \BibitemOpen
  \bibfield  {author} {\bibinfo {author} {\bibfnamefont {C.~L.}\ \bibnamefont
  {Kane}}\ and\ \bibinfo {author} {\bibfnamefont {E.~J.}\ \bibnamefont
  {Mele}},\ }\href {\doibase 10.1103/PhysRevLett.95.226801} {\bibfield
  {journal} {\bibinfo  {journal} {Phys. Rev. Lett.}\ }\textbf {\bibinfo
  {volume} {95}},\ \bibinfo {pages} {226801} (\bibinfo {year}
  {2005})}\BibitemShut {NoStop}%
\bibitem [{\citenamefont {Devakul}\ \emph {et~al.}(2021)\citenamefont
  {Devakul}, \citenamefont {Crépel}, \citenamefont {Zhang},\ and\
  \citenamefont {Fu}}]{devakul_2021}%
  \BibitemOpen
  \bibfield  {author} {\bibinfo {author} {\bibfnamefont {T.}~\bibnamefont
  {Devakul}}, \bibinfo {author} {\bibfnamefont {V.}~\bibnamefont {Crépel}},
  \bibinfo {author} {\bibfnamefont {Y.}~\bibnamefont {Zhang}}, \ and\ \bibinfo
  {author} {\bibfnamefont {L.}~\bibnamefont {Fu}},\ }\href@noop {} {} (\bibinfo
  {year} {2021}),\ \Eprint {http://arxiv.org/abs/2106.11954} {arXiv:2106.11954}
  \BibitemShut {NoStop}%
\bibitem [{\citenamefont {Po}\ \emph {et~al.}(2018)\citenamefont {Po},
  \citenamefont {Zou}, \citenamefont {Vishwanath},\ and\ \citenamefont
  {Senthil}}]{Po_2018}%
  \BibitemOpen
  \bibfield  {author} {\bibinfo {author} {\bibfnamefont {H.~C.}\ \bibnamefont
  {Po}}, \bibinfo {author} {\bibfnamefont {L.}~\bibnamefont {Zou}}, \bibinfo
  {author} {\bibfnamefont {A.}~\bibnamefont {Vishwanath}}, \ and\ \bibinfo
  {author} {\bibfnamefont {T.}~\bibnamefont {Senthil}},\ }\href {\doibase
  10.1103/PhysRevX.8.031089} {\bibfield  {journal} {\bibinfo  {journal} {Phys.
  Rev. X}\ }\textbf {\bibinfo {volume} {8}},\ \bibinfo {pages} {031089}
  (\bibinfo {year} {2018})}\BibitemShut {NoStop}%
\bibitem [{\citenamefont {Zhang}\ and\ \citenamefont
  {Senthil}(2019)}]{Zhang_2019}%
  \BibitemOpen
  \bibfield  {author} {\bibinfo {author} {\bibfnamefont {Y.-H.}\ \bibnamefont
  {Zhang}}\ and\ \bibinfo {author} {\bibfnamefont {T.}~\bibnamefont
  {Senthil}},\ }\href {\doibase 10.1103/PhysRevB.99.205150} {\bibfield
  {journal} {\bibinfo  {journal} {Phys. Rev. B}\ }\textbf {\bibinfo {volume}
  {99}},\ \bibinfo {pages} {205150} (\bibinfo {year} {2019})}\BibitemShut
  {NoStop}%
\bibitem [{\citenamefont {Yankowitz}\ \emph {et~al.}(2019)\citenamefont
  {Yankowitz}, \citenamefont {Chen}, \citenamefont {Polshyn}, \citenamefont
  {Zhang}, \citenamefont {Watanabe}, \citenamefont {Taniguchi}, \citenamefont
  {Graf}, \citenamefont {Young},\ and\ \citenamefont {Dean}}]{Yankowitz_2019}%
  \BibitemOpen
  \bibfield  {author} {\bibinfo {author} {\bibfnamefont {M.}~\bibnamefont
  {Yankowitz}}, \bibinfo {author} {\bibfnamefont {S.}~\bibnamefont {Chen}},
  \bibinfo {author} {\bibfnamefont {H.}~\bibnamefont {Polshyn}}, \bibinfo
  {author} {\bibfnamefont {Y.}~\bibnamefont {Zhang}}, \bibinfo {author}
  {\bibfnamefont {K.}~\bibnamefont {Watanabe}}, \bibinfo {author}
  {\bibfnamefont {T.}~\bibnamefont {Taniguchi}}, \bibinfo {author}
  {\bibfnamefont {D.}~\bibnamefont {Graf}}, \bibinfo {author} {\bibfnamefont
  {A.~F.}\ \bibnamefont {Young}}, \ and\ \bibinfo {author} {\bibfnamefont
  {C.~R.}\ \bibnamefont {Dean}},\ }\href {\doibase 10.1126/science.aav1910}
  {\bibfield  {journal} {\bibinfo  {journal} {Science}\ }\textbf {\bibinfo
  {volume} {363}},\ \bibinfo {pages} {1059} (\bibinfo {year}
  {2019})}\BibitemShut {NoStop}%
\bibitem [{\citenamefont {Zare}\ \emph {et~al.}(2013)\citenamefont {Zare},
  \citenamefont {Fazileh},\ and\ \citenamefont {Shahbazi}}]{Zare_2013}%
  \BibitemOpen
  \bibfield  {author} {\bibinfo {author} {\bibfnamefont {M.~H.}\ \bibnamefont
  {Zare}}, \bibinfo {author} {\bibfnamefont {F.}~\bibnamefont {Fazileh}}, \
  and\ \bibinfo {author} {\bibfnamefont {F.}~\bibnamefont {Shahbazi}},\ }\href
  {\doibase 10.1103/PhysRevB.87.224416} {\bibfield  {journal} {\bibinfo
  {journal} {Phys. Rev. B}\ }\textbf {\bibinfo {volume} {87}},\ \bibinfo
  {pages} {224416} (\bibinfo {year} {2013})}\BibitemShut {NoStop}%
\bibitem [{\citenamefont {Zare}\ \emph {et~al.}(2014)\citenamefont {Zare},
  \citenamefont {Mosadeq}, \citenamefont {Shahbazi},\ and\ \citenamefont
  {Jafari}}]{Zare_2014}%
  \BibitemOpen
  \bibfield  {author} {\bibinfo {author} {\bibfnamefont {M.~H.}\ \bibnamefont
  {Zare}}, \bibinfo {author} {\bibfnamefont {H.}~\bibnamefont {Mosadeq}},
  \bibinfo {author} {\bibfnamefont {F.}~\bibnamefont {Shahbazi}}, \ and\
  \bibinfo {author} {\bibfnamefont {S.~A.}\ \bibnamefont {Jafari}},\ }\href
  {http://stacks.iop.org/0953-8984/26/i=45/a=456004} {\bibfield  {journal}
  {\bibinfo  {journal} {J. Phys: Conden Matter}\ }\textbf {\bibinfo {volume}
  {26}},\ \bibinfo {pages} {456004} (\bibinfo {year} {2014})}\BibitemShut
  {NoStop}%
\bibitem [{\citenamefont {Luttinger}\ and\ \citenamefont
  {Tisza}(1946)}]{Luttinger_1946}%
  \BibitemOpen
  \bibfield  {author} {\bibinfo {author} {\bibfnamefont {J.~M.}\ \bibnamefont
  {Luttinger}}\ and\ \bibinfo {author} {\bibfnamefont {L.}~\bibnamefont
  {Tisza}},\ }\href {\doibase 10.1103/PhysRev.70.954} {\bibfield  {journal}
  {\bibinfo  {journal} {Phys. Rev.}\ }\textbf {\bibinfo {volume} {70}},\
  \bibinfo {pages} {954} (\bibinfo {year} {1946})}\BibitemShut {NoStop}%
\bibitem [{\citenamefont {Litvin}(1974)}]{Litvin_1974}%
  \BibitemOpen
  \bibfield  {author} {\bibinfo {author} {\bibfnamefont {D.}~\bibnamefont
  {Litvin}},\ }\href {\doibase https://doi.org/10.1016/0031-8914(74)90257-2}
  {\bibfield  {journal} {\bibinfo  {journal} {Physica}\ }\textbf {\bibinfo
  {volume} {77}},\ \bibinfo {pages} {205 } (\bibinfo {year}
  {1974})}\BibitemShut {NoStop}%
\bibitem [{\citenamefont {Li}\ \emph {et~al.}(2013)\citenamefont {Li},
  \citenamefont {Cao}, \citenamefont {Niu}, \citenamefont {Shi},\ and\
  \citenamefont {Feng}}]{Li_2013}%
  \BibitemOpen
  \bibfield  {author} {\bibinfo {author} {\bibfnamefont {X.}~\bibnamefont
  {Li}}, \bibinfo {author} {\bibfnamefont {T.}~\bibnamefont {Cao}}, \bibinfo
  {author} {\bibfnamefont {Q.}~\bibnamefont {Niu}}, \bibinfo {author}
  {\bibfnamefont {J.}~\bibnamefont {Shi}}, \ and\ \bibinfo {author}
  {\bibfnamefont {J.}~\bibnamefont {Feng}},\ }\href {\doibase
  10.1073/pnas.1219420110} {\bibfield  {journal} {\bibinfo  {journal}
  {Proceedings of the National Academy of Sciences}\ }\textbf {\bibinfo
  {volume} {110}},\ \bibinfo {pages} {3738} (\bibinfo {year}
  {2013})}\BibitemShut {NoStop}%
\bibitem [{\citenamefont {Sivadas}\ \emph {et~al.}(2015)\citenamefont
  {Sivadas}, \citenamefont {Daniels}, \citenamefont {Swendsen}, \citenamefont
  {Okamoto},\ and\ \citenamefont {Xiao}}]{Sivadas_2015}%
  \BibitemOpen
  \bibfield  {author} {\bibinfo {author} {\bibfnamefont {N.}~\bibnamefont
  {Sivadas}}, \bibinfo {author} {\bibfnamefont {M.~W.}\ \bibnamefont
  {Daniels}}, \bibinfo {author} {\bibfnamefont {R.~H.}\ \bibnamefont
  {Swendsen}}, \bibinfo {author} {\bibfnamefont {S.}~\bibnamefont {Okamoto}}, \
  and\ \bibinfo {author} {\bibfnamefont {D.}~\bibnamefont {Xiao}},\ }\href
  {\doibase 10.1103/PhysRevB.91.235425} {\bibfield  {journal} {\bibinfo
  {journal} {Phys. Rev. B}\ }\textbf {\bibinfo {volume} {91}},\ \bibinfo
  {pages} {235425} (\bibinfo {year} {2015})}\BibitemShut {NoStop}%
\bibitem [{\citenamefont {Liu}\ \emph {et~al.}(2016)\citenamefont {Liu},
  \citenamefont {Yu},\ and\ \citenamefont {Wang}}]{Liu_2016a}%
  \BibitemOpen
  \bibfield  {author} {\bibinfo {author} {\bibfnamefont {C.}~\bibnamefont
  {Liu}}, \bibinfo {author} {\bibfnamefont {R.}~\bibnamefont {Yu}}, \ and\
  \bibinfo {author} {\bibfnamefont {X.}~\bibnamefont {Wang}},\ }\href {\doibase
  10.1103/PhysRevB.94.174424} {\bibfield  {journal} {\bibinfo  {journal} {Phys.
  Rev. B}\ }\textbf {\bibinfo {volume} {94}},\ \bibinfo {pages} {174424}
  (\bibinfo {year} {2016})}\BibitemShut {NoStop}%
\bibitem [{\citenamefont {Inc.}()}]{Mathematica}%
  \BibitemOpen
  \bibfield  {author} {\bibinfo {author} {\bibfnamefont {W.~R.}\ \bibnamefont
  {Inc.}},\ }\href {https://www.wolfram.com/mathematica} {\enquote {\bibinfo
  {title} {Mathematica, {V}ersion 12.2},}\ }\bibinfo {note} {Champaign, IL,
  2020}\BibitemShut {NoStop}%
\bibitem [{\citenamefont {Wolf}\ \emph {et~al.}(2021)\citenamefont {Wolf},
  \citenamefont {Zilberberg}, \citenamefont {Blatter},\ and\ \citenamefont
  {Lado}}]{Wolf_2021}%
  \BibitemOpen
  \bibfield  {author} {\bibinfo {author} {\bibfnamefont {T.~M.~R.}\
  \bibnamefont {Wolf}}, \bibinfo {author} {\bibfnamefont {O.}~\bibnamefont
  {Zilberberg}}, \bibinfo {author} {\bibfnamefont {G.}~\bibnamefont {Blatter}},
  \ and\ \bibinfo {author} {\bibfnamefont {J.~L.}\ \bibnamefont {Lado}},\
  }\href {\doibase 10.1103/PhysRevLett.126.056803} {\bibfield  {journal}
  {\bibinfo  {journal} {Phys. Rev. Lett.}\ }\textbf {\bibinfo {volume} {126}},\
  \bibinfo {pages} {056803} (\bibinfo {year} {2021})}\BibitemShut {NoStop}%
\bibitem [{\citenamefont {Chubukov}(1991)}]{Chubukov_1991}%
  \BibitemOpen
  \bibfield  {author} {\bibinfo {author} {\bibfnamefont {A.}~\bibnamefont
  {Chubukov}},\ }\href {\doibase 10.1103/PhysRevB.44.392} {\bibfield  {journal}
  {\bibinfo  {journal} {Phys. Rev. B}\ }\textbf {\bibinfo {volume} {44}},\
  \bibinfo {pages} {392} (\bibinfo {year} {1991})}\BibitemShut {NoStop}%
\bibitem [{\citenamefont {Maksimov}\ \emph {et~al.}(2019)\citenamefont
  {Maksimov}, \citenamefont {Zhu}, \citenamefont {White},\ and\ \citenamefont
  {Chernyshev}}]{Maksimov_2019}%
  \BibitemOpen
  \bibfield  {author} {\bibinfo {author} {\bibfnamefont {P.~A.}\ \bibnamefont
  {Maksimov}}, \bibinfo {author} {\bibfnamefont {Z.}~\bibnamefont {Zhu}},
  \bibinfo {author} {\bibfnamefont {S.~R.}\ \bibnamefont {White}}, \ and\
  \bibinfo {author} {\bibfnamefont {A.~L.}\ \bibnamefont {Chernyshev}},\ }\href
  {\doibase 10.1103/PhysRevX.9.021017} {\bibfield  {journal} {\bibinfo
  {journal} {Phys. Rev. X}\ }\textbf {\bibinfo {volume} {9}},\ \bibinfo {pages}
  {021017} (\bibinfo {year} {2019})}\BibitemShut {NoStop}%
\bibitem [{\citenamefont {Mucciolo}\ \emph {et~al.}(2004)\citenamefont
  {Mucciolo}, \citenamefont {Castro~Neto},\ and\ \citenamefont
  {Chamon}}]{Mucciolo_2004}%
  \BibitemOpen
  \bibfield  {author} {\bibinfo {author} {\bibfnamefont {E.~R.}\ \bibnamefont
  {Mucciolo}}, \bibinfo {author} {\bibfnamefont {A.~H.}\ \bibnamefont
  {Castro~Neto}}, \ and\ \bibinfo {author} {\bibfnamefont {C.}~\bibnamefont
  {Chamon}},\ }\href {\doibase 10.1103/PhysRevB.69.214424} {\bibfield
  {journal} {\bibinfo  {journal} {Phys. Rev. B}\ }\textbf {\bibinfo {volume}
  {69}},\ \bibinfo {pages} {214424} (\bibinfo {year} {2004})}\BibitemShut
  {NoStop}%
\bibitem [{\citenamefont {Bauer}\ \emph {et~al.}(2011)\citenamefont {Bauer},
  \citenamefont {Carr}, \citenamefont {Evertz}, \citenamefont {Feiguin},
  \citenamefont {Freire}, \citenamefont {Fuchs}, \citenamefont {Gamper},
  \citenamefont {Gukelberger}, \citenamefont {Gull}, \citenamefont {Guertler},
  \citenamefont {Hehn}, \citenamefont {Igarashi}, \citenamefont {Isakov},
  \citenamefont {Koop}, \citenamefont {Ma}, \citenamefont {Mates},
  \citenamefont {Matsuo}, \citenamefont {Parcollet}, \citenamefont
  {Paw{\l}owski}, \citenamefont {Picon}, \citenamefont {Pollet}, \citenamefont
  {Santos}, \citenamefont {Scarola}, \citenamefont {Schollwöck}, \citenamefont
  {Silva}, \citenamefont {Surer}, \citenamefont {Todo}, \citenamefont {Trebst},
  \citenamefont {Troyer}, \citenamefont {Wall}, \citenamefont {Werner},\ and\
  \citenamefont {Wessel}}]{Bauer_2011}%
  \BibitemOpen
  \bibfield  {author} {\bibinfo {author} {\bibfnamefont {B.}~\bibnamefont
  {Bauer}}, \bibinfo {author} {\bibfnamefont {L.~D.}\ \bibnamefont {Carr}},
  \bibinfo {author} {\bibfnamefont {H.~G.}\ \bibnamefont {Evertz}}, \bibinfo
  {author} {\bibfnamefont {A.}~\bibnamefont {Feiguin}}, \bibinfo {author}
  {\bibfnamefont {J.}~\bibnamefont {Freire}}, \bibinfo {author} {\bibfnamefont
  {S.}~\bibnamefont {Fuchs}}, \bibinfo {author} {\bibfnamefont
  {L.}~\bibnamefont {Gamper}}, \bibinfo {author} {\bibfnamefont
  {J.}~\bibnamefont {Gukelberger}}, \bibinfo {author} {\bibfnamefont
  {E.}~\bibnamefont {Gull}}, \bibinfo {author} {\bibfnamefont {S.}~\bibnamefont
  {Guertler}}, \bibinfo {author} {\bibfnamefont {A.}~\bibnamefont {Hehn}},
  \bibinfo {author} {\bibfnamefont {R.}~\bibnamefont {Igarashi}}, \bibinfo
  {author} {\bibfnamefont {S.~V.}\ \bibnamefont {Isakov}}, \bibinfo {author}
  {\bibfnamefont {D.}~\bibnamefont {Koop}}, \bibinfo {author} {\bibfnamefont
  {P.~N.}\ \bibnamefont {Ma}}, \bibinfo {author} {\bibfnamefont
  {P.}~\bibnamefont {Mates}}, \bibinfo {author} {\bibfnamefont
  {H.}~\bibnamefont {Matsuo}}, \bibinfo {author} {\bibfnamefont
  {O.}~\bibnamefont {Parcollet}}, \bibinfo {author} {\bibfnamefont
  {G.}~\bibnamefont {Paw{\l}owski}}, \bibinfo {author} {\bibfnamefont {J.~D.}\
  \bibnamefont {Picon}}, \bibinfo {author} {\bibfnamefont {L.}~\bibnamefont
  {Pollet}}, \bibinfo {author} {\bibfnamefont {E.}~\bibnamefont {Santos}},
  \bibinfo {author} {\bibfnamefont {V.~W.}\ \bibnamefont {Scarola}}, \bibinfo
  {author} {\bibfnamefont {U.}~\bibnamefont {Schollwöck}}, \bibinfo {author}
  {\bibfnamefont {C.}~\bibnamefont {Silva}}, \bibinfo {author} {\bibfnamefont
  {B.}~\bibnamefont {Surer}}, \bibinfo {author} {\bibfnamefont
  {S.}~\bibnamefont {Todo}}, \bibinfo {author} {\bibfnamefont {S.}~\bibnamefont
  {Trebst}}, \bibinfo {author} {\bibfnamefont {M.}~\bibnamefont {Troyer}},
  \bibinfo {author} {\bibfnamefont {M.~L.}\ \bibnamefont {Wall}}, \bibinfo
  {author} {\bibfnamefont {P.}~\bibnamefont {Werner}}, \ and\ \bibinfo {author}
  {\bibfnamefont {S.}~\bibnamefont {Wessel}},\ }\href {\doibase
  10.1088/1742-5468/2011/05/p05001} {\bibfield  {journal} {\bibinfo  {journal}
  {Journal of Statistical Mechanics: Theory and Experiment}\ }\textbf {\bibinfo
  {volume} {2011}},\ \bibinfo {pages} {P05001} (\bibinfo {year}
  {2011})}\BibitemShut {NoStop}%
\bibitem [{\citenamefont {Thesberg}\ and\ \citenamefont
  {S\o{}rensen}(2014)}]{Thesberg_2014}%
  \BibitemOpen
  \bibfield  {author} {\bibinfo {author} {\bibfnamefont {M.}~\bibnamefont
  {Thesberg}}\ and\ \bibinfo {author} {\bibfnamefont {E.~S.}\ \bibnamefont
  {S\o{}rensen}},\ }\href {\doibase 10.1103/PhysRevB.90.115117} {\bibfield
  {journal} {\bibinfo  {journal} {Phys. Rev. B}\ }\textbf {\bibinfo {volume}
  {90}},\ \bibinfo {pages} {115117} (\bibinfo {year} {2014})}\BibitemShut
  {NoStop}%
\bibitem [{\citenamefont {Mosadeq}\ \emph {et~al.}(2011)\citenamefont
  {Mosadeq}, \citenamefont {Shahbazi},\ and\ \citenamefont
  {Jafari}}]{Mosadeq_2011}%
  \BibitemOpen
  \bibfield  {author} {\bibinfo {author} {\bibfnamefont {H.}~\bibnamefont
  {Mosadeq}}, \bibinfo {author} {\bibfnamefont {F.}~\bibnamefont {Shahbazi}}, \
  and\ \bibinfo {author} {\bibfnamefont {S.~A.}\ \bibnamefont {Jafari}},\
  }\href {\doibase 10.1088/0953-8984/23/22/226006} {\bibfield  {journal}
  {\bibinfo  {journal} {Journal of Physics: Condensed Matter}\ }\textbf
  {\bibinfo {volume} {23}},\ \bibinfo {pages} {226006} (\bibinfo {year}
  {2011})}\BibitemShut {NoStop}%
\bibitem [{\citenamefont {Zhang}\ \emph {et~al.}(2021)\citenamefont {Zhang},
  \citenamefont {Devakul},\ and\ \citenamefont {Fu}}]{zhang_2021}%
  \BibitemOpen
  \bibfield  {author} {\bibinfo {author} {\bibfnamefont {Y.}~\bibnamefont
  {Zhang}}, \bibinfo {author} {\bibfnamefont {T.}~\bibnamefont {Devakul}}, \
  and\ \bibinfo {author} {\bibfnamefont {L.}~\bibnamefont {Fu}},\ }\href@noop
  {} {} (\bibinfo {year} {2021}),\ \Eprint {http://arxiv.org/abs/2107.02167}
  {arXiv:2107.02167} \BibitemShut {NoStop}%
\end{thebibliography}%
	%%%%%%%%%%%%%%%%%%%%%%%%%%%%%%%%%%%%%%
	%%%%%%%%%%%%%%%%%%%%%%%%%%%%%%%%%%%%%%

	%\bibliography{main}
\end{document}